\newif\ifmnras
\mnrastrue
\ifmnras
	\documentclass[a4paper,fleqn,usenatbib]{mnras}
\else
	\documentclass[apj,numberedappendix]{emulateapj}
\fi

\usepackage{graphics,epsf}
\usepackage{amsmath}                
\usepackage{amsfonts}               
\usepackage{amssymb}                
\usepackage{epsfig}                 
\usepackage{graphicx}
\usepackage{adjustbox}
\usepackage{subfig}
\usepackage{booktabs}
\usepackage{xfrac}

\usepackage{tikz}
\usepackage{pgfplots}
\usepackage{pgfplotstable}
\pgfplotsset{compat=1.16} 

\usepackage{xcolor}
\definecolor{plot1}{HTML}{003f5c}
\definecolor{plot2}{HTML}{7a5195}
\definecolor{plot3}{HTML}{ef5675}
\definecolor{plot4}{HTML}{ffa600}

\def \cm{~\rm{cm}}

\def \g{~\rm{g}}

\def \msun{{M$_\odot$}}
\def \octo{{\sc Octo-Tiger}}

\def \flower{{\sc Flow-er}}
\def \snsph{{\sc SNSPH}}

\ifmnras
	\def \aap{A\&A}
	
	\def \apj{ApJ}
	\def \apjl{ApJ}
	\def \apjs{ApJS}
	\def \jaavso{JAAVSO}

	\def \mnras{MNRAS}

	\def \prd{PhRvD}
	
\fi

\usepackage{xcolor}
\definecolor{redak}{rgb}{0.9,0.15,0.05}

\usepackage{tikz}

\newcommand{\revisionII}[1]{\textcolor{black}{#1}}

\ifmnras
	\title[WD mergers and the origin of RCB stars]{Hydrodynamic simulations of white dwarf-white dwarf mergers and the origin of R Coronae Borealis stars} 

	\author[Shiber et al.]{Sagiv Shiber$^{1}$\thanks{E-mail: \href{sshiber1@lsu.edu}{sshiber1@lsu.edu}}, Orsola De Marco$^{2, 3}$, Patrick M. Motl$^4$, Bradley Munson$^1$,  \newauthor Dominic C. Marcello$^{1,5}$, Juhan Frank$^1$, Patrick Diehl$^{1,5,6}$, Geoffrey C. Clayton$^1$, \newauthor 
	Bennett N. Skinner$^{7,8,9}$, Hartmut Kaiser$^{5,10}$, Gregor Dai\ss$^{5,11}$, Dirk Pfl\"uger$^{11}$,
 \newauthor
 and Jan E. Staff$^{12}$\thanks{We are saddened to report the passing of our esteemed colleague and friend, Jan Staff.} \\
\\
$^{1}$Department of Physics and Astronomy, Louisiana State University, Baton Rouge, LA, 70803 USA \\
$^{2}$School of Mathematical and Physical Sciences, Macquarie University, Sydney, NSW 2109, Australia \\
$^{3}$Astronomy, Astrophysics and Astrophotonics Research Centre, Macquarie University, Sydney, NSW 2109, Australia \\
$^{4}$The School of Sciences, Indiana University Kokomo, Kokomo, Indiana 46902, USA\\
$^{5}$LSU Center for Computation \& Technology, Louisiana State University, Baton Rouge, LA, 70803 USA\\
$^{6}$Applied Computer Science (CCS-7), Los Alamos National Laboratory, Los Alamos, NM 87545
$^{7}$NSF-supported REU Student at LSU\\
$^{7}$IPVS, University of Stuttgart,
Stuttgart, 70174 Stuttgart, Germany\\
$^{9}$Division of Computer Science and Engineering, Louisiana State University, Baton Rouge, LA, 70803 USA \\
$^{6}$Applied Computer Science (CCS-7), Los Alamos National Laboratory, Los Alamos, NM 87545 \\
$^{12}$Department of Space, Earth and Environment, Chalmers University of Technology, SE-41296 Gothenburg, Sweden\\
}
	\date{Accepted XXX. Received YYY; in original form ZZZ}

	\pubyear{2024}
\fi

\begin{document}
\label{firstpage}

\ifmnras
	\pagerange{\pageref{firstpage}--\pageref{lastpage}}
	\maketitle
\fi

\begin{abstract}
We study the properties of double white dwarf (DWD) mergers by performing hydrodynamic simulations using the new and improved adaptive mesh refinement code \octo. We follow the orbital evolution of DWD systems of mass ratio $q=0.7$ for tens of orbits until and after the merger to investigate them as a possible origin for R Coronae Borealis (RCB) type stars.   
We reproduce previous results, finding that during the merger, the helium WD donor star is tidally disrupted {within 20-80 minutes since the beginning of the simulation} onto the accretor carbon-oxygen WD, creating a high temperature shell around the accretor. We investigate the possible helium burning in this shell and the merged object's general structure. Specifically, we are interested in the amount of oxygen-16 dredged-up \revisionII{from the accretor} to the hot shell and the amount of oxygen-18 produced. This is critical as the discovery of very low oxygen-16 to oxygen-18 ratios in RCB stars pointed out the merger scenario as a favorable explanation for their origin. 
A small amount of hydrogen in the donor may help keep the oxygen-16 to oxygen-18 ratios within observational bounds, even if moderate dredge-up {from the accretor occurs}.
In addition, we perform a resolution study to reconcile the difference found in the amount of oxygen-16 dredge-up between smoothed-particle hydrodynamics and grid-based simulations.
\end{abstract}


\section{Introduction}
\label{sec:intro}

{R Coronae Borealis (RCB)} stars are low-mass, hydrogen-deficient, { carbon-rich} giants, primarily made of helium
\citep{2012JAVSO..40..539C}. They are almost indistinguishable from a second class of stars, known as the hydrogen deficient, carbon (HdC) stars, with the exception that RCB stars are known to exhibit irregular and dramatic light variability in the form of deep declines that can leave the star at minimum for years before recovery is observed \citep{2022A&A...667A..83T,2023MNRAS.521.1674C}. 

For a long time, two scenarios were considered for their formation, a final helium shell flash in a single post-AGB star, or a merger of two WDs {(e.g., \citealt{Iben1984}; \citealt{2012JAVSO..40..539C})}. The discovery of very { low $^{16}$O/$^{18}$O ratios} \citep{2007ApJ...662.1220C,2022A&A...667A..84K}, \revisionII{much less than the solar value 
$\sim 500$,} along with the {likely average mass of $\sim~$0.9~\msun\ for these stars} \citep[][]{1998MNRAS.296.1019H,2015ApJ...809..184K}{} effectively eliminated the final flash scenario in favor of the merger one. Merging WDs are known to create $^{18}$O, which, under certain circumstances, can be brought to the surface, making the $^{16}$O/$^{18}$O ratio as low as unity \citep[e.g.,][]{2020MNRAS.498.2912C,munson21}. 

To determine whether these observations could be understood in the context of a WD merger, \citet{Staff2012} simulated WD-WD mergers of different mass ratios, $q=M_2/M_1=0.5-1.0$ with the hydrodynamic code \flower\ 
\citep{D_Souza_2006, Motl2007}. 
In that paper, it was found that the correct oxygen ratio could only be achieved under specific conditions, and that mergers with mass ratio, $q=0.7$, are the most likely to produce this ratio.
{A similar enrichment of $^{18}$O has been found by \cite{Longland2011}, when they analyzed a merger of a $0.8~M_{\rm \odot}$ CO WD and a $0.4~M_{\rm \odot}$ He WD ($q=0.5$) that had been simulated using a smoothed-particle hydrodynamics (SPH) code.}
Later on, \citet{Staff2018} carried out several additional simulations, using, in addition to the unigrid technique used by \citet{Staff2012}, also an early version of the AMR technique code \octo\ \citep{Marcello2021}, and the SPH code \citep[SNSPH;][]{Fryer2006}. 
In that paper, they addressed the question of how much $^{16}$O is dredged up from the accretor, CO WD, out to the surface of the merged star. The problem with too much dredge-up is that no matter how much $^{18}$O is fused during the merger, its abundance will be greatly diluted by the dredged-up material. To reduce the dredge-up, \cite{Staff2018} also considered accretors that are hybrid WDs, having a thick shell of $>0.1$~M$_\odot$ of helium on top of the CO core. They concluded that while \flower\ and \octo\ agreed, \snsph\ produced a smaller amount of dredge up. Whether the accretor was a hybrid WD or not, had little effect on the results. The conclusion was that if \octo\ results were to be believed, we might be unable to reproduce the low $^{16}$O/$^{18}$O isotopic ratio observed.

Following the work by \citet{Staff2012}, \citet{Menon2013,2019MNRAS.482.2320M} carried out a study using a 1D implicit code to evolve a post-merger star and determine the abundance patterns. They used information from the hydrodynamics simulation in the form of conditions such as temperature in four radial zones. By applying a very specific mixing prescription, they could reproduce several observed characteristics of RCB star, including the $^{16}$O/$^{18}$O ratio. 

Later, \citet{2020MNRAS.498.2912C} and \citet{munson21} carried out very similar studies. The latter study was based on a later 3D \octo\ WD merger simulation with a mass ratio $q=0.6$, which they carried out specifically for their research. Instead of using a 4-zone system, as done by \citet{Menon2013}, they mapped the 3D simulation into the 1D implicit code, subject to a stabilization phase to bring the object into equilibrium. They observed that it is much harder to obtain the correct observed abundance values in this way.

In this paper, we will use our new and improved \octo\ \citep{Marcello2021} to calculate several simulations of the $q=0.7$ WD merger, similarly to what was done by \citet{Staff2018}. This mass ratio is not only appropriate for an RCB star, but by using the same mass ratio as done previously it allows a greater ability to compare simulations.
We carry out seven simulations, five \octo\ simulations, and two additional \flower\ simulations for code comparison. These simulations bracket in resolution the simulation of \citet{Staff2018} and allow us to study the convergence properties of the simulations and {the amount of $^{16}$O that is dredged up from the accretor}. {A second aim of this paper is to determine the temperature in the ``Shell of Fire" (SoF; \citealt{Staff2012}), the region in which partial helium burning can take place. This directly informs the amount of $^{18}$O that is generated in the merger.} 

\revisionII{Finally, it is worth emphasizing that the existence of short-period double WD is beyond doubt and confirmed by numerous observations. Two commonly accepted evolutionary scenarios for the origin of these systems have been extensively explored in the literature \citep{IbenTutu86, LipuPost88, TutuYung94, Lietal19}. For additional references to both observations and theory see \citet{Chenetal22}.  The most likely origin for the systems we consider in this paper is through mass transfer from a red giant onto a CO WD \citep{Lietal19}.}

In Section~\ref{sec:prev_sims} we discuss details of the simulations that have been carried out in the past with $q\sim 0.7$. In Section~\ref{sec:methods} we describe our simulations' setup. Section~\ref{sec:results} shows in detail the time evolution of our reference simulation and discusses the numerical properties of all of our simulations. We investigate in Section~\ref{sec:post_merger} the properties of the merged object such as its spin, the nuclear reactions, and the $^{16}$O dredge-up,
and finally, in Section~\ref{sec:concl}, we conclude.

\section{Previous Related Simulations}
\label{sec:prev_sims}

In this section, we discuss previous WD-WD merger simulations of mass ratio $q=0.7$, intending to integrate their results with ours. All previous comparable efforts have been summarized in the first part of Table~\ref{tab:past_sims}. 

\begin{table*}
 \caption{Overview of all considered Q0.7 simulations. {\sc L12} is the simulation with 12 levels of refinement; the suffix {\sc ND} means non-driven, and the suffix {\sc I} means that the simulation was conducted in the inertial instead of the rotating frame; Simulations prefixed with {\sc FL} are carried out with \flower. }
 \label{tab:past_sims}
 \centering
 \begin{tabular}{clllllllllll}
 \toprule
 Source & Code & $M_{\rm tot}$ & EoS & $T_{\rm drv}$ & $T_{\rm merge}$ & Resolution; $\Delta x_{\rm min}$ ($R_{\rm \odot}$)& donor res & $a_0$ & $P_0$ & Hybrid \\
 &&(\msun) &&($P_0$)&($P_0$)&(cells/particles)$^b$&(radial cells)& ($10^9$ cm) & (sec) & \\
 \midrule
 Staff 2012 & Flow-er & 0.9 & ZTWD & 2.0 & 10.2 & 226x146x256=8.4M & 150&?&?& No\\
 Staff 2018$^a$ & Octo &  0.89 & ZTWD & 4 & $\approx 22$& ?; $6.5\times10^{-4}$&?&3.48&?& No\\
 Staff 2018 & SNSPH &  0.9 &  ZTWD & 6?&?& 20M particles &8.2M &?&?& No \\
 Motl 2017 & Flow-er & 0.9 & poly & 2.3 & 9.7 & 162x98x256=4M & 70 &3.414&114.0&No\\
 Motl 2017 & Flow-er & 0.9 & poly & 1.7 & 21.0 & 162x98x256=4M & 58 &3.414&114.0& No\\
 Motl 2017 & SNSPH & 0.9 & poly &  1.0 & 11.5 &100k& 85k&3.410&116.0& No\\
 Motl 2017 & Flow-er & 0.9 & ideal & 2.3 & 12.4 & 162x98x256=4M & 70 &3.414&114.0& No\\
 Motl 2017 & SNSPH & 0.9 & ideal &  1.0 & 10.0 &100k& 85k &3.410&116.0& No \\
 Diehl 2021 & Octo & 0.85 & ideal & 2 & 6.7 &0.5M; $1.5\times10^{-3}$ & 27 &3.27&111.0& Yes\\
 Diehl 2021 & Octo & 0.85 & ideal & 2 & 13.4 &2.2M; $7.4\times10^{-3}$ & 54 &3.27&111.0&  Yes \\
 Diehl 2021 & Octo & 0.85 & ideal & 2 & 15.5 &3.1M; $3.7\times10^{-4}$ & 108 &3.27&111.0& Yes\\
 Diehl 2021 & Octo & 0.85 &  ideal & 2 & 16.8 &11.4M; $1.85\times10^{-4}$ & 215 &3.27& 111.0& Yes\\
 \midrule
 \multicolumn{10}{c}{This work}\\
 \midrule
 L10ND & Octo (Bd) & 0.9 & ZTWD &  0 & 36.8 & 1.7M; $1.2\times10^{-3}$ & 36 &3.453&116.0& Yes\\
 L11 & Octo (Bd) & 0.9 & ZTWD &  1.3 & 25.0 & 2.5M; $5.73\times10^{-4}$ & 73 &3.413&114.0& Yes\\
 L12 & Octo (Bd) & 0.9 & ZTWD &  1.3 & 38.7 & 5.3M; $2.86\times10^{-4}$& 145 &3.403&113.6& Yes\\
 L12I& Octo (Bd) & 0.9 & ZTWD & 0 & 11.9 & 5.3M; $2.86\times10^{-4}$& 145 &3.403&113.6& Yes\\
 L13 & Octo (Bd) & 0.9 & ZTWD &  2.3 & 9.9 & 20.1M; $1.43\times10^{-4}$& 290 &3.401&113.4& Yes\\
 FL-1& Flow-er & 0.9 & ZTWD &  1 & 37 & 322x258x256=21.3M& 130 &3.413&114.0& No\\
 FL-2& Flow-er & 0.9 & ZTWD &  2 & 16.9 & 322x258x256=21.3M& 130 &3.413&114.0& No\\
 \bottomrule
 \end{tabular}
 
 \footnotesize{$^a$ See also Montiel 2015. 
 $^b$ For the simulations done with \octo, in the Resolution column, we list the {\textit{initial}} number of cells in the adaptive-mesh grid as well as the minimal cell width (in solar radii). $^c$ We have run, additionally, two non-driven simulations, {\sc L11ND} and {\sc L12ND}, not listed here, as they resulted in the accretor's expansion due to numerical instabilities (see Appendix \ref{app-nondriven} for more details). {A question mark represents data that cannot be retrieved from available information.}}
\end{table*}

The simulations that appear in \cite{Staff2012}, \cite{Staff2018}, \cite{Motl2017}, and \cite{diehl2021performance}, are listed in the first part of the table. The \octo\ simulation from \cite{Staff2018} also appear in \cite{Montiel2015}. Three of the simulations were carried out with the SPH code \snsph; the rest were carried out using the finite-volume codes, \flower\ (4 simulations) and a previous version of \octo\ (5 simulations, listed as 'Octo' in the second column). The system's initial properties, initial total mass ($M_{\rm T}$), initial orbital separation ($a_0$), and initial orbital period ($P_0$) are also listed. The equation of state (EoS) used in each simulation is shown in column $4$, where ``ZTWD", ``poly", and ``ideal" stands for zero-temperature white dwarf, polytropic, and ideal gas equation of state, respectively. When using a polytropic EoS or an ideal gas EoS, there is one degree of freedom in converting the simulation units (from code units) to c.g.s. Therefore the simulation units (and as such the system properties) can be scaled in those simulations, \emph{e.g.}, the simulations of \cite{Motl2017} and \cite{diehl2021performance}. Using a ZTWD EoS, on the other hand, introduces two additional physical constants and thus the code units in these simulations cannot be scaled arbitrarily, and their conversion to c.g.s units is fixed. In Table~\ref{tab:past_sims}, we scaled the initial orbital separation (column 9) and initial orbital period (column 10) of \cite{Motl2017} and \cite{diehl2021performance} to match the initial system properties of the ZTWD simulations closely. 

In all previous $q=0.7$ simulations the donor star was initially forced into contact with its Roche lobe by a ``driving" mechanism, specifically, by an artificial removal of angular momentum from the system at a rate of 1\% per orbit. This initiates a mass transfer that can be well resolved by the code early in the simulation time. Usually, driving was applied for several initial orbits. We list the exact driving duration for each simulation in column $T_{\rm drv}$ in Table~\ref{tab:past_sims}. A longer driving time causes the system to achieve a deeper contact, which can alter the remaining evolution of the simulation.  

The resolution of past simulations spans a relatively wide range. Comparing resolution across different techniques is not straightforward. For \flower, a uniform (non-AMR) cylindrical grid, we list the number of cells across each dimension, $N_R\times N_z \times N_\theta$, and the total number of cells, where $R$, $z$, and $\theta$, denote the cylindrical radial, vertical, and azimuthal axes, respectively. For \octo, an AMR grid, we can list only the initial number of cells.  
This is a minimum value, as the number of cells increases as the simulation evolves in time. We also list in parentheses the smallest cell size in solar radii for \octo.

The AMR grid extends several times beyond the 
dimensions of the non-AMR grid and not all regions are equally resolved. A better comparison between the non-AMR and AMR codes is the number of cells across the donor's diameter, which we specify in the `donor res' column of Table~\ref{tab:past_sims}. For \snsph, a SPH code, we list the total number of particles in the simulation (in the Resolution column), and the number of particles consisting of the donor only (in the `donor res' column). The high number of particles in the donor relative to the total number of particles is meant to better resolve the mass transfer in those simulations. 

In column $11$ we specify whether the accretor is a hybrid CO-He WD rather than a CO-only WD. As mentioned, this can affect the amount of $^{16}$O in the burning regions and consequently the ratio of $^{16}$O to $^{18}$O produced by the burning. We note that in \cite{diehl2021performance} the mass of the helium shell was $\sim 0.045~{\rm M_{\odot}}$, while in the \octo\ simulations that we present in this paper (see Section~\ref{ssec:octo_setup}) the helium shell is $0.13~{\rm M_{\odot}}$. 
Lastly, we list the merger time, $T_{\rm merge}$, which is the time from the beginning of the simulation until a merger occurred in the units of initial orbital periods. 

From this table, it is notable that both the amount of driving and the simulation resolution affect the merging time. Specifically, the 4M-cell \flower\ run with polytropic EoS that has been driven for $2.3$ orbits merged much faster compared to the other \flower\ run with the same EoS and the same resolution that has been driven only for $1.3$ orbits. Alternatively, the simulations of \cite{diehl2021performance} were driven by the same amount. However, the merging time increases with finer resolution. 
This can be explained by the fact that longer driving pushes the system into deeper contact and higher initial mass transfer. As a result, the system evolves faster and eventually merges earlier. This is also true for higher resolution runs. The finer resolution promotes a lower mass transfer rate, and the evolution is slower. It is possible that at some fine resolution, the mass transfer rate could not be decreased below a certain value, which would represent the actual physical mass transfer rate, and the merging time will converge concerning resolution. 

Besides affecting the merging time, resolution can have an effect on the temperature and on the different compositions of different atmospheric layers of the merger, in particular when considering a hybrid WD accretor. The temperature is critical for helium burning and the production of oxygen-18, while the composition is important for the mixing of oxygen-16 with oxygen-18. Examining the effects of resolution, together with extending to higher resolution, are some of the main goals of the current paper.

\section{simulation setups}
\label{sec:methods}

In this section we give details of the setups for all our new simulations, starting with those carried out with \octo\ and finishing with those carried out with \flower.

\subsection{The \octo\ simulation setups}
\label{ssec:octo_setup}

We carried out five simulations using the benchmarked version of \octo\ \citep[][labled ``Bd" in Table~\ref{tab:past_sims}]{Marcello2021} with four different resolutions corresponding to 10, 11, 12, and, 13 levels of refinement (L10 to L13 in Table~\ref{tab:past_sims}). Simulation L13 has the highest resolution used for an \octo\ simulation, with almost twice the number of cells across the donor star than in the simulation by \cite{Staff2012}. This simulation also has almost twice the initial number of cells than in the high resolution simulation of \cite{diehl2021performance}. The initial number of cells and the smallest cell size in each run are listed in Table~\ref{tab:past_sims} ($7$th column), as well as the number of cells across the donor WD ($8$th column). The simulation domain extends up to $\sim 200$ times the initial separation, which allows us to follow the outflowing gas from the system up to a significant distance and for a longer time.

Each simulation starts off with a WD-WD binary with a total mass of $M_{\rm T}=0.9~{\rm M_\odot}$. The primary (accretor star) and secondary (donor star) initial masses are $0.53~{\rm M_\odot}$ and $0.37~{\rm M_\odot}$, respectively. {We choose this value for the total mass because of observational indications (e.g., \citealt{2015ApJ...809..184K}), and a mass ratio of $q=M_{\rm donor}/M_{\rm accretor}$=$0.7$ is chosen because in previous simulations this ratio yielded the most suitable SoF temperature for incomplete helium burning (see \citealt{Staff2012}). The binary's initial conditions are calculated using a self-consistent field (SCF) technique as described in \citealt{Marcello2021}), except that a different equation of state, the zero-temperature white dwarf equation of state, is being used (see more details in Section~\ref{ssec:eos}). The donor star initially fills its Roche lobe and both stellar spins are synchronized to the orbital frequency. This results in an initial orbital separation of $0.05 ~{\rm R_\odot}$ and in an orbital period of $\sim$ 2 minutes. In all of the simulations except {\sc L12I}, the grid is set to constantly rotate at the same frequency as the initial binary frequency. We carry out simulation {\sc L12I} in an inertial frame of reference to study the impact of the chosen frame of reference on the system's evolution and on the post-merger object's properties. }

The accretor is a hybrid He/CO WD, consisting of a $0.4~{\rm M_\odot}$, CO core and an outer layer of $0.13~{\rm M_{\odot}}$ of helium, while the donor is a helium WD. {The CO WD core and its He envelope are evolved as two distinct fluids (or species) by \octo\ and the donor star is evolved as a third specified, distinct fluid. Each species has a mean molecular weight. The accretor CO core fluid is initialized with a mean molecular weight corresponding to an equal mixture of ionized carbon and oxygen. The mean molecular weights of the CO WD atmosphere and of the helium WD are $\mu=4/3$, corresponding to ionized helium. The average mean molecular weight in a cell is used then to calculate the temperature.}  

\begin{figure*}
    \centering
    \includegraphics[scale=0.4]{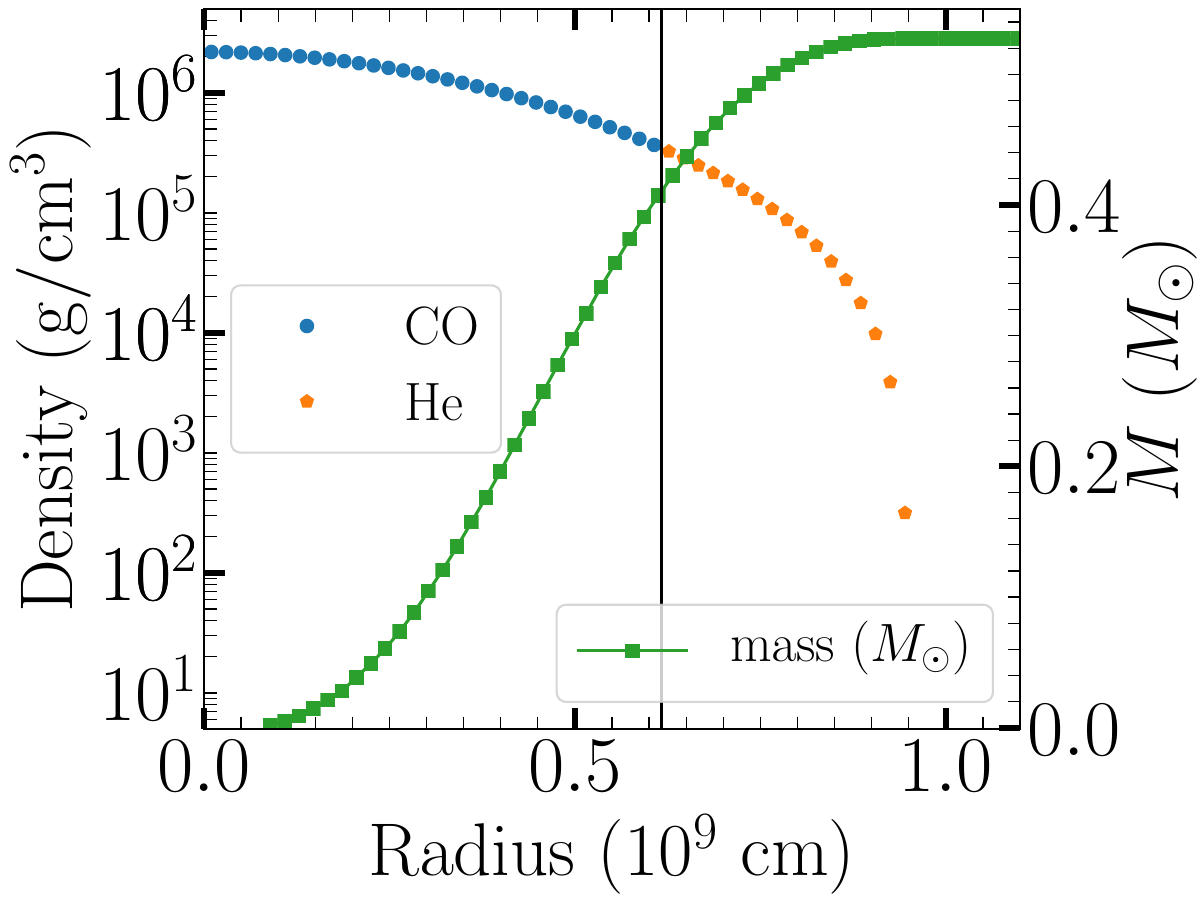}
    \includegraphics[scale=0.4]{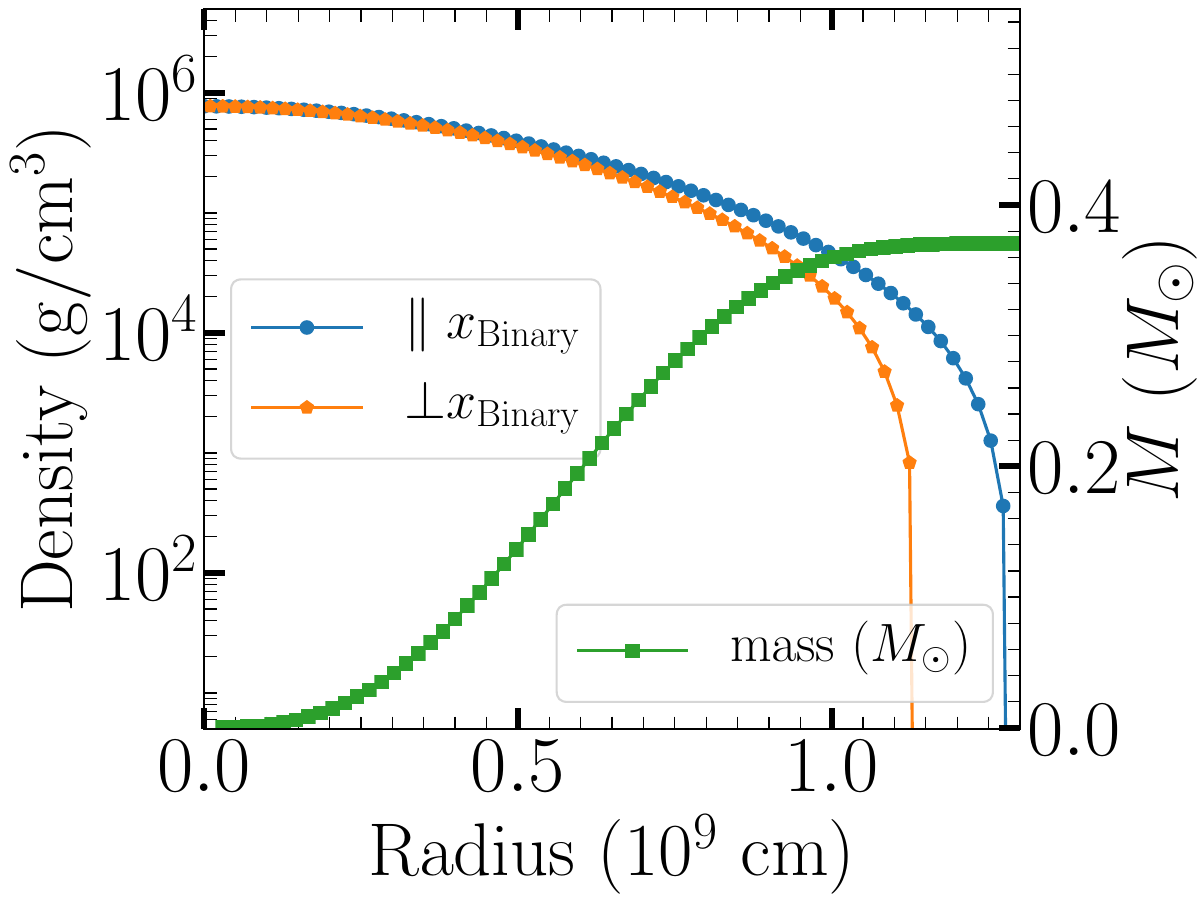}      
    \caption{{Initial density and mass profiles of the binary in simulation {\sc L12}. Left: the hybrid accretor star, which is composed of a $0.4~M_{\rm \odot}$ inner Carbon-Oxygen (CO) core and a $0.13~M_{\rm \odot}$ outer Helium (He) shell. Right: the donor star structure, where $x_{\rm binary}$ is the axis connecting the stars' centers of mass }}
    \label{fig:q07_level12_set-up}
\end{figure*}

{In Figure~\ref{fig:q07_level12_set-up} we plot the resultant density and mass profiles of the accretor (left panel) and donor (right panel) stars, after the SCF initialization and for simulation {\sc L12}.} 
{The inner $0.4~M_{\rm \odot}$ of the accretor star (3/4 of the accretor's mass) is the accretor CO core and it is surrounded by a He shell. The thickness of the accretor's helium shell depends on resolution and is 5, 10, 18, and 36 cells, for the simulations with 10, 11, 12, and 13 levels of refinement, respectively. This shell serves as a buffer region between the CO core and the SoF, and therefore can potentially abate the dredge-up of $^{16}$O from the core. Such buffer regions have been considered in several previous numerical studies (e.g., \citealt{Staff2018}), and we choose a relatively massive shell to maximize its effect. However, in running our simulations we observe that this helium shell is almost entirely mixed in with the CO core well before the binary merges and therefore has only a minor effect on the resulting dredge-up of $^{16}$O. This is mainly due to a numerical limitation of our code that cannot maintain a steep abundance gradient over an extended amount of time.}

Lastly, to initiate a higher mass transfer rate which ultimately results in a faster merger, we extract angular momentum from the system's orbit in simulations {\sc L11}, {\sc L12} and {\sc L13} during the first $1.3$, $1.3$, and $2.3$ orbits, respectively (fifth column of Table~\ref{tab:past_sims}). This is done to save computer time. We drove {\sc L13} longer because in this highest resolution simulation, the initial mass transfer would be the lowest without enough driving and the computed time step is also very small.
The extraction of angular momentum is done by adding adequate source terms in the momentum equations as explained by \cite[][their equation 44]{Marcello2021} and at a rate of 1\% of the orbital angular momentum per (initial) orbital period.

\subsubsection{The equation of state}
\label{ssec:eos}

The construction of the stars by the SCF method uses a zero temperature white dwarf equation of state (ZTWD EoS), rather than a polytropic equation of state with an index $n=\sfrac{3}{2}$. The ZTWD pressure is obtained by: 
\begin{equation}
P_\mathrm{ZTWD} := A\left[x\left(2 x^2 - 3\right)\left(x^2+1\right)^\frac{1}{2} + 3 \mathrm{sinh}^{-1} x \right]
\label{eq:ztwd}
\end{equation}
where $x:=\left(\sfrac{\rho}{B}\right)^{\frac{1}{3}}$, $A$ and $B$ are constants, and $\rho$ is mass density. Note that like the polytropic equation of state, Equation \ref{eq:ztwd} is also barotropic, allowing it to be incorporated into an SCF solver in a manner similar to what was done by \cite{Even2009}. 

After initialization, we evolve the simulations with a combined EoS of both ZTWD EoS and ideal gas \citep{Staff2018}: 
\begin{equation}
 P = P_{\rm deg} + \left(\gamma - 1 \right)E_{\rm th},
 \label{P}    
\end{equation} 
Where $E_{\rm th}$ is the thermal energy density, and $\gamma=5/3$ is the adiabatic index. The only difference with the method of \citet{Staff2018} is that we compute the local molecular weight for each cell rather than assuming it to be $2$ for the temperature calculation, which results in lower temperatures compared to \citet{Staff2018}. A more detailed description of this EoS and the exact calculation of temperature appears in Appendix~\ref{app-eos}.  

\subsection{The \flower\ simulation setups}
\label{ssec:flower_setup}

In order to compare the simulations' outcome with a different code we use the Eulerian uniform and cylindrical grid code \flower\ (\citealt{D_Souza_2006, Motl2007}). We have carried out two additional simulations {\sc FL-1} and {\sc FL-2} with \flower. Both are evolved in the rotating frame and have identical resolutions. Simulations {\sc FL-1} and {\sc FL-2} are driven, by the removal of angular momentum uniformly at a rate of 1~percent, for 1 and 2 (initial) orbital periods, respectively (Table~\ref{tab:past_sims}).

The density from the \octo\ SCF model is interpolated onto a uniform cylindrical mesh using cubic interpolation. Parameters from the \octo\ SCF model such as the angular frequency of the initial data are used with the same value as in \octo. The interpolation causes some initial oscillation in the density not present in the \octo\ runs but those perturbations decay well before the merger.

The spatial resolution has 322 radial zones, 258 vertical zones, and 256 azimuthal zones. The total number of cells is $\approx$20 M, larger than the 4 M cells resolution of the \flower\ simulations of \citet{Motl2017}. 
The grid extends in radius to $6.4 \times 10^{9}$~cm and the spacing length in the radial direction equals to the spacing length in the vertical direction.

\section{Orbital evolution and merger}
\label{sec:results}

In this section, we start by describing the behavior of the \octo\ simulation carried out with 12 levels of refinement in the rotating frame ({\sc L12} in Table~\ref{tab:past_sims}). We later compare it to the other simulations.

\subsection{{Binary evolution leading up to merger}}
\label{ssec:evolution}

As mentioned before, in order to save computational time, the initial phases of simulation {\sc L12} were driven by continuously extracting angular momentum for the first 1.3 orbits at a rate of one percent per orbit. Still, the binary took 38.7 initial orbital periods, $P_0$, until it merged. We simulated an additional 5~$P_0$ past-merger to allow the merger enough time to relax and to become more axially symmetric.

\begin{figure*}
    \centering
    \includegraphics[scale=0.28]{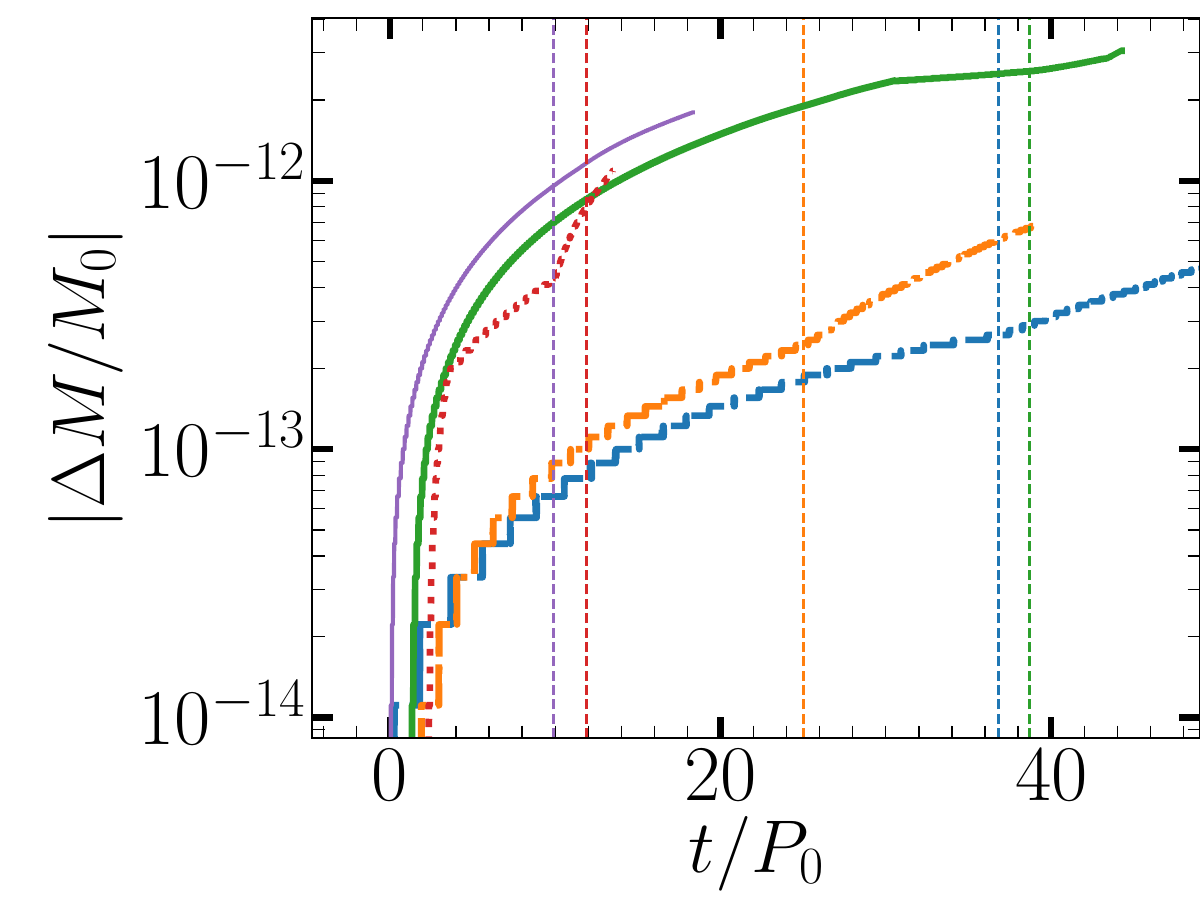}
    \includegraphics[scale=0.28]{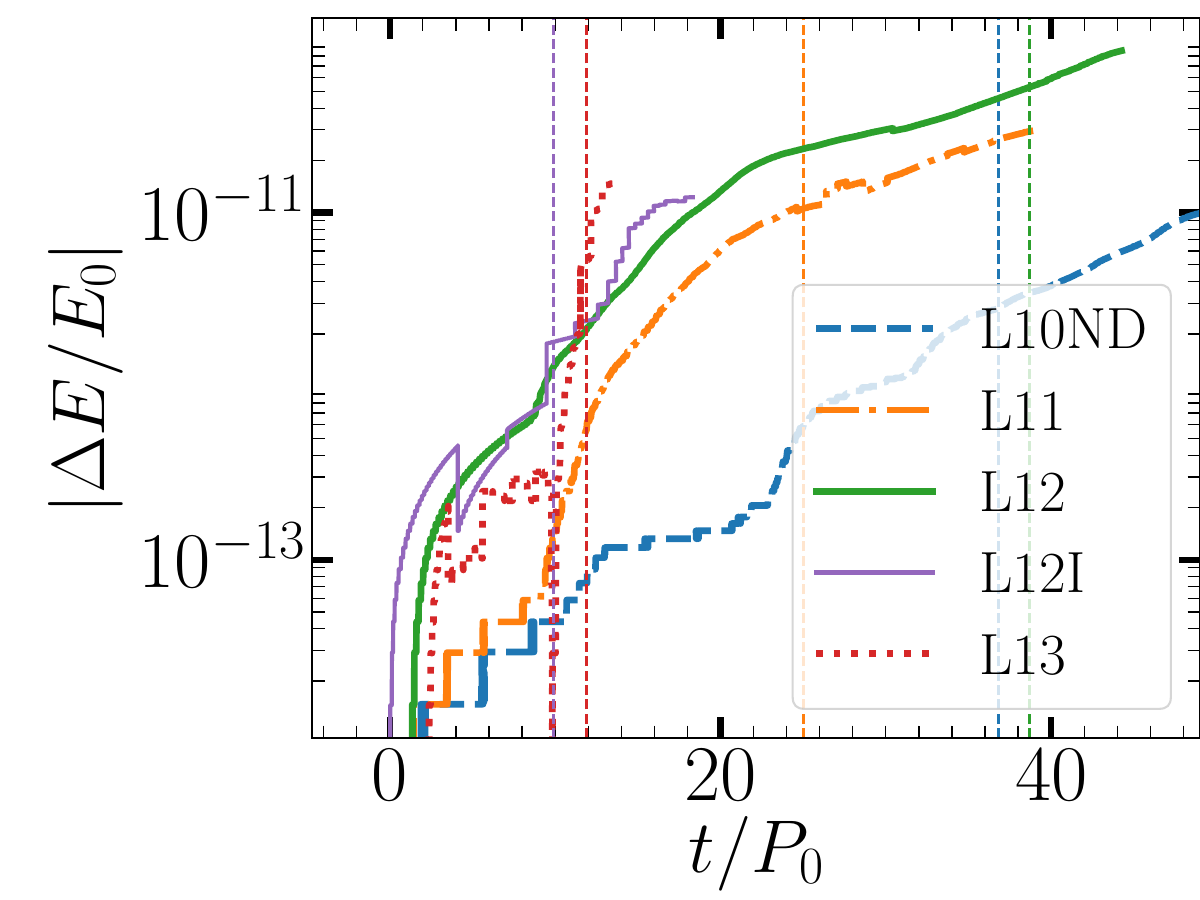}
      \includegraphics[scale=0.28]{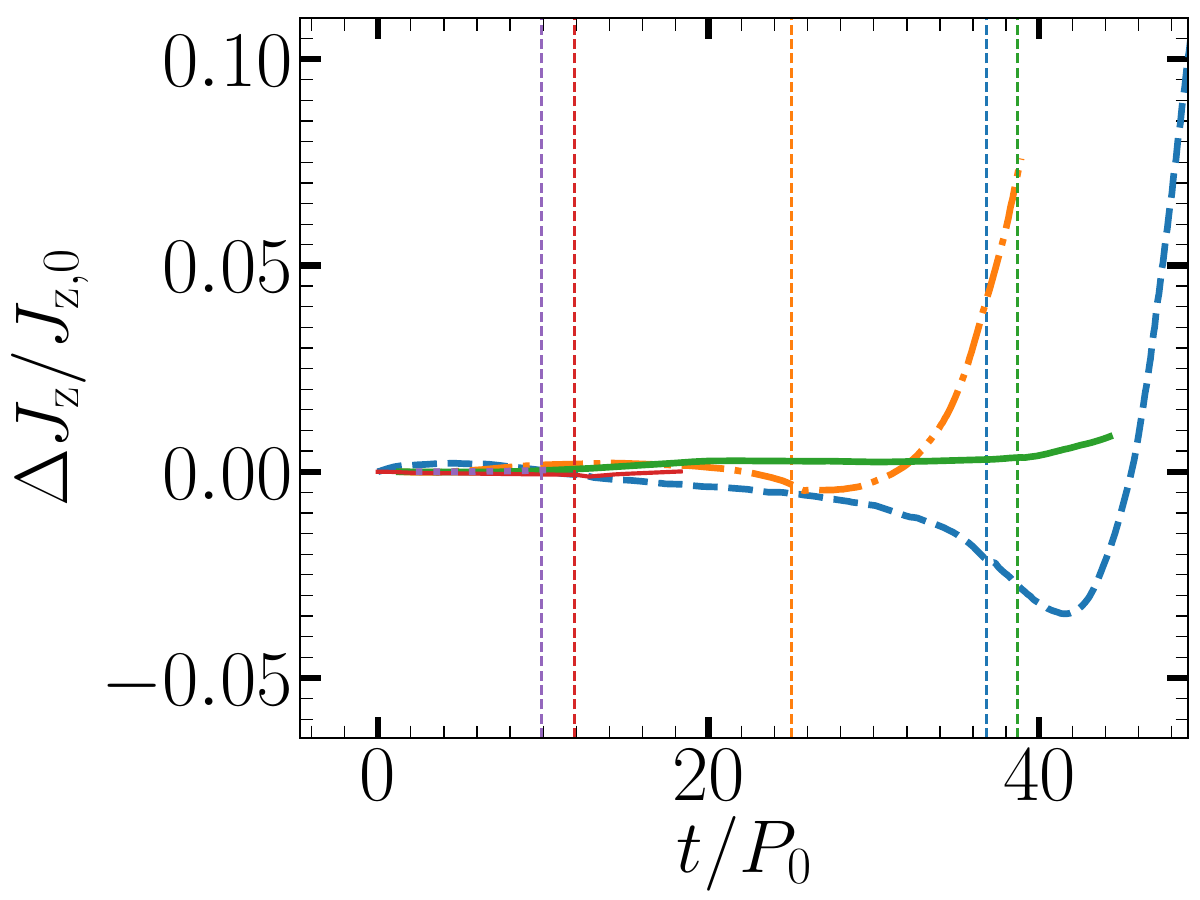}
    \caption{Conservation of mass, energy, and angular momentum in our simulations for 
    $t\ge T_{\rm drv}$. 
    Vertical dashed lines denote the merger time. 
    Note that in the angular momentum plot, we do not take into account angular momentum losses due to gas that flows out of the grid. {$P_0$ is the initial orbital period of the simulation (between $113-116~{\rm s}$, depending on the specific simulation) as listed in Table~\ref{tab:past_sims}}.}
    \label{fig:q07_level12_rot_cons}
\end{figure*}

In Figure~\ref{fig:q07_level12_rot_cons} we present, three quantities that should be conserved, considering conservation properties only after the driving phase. The quantities $\Delta M$, $\Delta E$, and $\Delta J_z$ are computed as in \cite{Marcello2021}, and they include mass and energy that flows out of the grid, but not angular momentum outflows. For ({\sc L12}), mass and energy were conserved to better than one part in $10^{-11}$ and $10^{-10}$, respectively. Any inaccuracies stem solely from using a ``density floor", which prevents vacuum conditions in the low-density regions outside the star. We show in these plots the conservation values of simulations of different resolutions for comparison. We will discuss numerical effects like resolution in Section~\ref{ssec:numeric_eff}. The angular momentum is conserved at the $0.5$\% level for L12, where most of the non-conservation stems from numerical viscous torques during and after the merger (as described in \citealt{Marcello2021}). 

\begin{figure*}
\centering   
    \includegraphics[scale=0.55, trim={0 2.8cm 0 2.4cm}, clip]{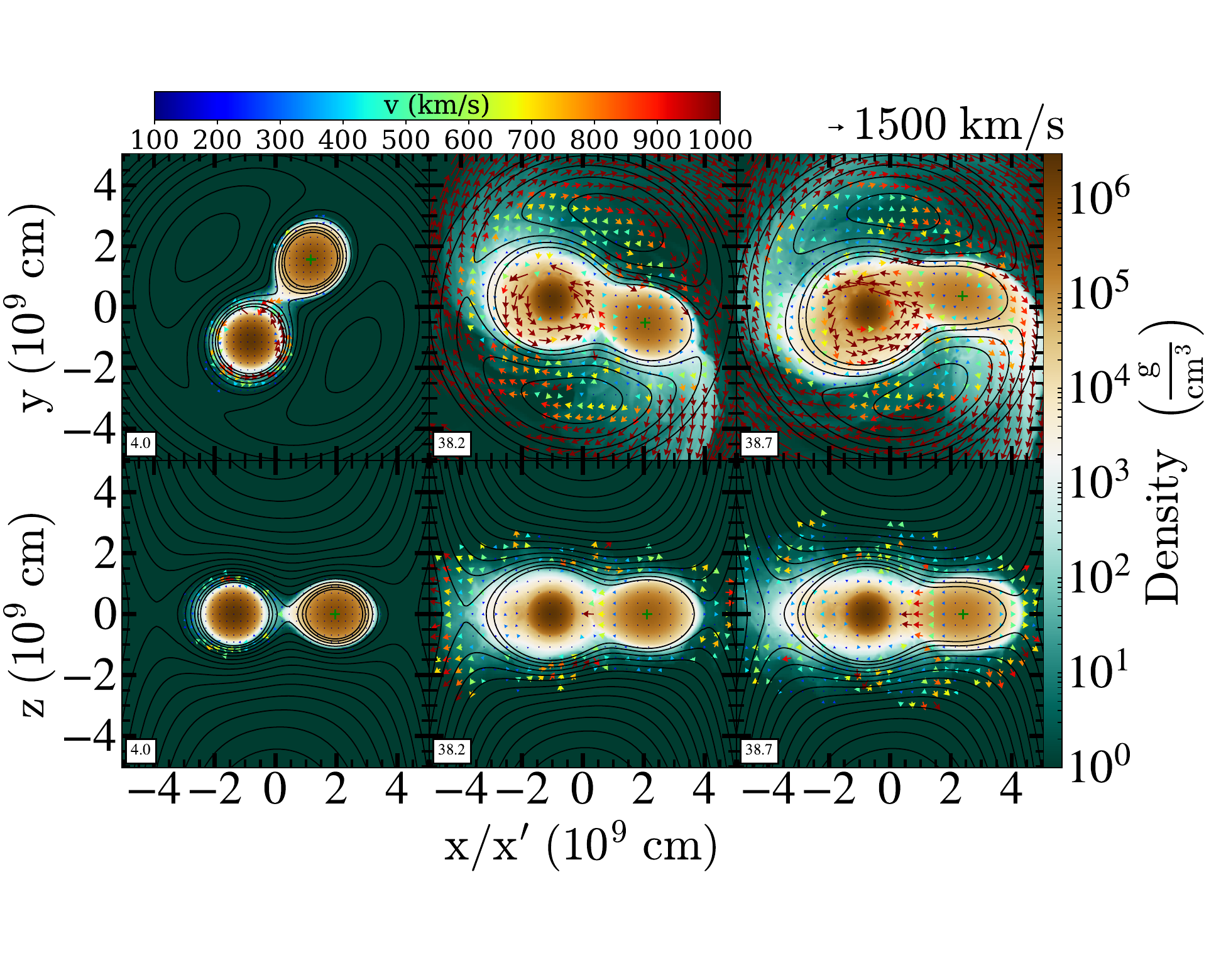}    
    \caption{Simulation {\sc L12}. Density slices at the equatorial plane (top row) and at the meridional plane (bottom row) with contours of $\Phi_{\rm eff}$. Arrows show magnitude and direction of velocities with respect to the instantaneous rotating frame, for gas with density greater than $1 \g \cm{}^{-3}$. {The time (in initial orbital periods, $P_0=113.6~{\rm s}$) is shown on the lower-left corner of each panel.}}
\label{fig:l12_ev_dens}
\end{figure*}

\begin{figure*}
\centering   
    \includegraphics[scale=0.55, trim={0 2.8cm 0 3.9cm}, clip]{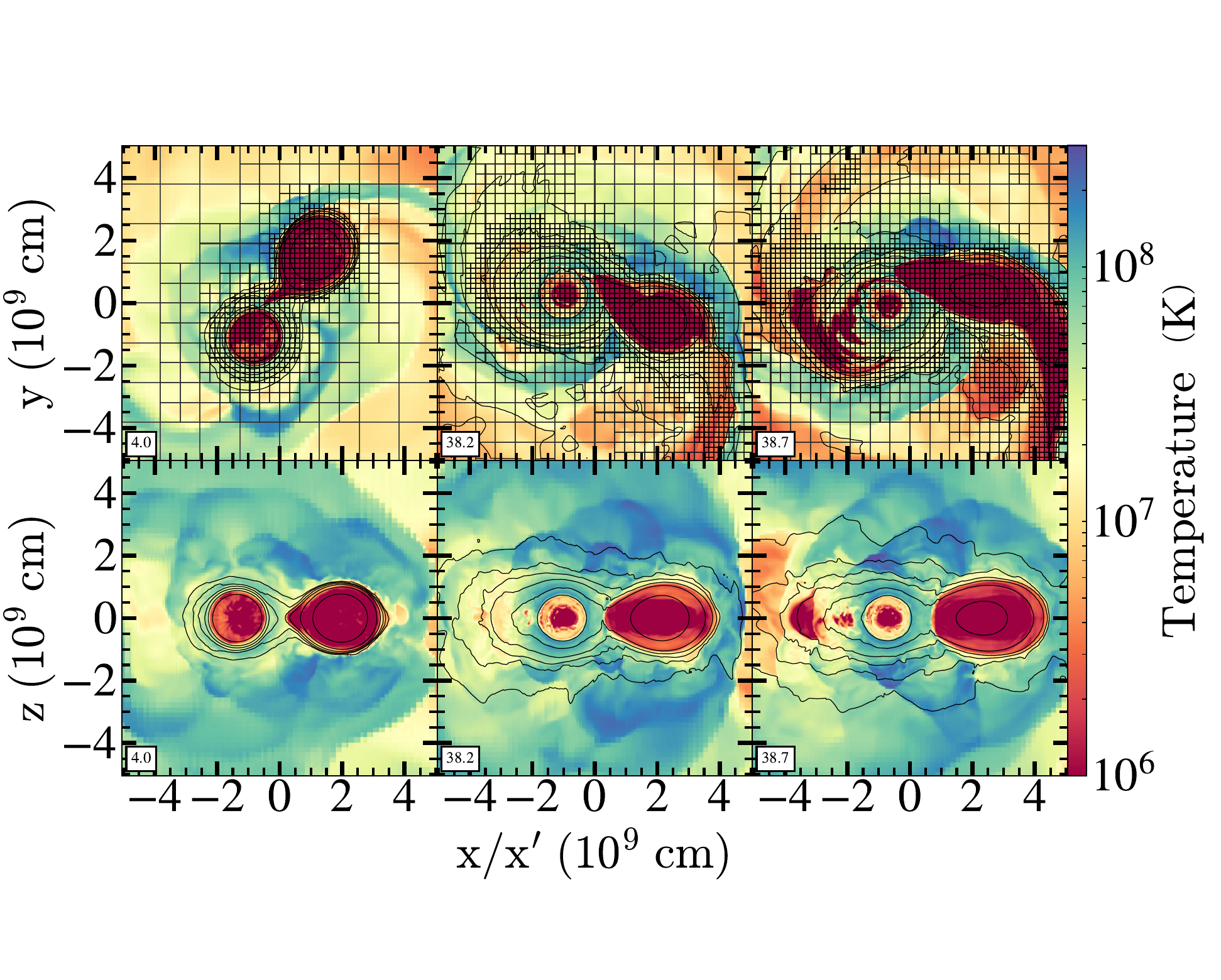}
    \caption{Simulation {\sc L12}. Temperature slices at the equatorial plane (top row) and at the meridional plane (bottom row) with, overlaid, the grid structure (top row) and density contours (bottom row). {The time (in initial orbital periods, $P_0=113.6~{\rm s}$) is shown on the lower-left corner of each panel.}}
\label{fig:l12_ev_temp}
\end{figure*}

In Figures~\ref{fig:l12_ev_dens} and~\ref{fig:l12_ev_temp} we show density and temperature maps, respectively, of the equatorial (top row) and meridional (bottom row) planes at three times, representing the beginning of the evolution (left panel), the time just before the merger (central panel), and the time of the merger (right panel). We also overlay velocity vectors (in the rotating frame) for gas with densities greater than $1 \g \cm^{-3}$ and equipotential contours. On the equatorial temperature slices, we overlay instead the grid structure (each square represents a subgrid of 8 by 8 by 8 cells) while on the meridional slices, we overlay density contours\footnote{Movies of the simulations can be obtained via \href{https://youtube.com/playlist?list=PLkUKbicT-DWSCYxLlKtycnHT7ohcZog9_&si=hjkH5I7iHpDIAMQW}{this link}}. 

\begin{figure}
    \centering
    \includegraphics[scale=0.4]{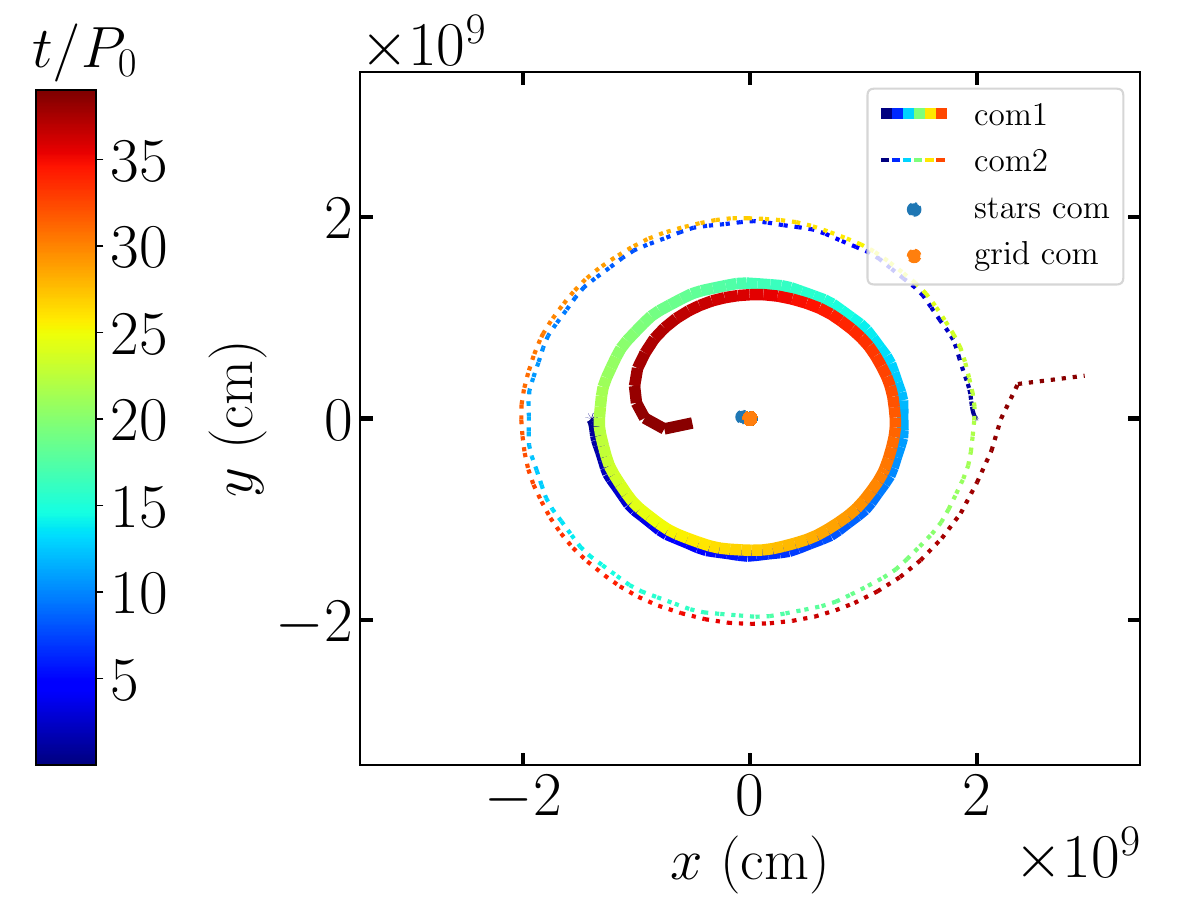}
    \caption{The orbital evolution of the system in simulations {\sc L12}, showing the centers of mass of the individual stars until shortly before the merger, as well as the centers of mass of the system and of the grid - expected to be stationary. {The stars merge at $t=T_{\rm merge}=38.7P_0\simeq4400~{\rm s}\simeq73~{\rm min}$.}}
    \label{fig:q07_level12_rot_orbital}
\end{figure}

Shortly after the simulation begins, a stream of gas from the donor (less massive, larger star) flows through the L1 Lagrange point and flows around the accretor mainly along the equator. As the simulation progresses, mass starts flowing through the two outer Lagrange points as well and eventually, the donor is tidally stretched and wrapped onto the accretor. The cold donor material is heated by shocks when it impacts the accretor's surface, and the temperature reaches helium-burning levels in a shell around the accretor. This SoF and its properties together with the potential nuclear fusion within it will be discussed in Section~\ref{ssec:SoF}.  
In Figure~\ref{fig:q07_level12_rot_orbital} we plot the overall orbital evolution on the orbital plane. The center of mass position of each star is plotted, the accretor's as a thick solid line and the donor's as a thin dotted line. Colors represent time in units of the initial orbital period, $P_0$ (blue is the earliest, and dark red is the latest). To identify the cells of each star, we use the technique described in \cite{Marcello2021}. We plot the evolution until just before the merger, when this technique no longer works because the two individual stars are no longer well-defined. {We typically find that the stars become indistinguishable by this technique when the binary orbital frequency, $\Omega_{\rm orb} = j_{z, {\rm orb}}/a^2$, lags behind the Keplerian frequency, $\Omega_{\rm kep}=\sqrt{G(M_1 + M_2)/a^3}$, and we set the time of the merger, $T_{\rm merge}$, according to the earliest time this occurs. For simulation {\sc L12} we obtain $T_{\rm merge}=38.7P_0=73~{\rm min}$. }

Additionally, we plot in Figure~\ref{fig:q07_level12_rot_orbital} the center of mass of the entire grid in orange, and the combined center of mass of the two stars in blue. The grid center of mass remains in place, which indicates again a good degree of linear momentum conservation. The combined center of mass of the two stars slightly shifts during the merger, implying slightly asymmetrical outflows through the outer Lagrange points.

\begin{figure*}
    \centering
    \includegraphics[scale=0.33]{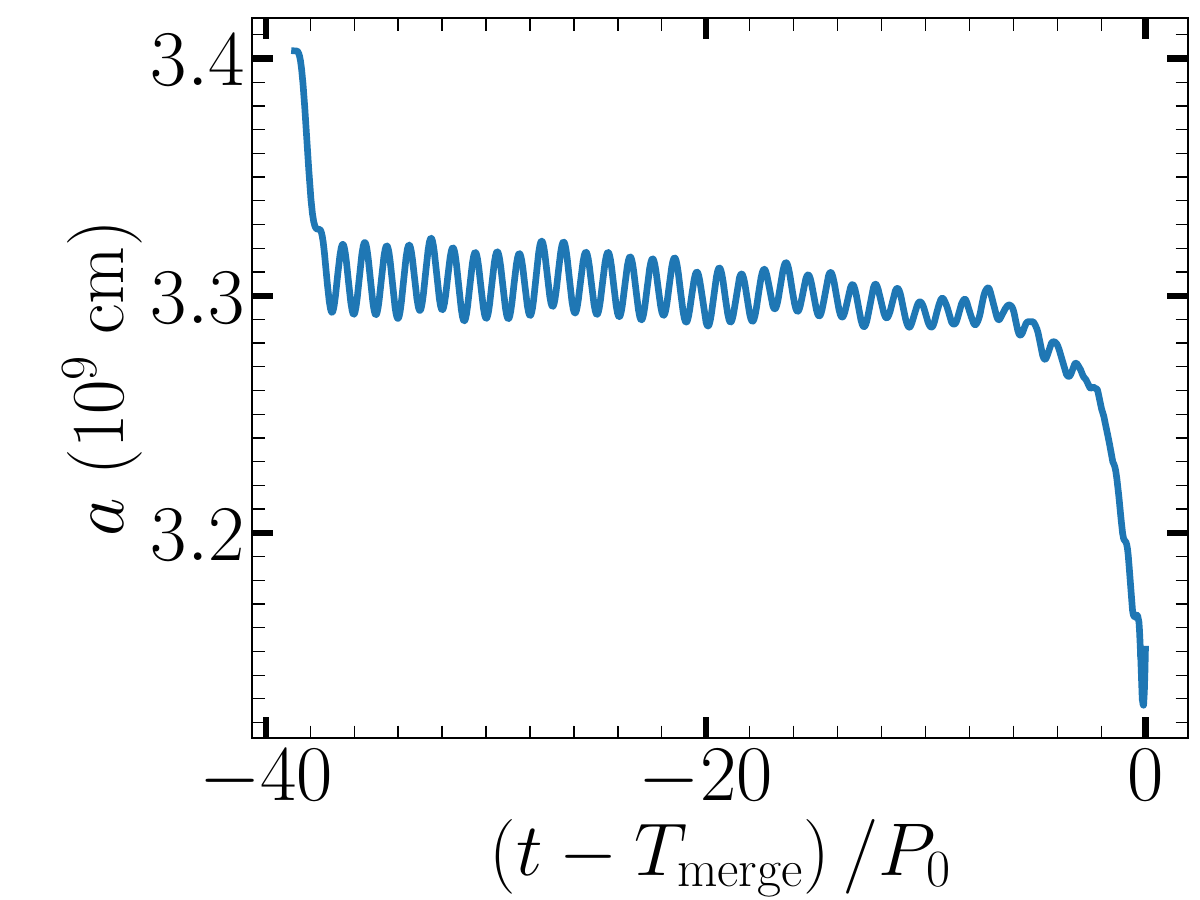}
    \includegraphics[scale=0.33]{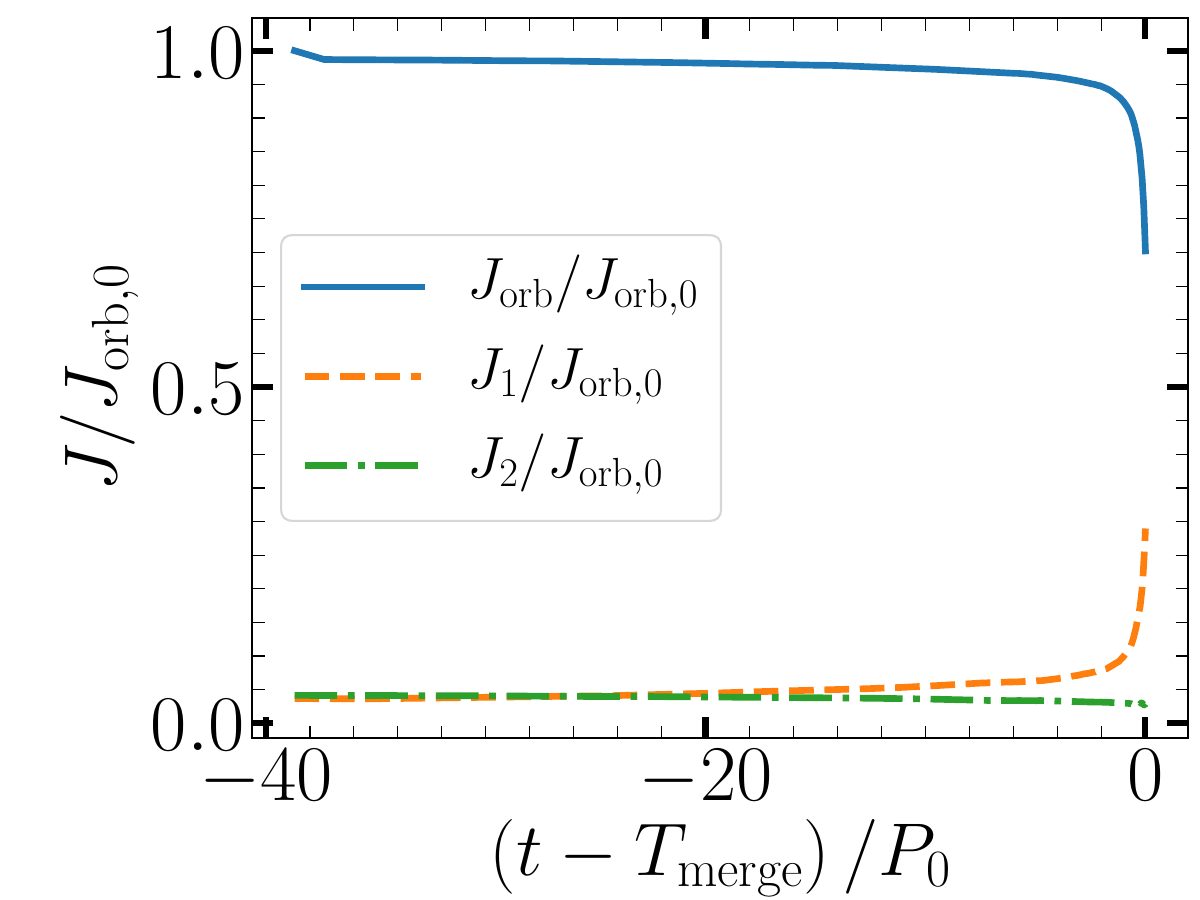}
    \includegraphics[scale=0.33]{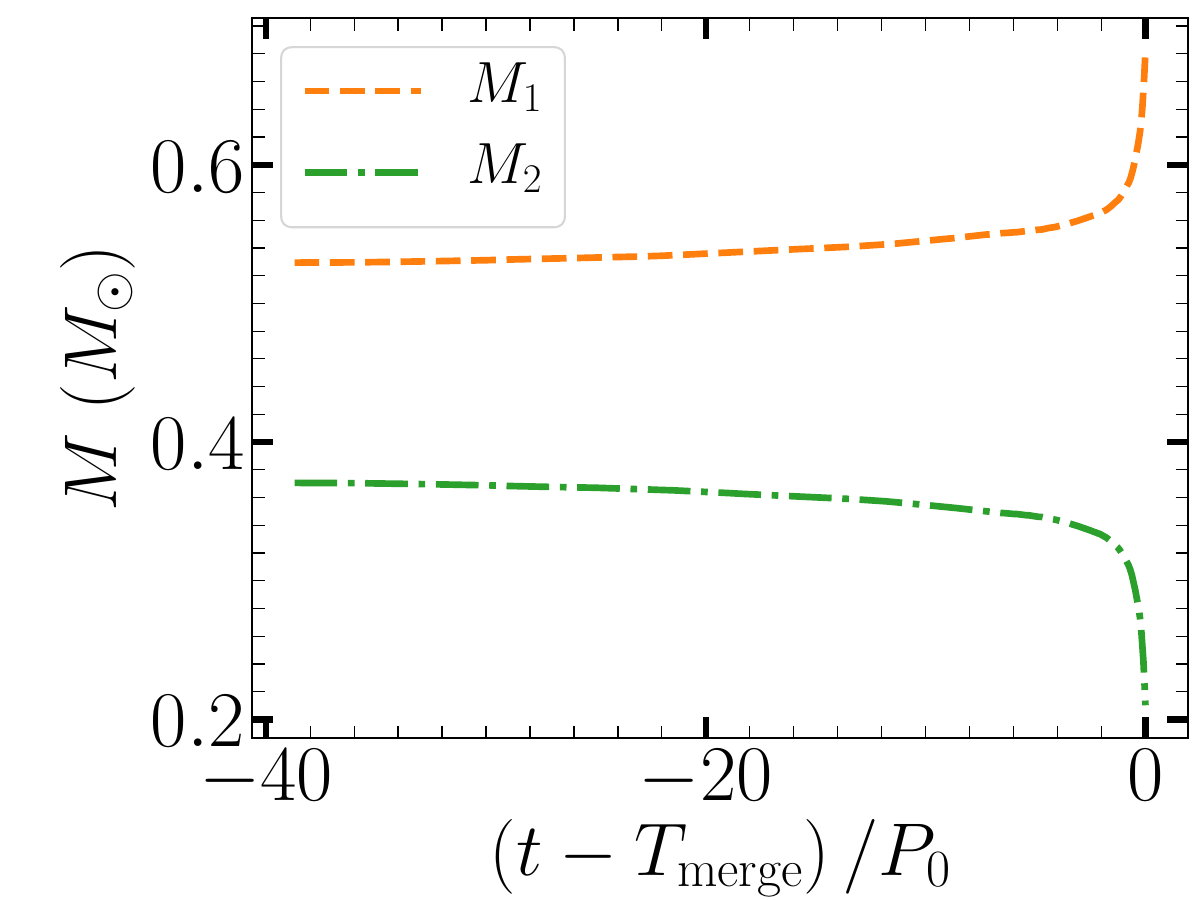}
    \includegraphics[scale=0.33]{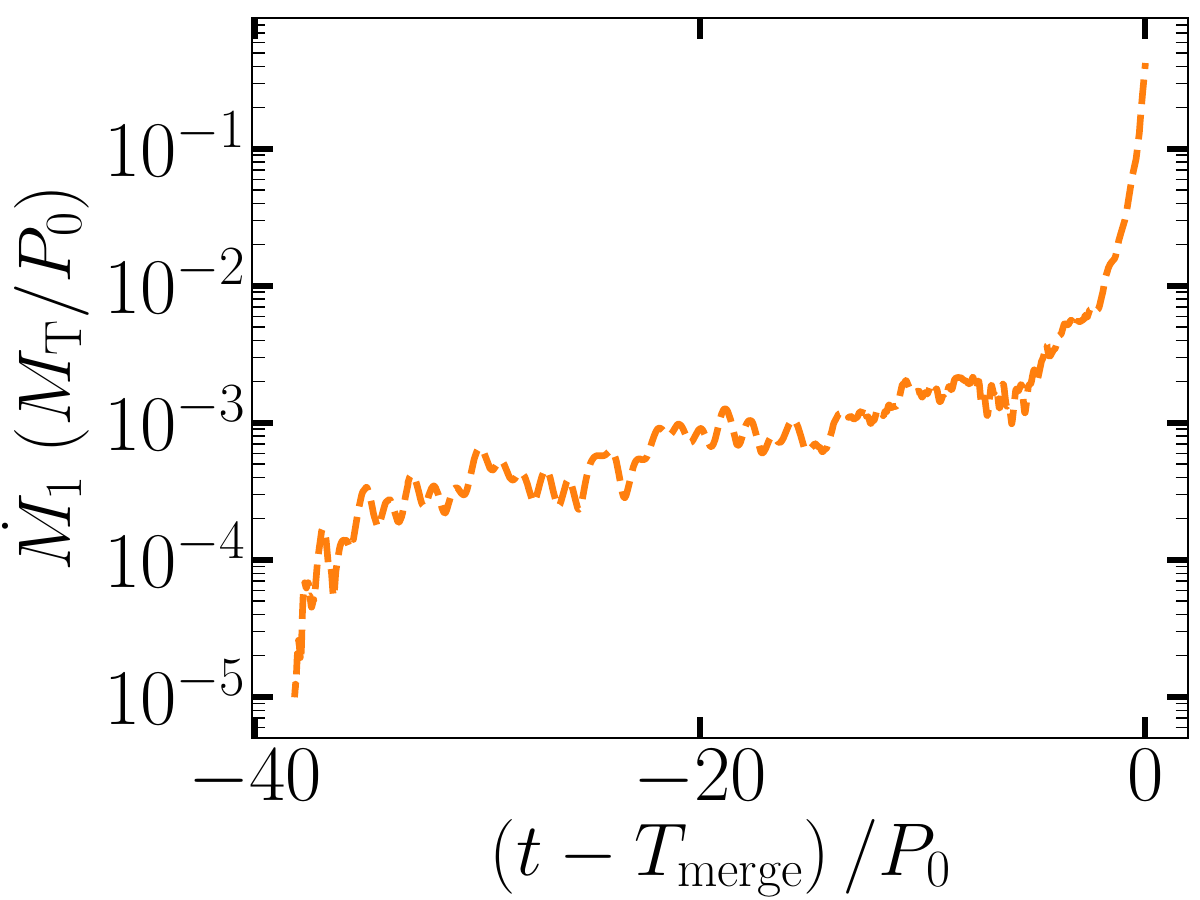}
    \caption{The orbital separation, angular momenta, masses, and mass transfer rates for the {\sc L12} simulation (see Table~\ref{tab:past_sims}), all as a function of time. The system was driven for 1.3 orbits at a rate of one percent of its orbital angular momentum per initial orbit and eventually merged at $T_{\rm merge}=38.7$. Note that the mass transfer is measured in units of total mass $M_T=0.9~{\rm M_{\odot}}$ per initial orbit, $P_0=113.6~{\rm sec}$ for this simulation.}
    \label{fig:l12_diags}
\end{figure*}

In Figure \ref{fig:l12_diags} we show the orbital separation, the orbital angular momentum, the angular momentum of the primary (star 1) and the secondary (star 2), the mass of each star, and the mass transfer rate (in units of the total mass per initial orbital period), all as a function of time. During the first 1.3 orbits both the separation and the orbital angular momentum decrease, as expected due to the driving, while the orbital frequency increases and the binary axis rotates counterclockwise in the rotating frame. 

After this initial driving phase, mass transfer onto the more massive, accretor star, acts to increase the separation. {At this point, the mass transfer rate is $10^{-4}$ $M_{\rm T}/P_0\simeq 3\times10^{-3}~M_{\rm \odot} / {\rm hr}$ (lower-right panel)}. However, the accretor also drains angular momentum from the orbit, {at a rate of $10^{-3} J_{{\rm orb},0} / P_0$ (upper-right panel)}, which counteracts the effect of mass transfer and tends to shrink the orbit. {Overall the orbital separation remains fairly constant (upper-left panel).} The mass transfer rate continues to increase, while the orbital angular momentum becomes angular momentum of the accretor star, and eventually, the donor star is tidally disrupted onto the accretor. { This picture of unstable mass transfer that increases with time while the orbital angular momentum decreases and that leads to the tidal disruption of the donor star is consistent with what is described in \citealt{Motl2017} and is termed there a generic `case A' evolution for WD-WD binaries. This is to be distinguished from a `case B' evolution where the mass transfer stabilizes or decays with time, which leads to a dynamically stable binary configuration}. We further discuss quantitatively the mass transfer and verify it against an analytic expression in {Section~\ref{ssec:mass-transfer} and in Appendix~\ref{app-verification}}.

\begin{figure*}
    \centering
    \includegraphics[scale=0.33]{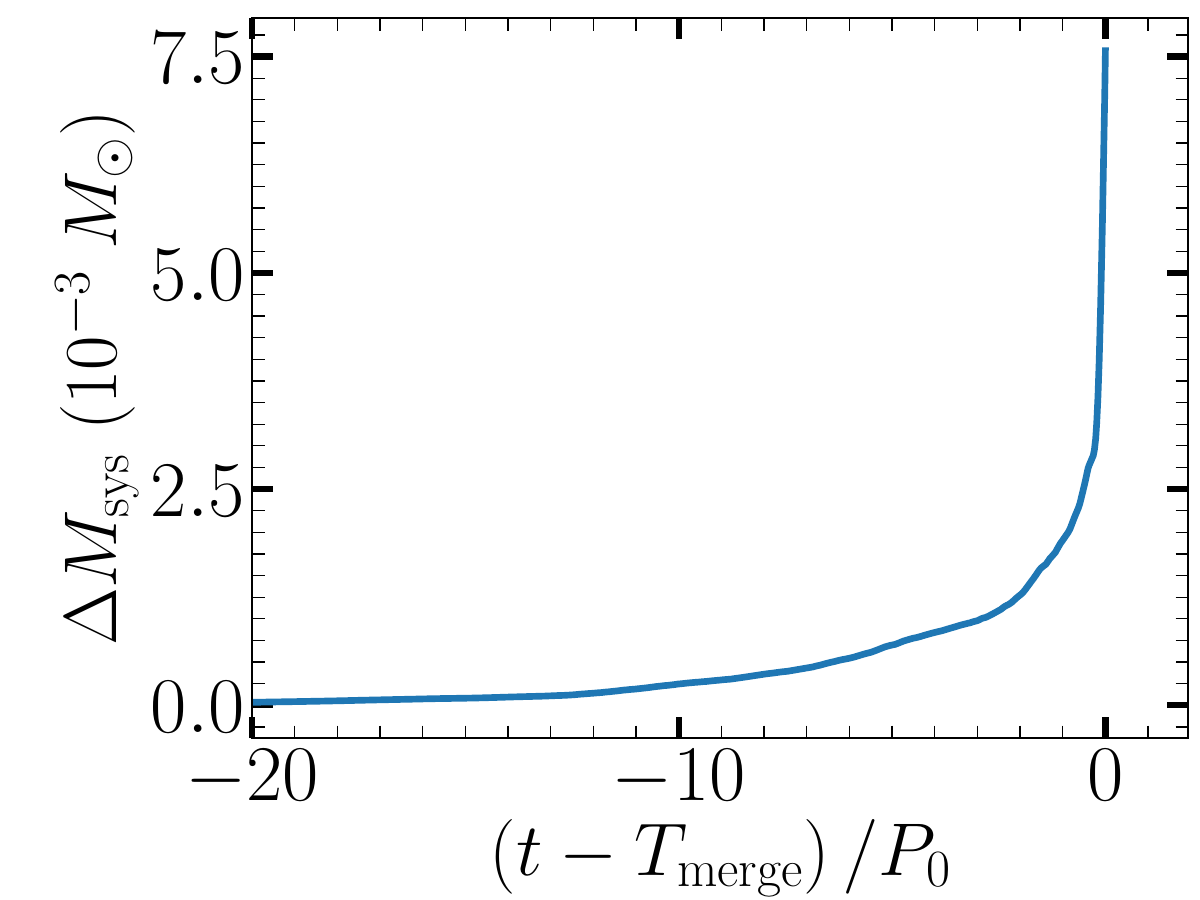}
    \includegraphics[scale=0.33]{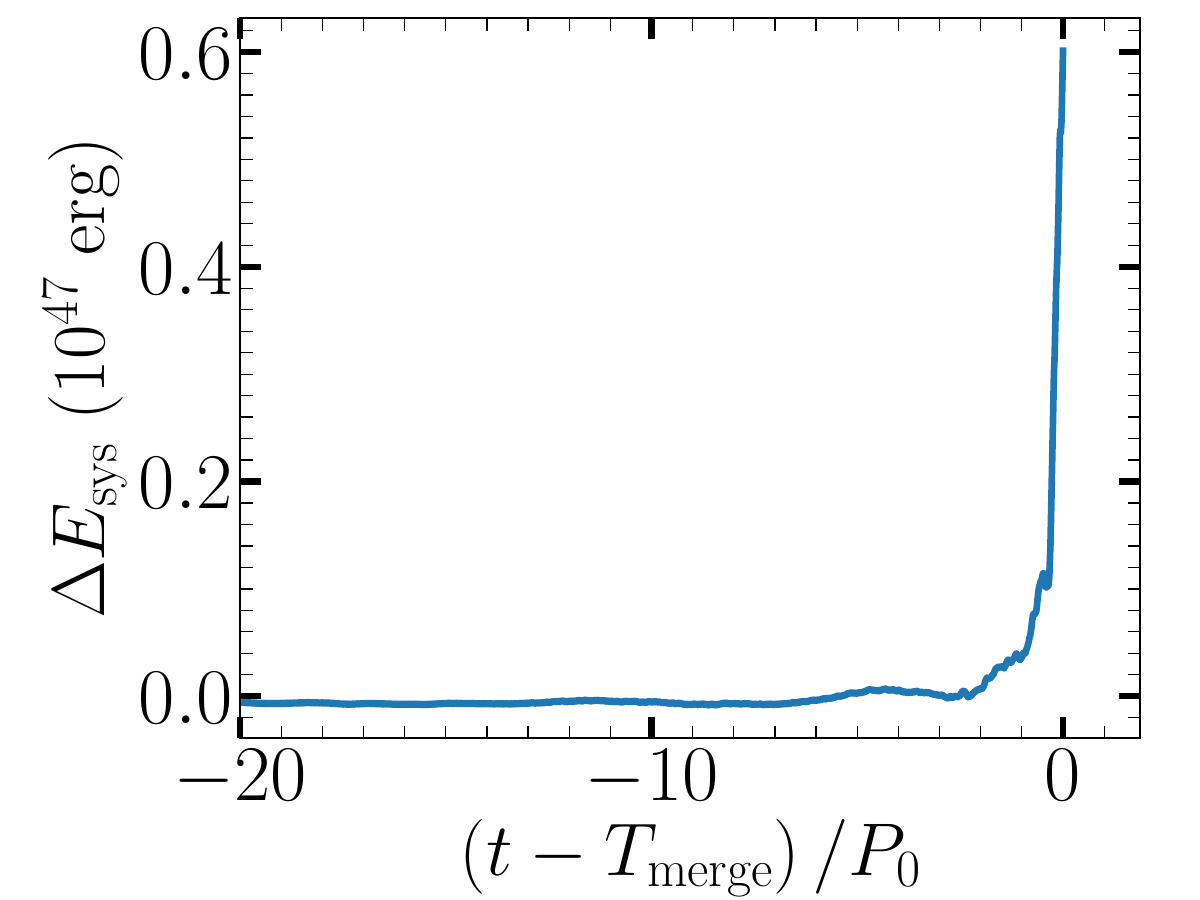}
    \caption{Mass (left panel) and energy (right panel) that flows out of the outer Lagrange points L2 and L3 in simulation {\sc L12}. The energy of the merged object decreases as a result of the L2 and L3 outflow. }
    \label{fig:l12_losses}
\end{figure*}

Just prior, but mostly during the merger, mass is flowing through the outer Lagrange points, L2 and L3 at an increasing rate. Because our diagnostics method accounts gas as belonging to a star only if it is inside the star's Roche lobe, we can quantify the amount of mass and energy lost from L2 and L3, $\Delta M$ and $\Delta E$, as $\Delta M = (M_1 + M_2)|_{t=0} - (M_1+M_2)|_t$, and $\Delta E= (E_1+E_2)|_{t=0} - (E_1+E_2)|_t$, respectively. In Figure~\ref{fig:l12_losses}, we plot these quantities as a function of time. We find that a few percent of the total mass is flowing out of L2 and L3, {and that the system's total energy becomes a few percent more negative close to the merger}. Therefore the energy outflow through L2 and L3 is positive. {Most of the outflowing mass remains bound and ends up being accreted by the merged object at late times.
For simulation {\sc L12} we find $\sim 1\times 10^{-3}~{\rm M_{\odot}}$ of unbound mass, although we stopped the simulation before the amount of unbound mass reached a steady value. The amount of unbound mass decreases with resolution, and therefore the maximal value of unbound mass we find is in our lowest resolution run. We conclude that only a small fraction of the total mass becomes unbound as a result of the merger ($\lesssim 5\times 10^{-3}~{\rm M_{\odot}}$). }

\begin{figure*}
    \centering
    \includegraphics[scale=0.6]{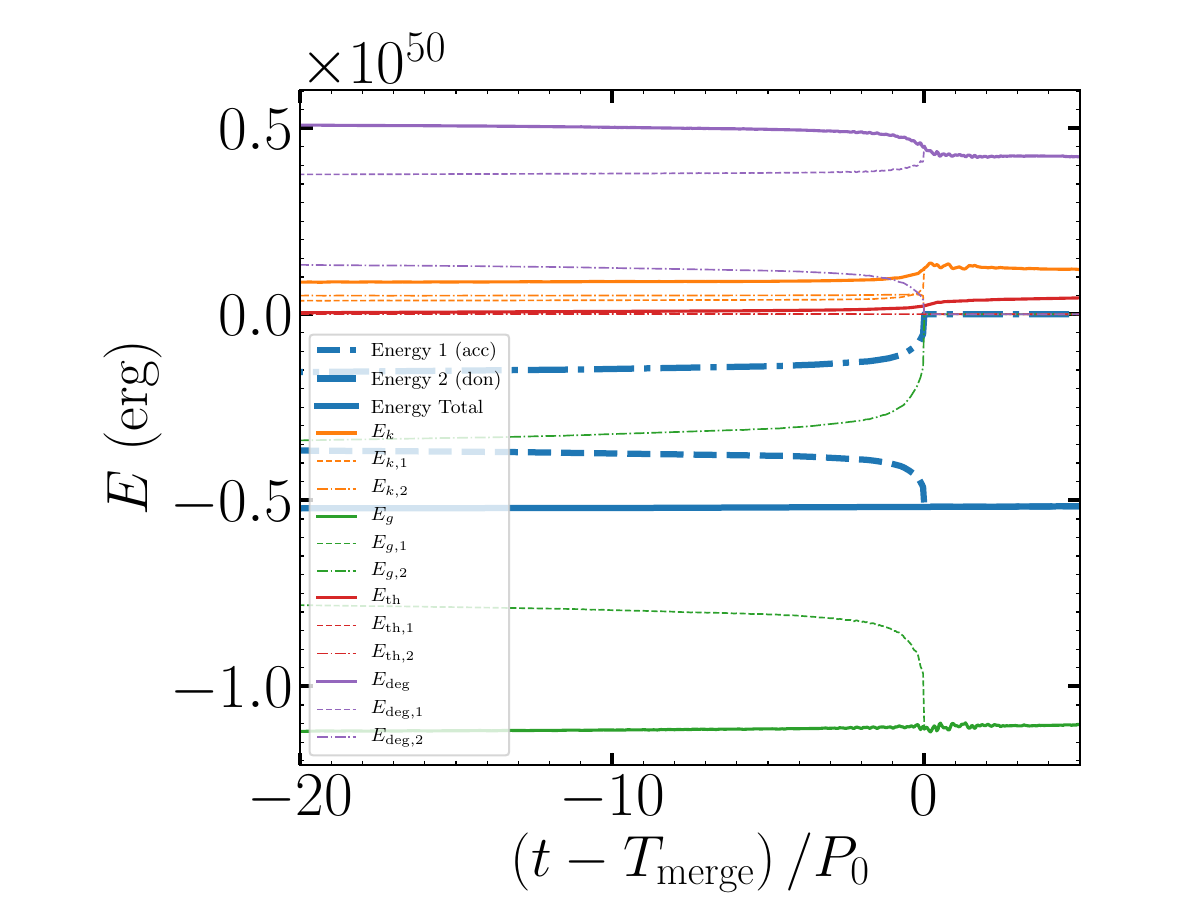}
    \caption{The evolution of the kinetic (orange), gravitational (green), thermal (red), degeneracy (magenta), and total (thick blue) energies for accretor (dashed), donor (dot-dashed), and for the binary system (solid) according to simulation L12. Note how the contributions of the donor vanish, and those of the accretor become the system's at the merger.
    }
    \label{fig:l12_energy}
\end{figure*}

To understand better the energy distribution prior to and during the merger, we plot in
Figure~\ref{fig:l12_energy}
the four energy components for each star and the system: kinetic ($E_k$, orange), gravitational ($E_g$, green), degenerate internal ($E_{\rm deg}$, magenta), and thermal internal ($E_{\rm th}$, red). The total for each of these four energy components for the system is shown as solid lines of the color assigned above. The quantities due to the accretor (denoted by 1) are shown as dashed lines, while the corresponding contributions by the donor (denoted by 2) are shown as dash-dotted lines. The total energy contributions for each component and the system are shown as thick blue lines. Note that the degenerate internal energy is calculated directly from the density using the corresponding expression for the ZTWD, while the thermal internal energy is calculated by a procedure described in detail in Appendix~\ref{app-eos}. All of these energy quantities are measured in the inertial frame. 
The material flowing out of the donor is heated when it impacts the accretor at a high speed. This translates to a decline in the donor's degenerate energy and an increase in the accretor's thermal and kinetic energies. Eventually, all the energy contributions of the donor vanish at the merger. Interestingly, the donor material accreted onto the accretor still possesses some degenerate energy, and the total degenerate energy of the system after the merger is slightly greater than the initial accretor's degenerate energy. 

\begin{figure}
    \centering    
    \includegraphics[scale=0.36]{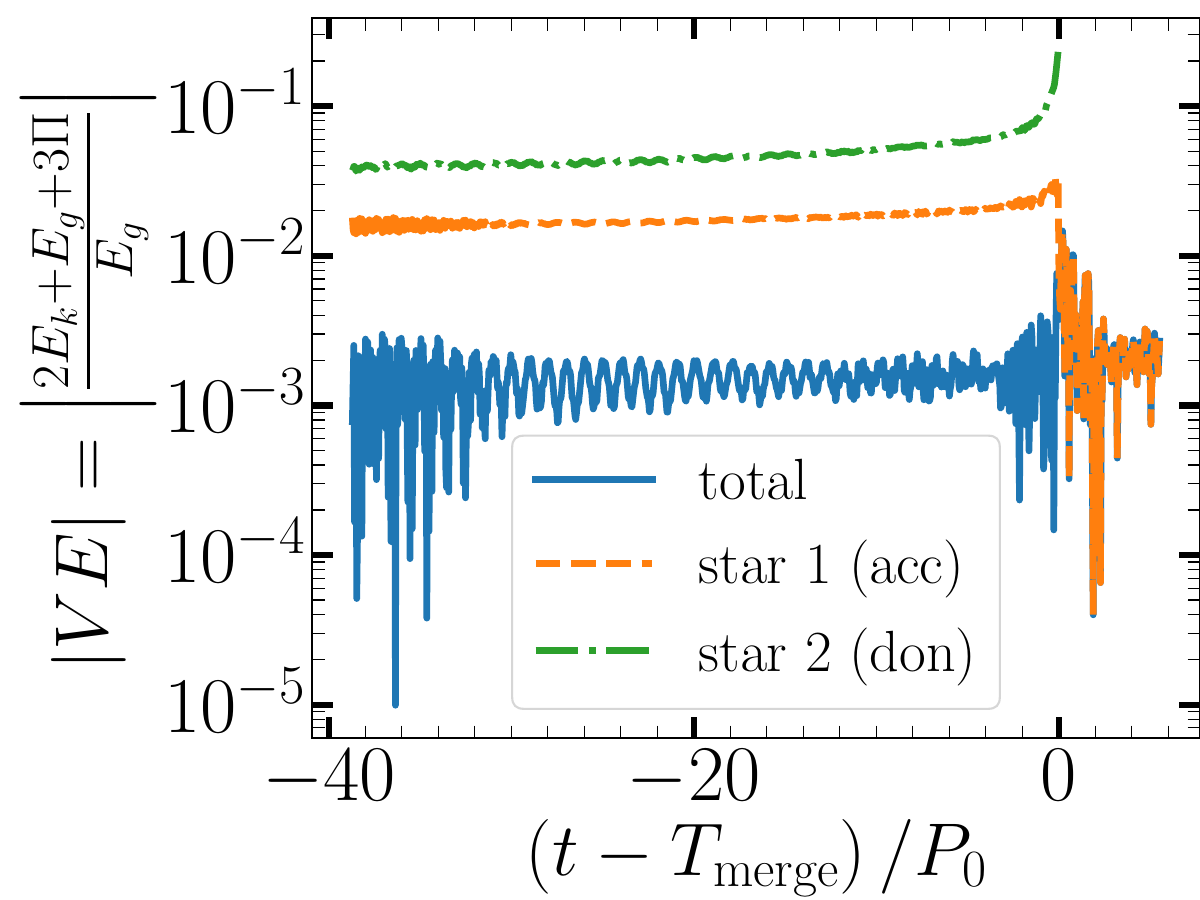}
    \caption{The virial error (VE) for the accretor (dashed, orange), the donor (dot-dashed, green), and the binary system (solid, blue). At the merger the VE for the donor diverges as $E_g$ vanishes, and the VE for the accretor is essentially that of the system.
    }
    \label{fig:l12_ve}
\end{figure}

Figure \ref{fig:l12_ve} shows the ``virial error", defined as ${\rm VE} =\left|2E_k + E_g + 3\Pi\right|/\left|E_g \right|$, where $E_k$, and $E_g$, are the kinetic and gravitational energy, respectively, while $\Pi=\int P dV$ is the total pressure integrated over volume. The closer this quantity is to zero, the more accurate the numerical representation of the system. We plot the virial error as a function of time for simulation {\sc L12}, although all other runs show the same trend as well. Initially, both stars individually obey the virial relation only approximately, because they are not isolated systems. The error is greatest for the donor (green dashed-dotted line) since we are ignoring the more massive companion. As the donor approaches disruption, the relative virial error increases. In contrast, the virial error for the system is $\sim 10^{-3}$ throughout the evolution, and even after the merger (blue solid line), suggesting that our numerical representation of the system is very precise.

\subsection{Numerical, physical and conservation properties of the simulations}
\label{ssec:numeric_eff}
In this section, we discuss the results in light of the numerical properties of the simulations. {As a first step, we have analyzed our results during the early mass transfer to conclude the nature of the mass transfer (Section~\ref{ssec:mass-transfer})}. Consecutively, we have tested several numerical properties of our simulations such as the effects of resolution (Section~\ref{ssec:res}), of carrying out the simulation in the rotating and inertial frames (Section~\ref{ssec:inertial}), and ultimately the amount of driving during the early part of the interaction (Section~\ref{ssec:nondriven_sims}). {For another numerical test, where we have found remarkably similar results across the different codes, \octo\ and \flower, we refer the reader to Appendix~\ref{app-flower}}.

\subsubsection{{Instability of mass transfer}}
\label{ssec:mass-transfer}

The total angular momentum of a simple binary model, in which the component stars have constant masses and moments of inertia, and are tidally locked to the orbit, assumed circular, is the sum of orbital and spin terms at the same angular velocity. If the orbital-spin frequency is given by the usual point-mass Keplerian approximation, then the total angular momentum depends only on the binary separation $a$, and has a minimum at a separation $a_{\rm min}^2=3(I_1+I_2)/\mu$, where $\mu=M_1M_2/(M_1+M_2)$ is the two-body reduced mass \citep{Rasio1995}. With the initial moments of inertia of the equilibrium model binary, we find $a_{\rm min}\simeq 0.48 a_0$.  

If angular momentum losses drive such a binary to the above minimum, no further reduction of total angular momentum and separation is possible unless synchronism is broken, and one or both components spin slower than the orbital frequency. But then tides will attempt to synchronize the lagging spins, reducing further the orbital angular momentum, and we have a run-away process, known as the Darwin instability \citep{Darwin1879}. Therefore we expect that in a system driven by the Darwin instability, both orbital and spin frequencies increase with time, while the spin frequencies remain lower than the orbital. In all of our simulations the unstable behavior begins well before $a_{\rm min}$ is reached (see below). 

The simple arguments given above require revision if mass transfer begins at a separation exceeding $a_{\rm min}$, if the mass is lost from the system, or if the tidal distortions of the stars are significant \citep{LRS1994}. All of these effects come into play in our simulations. For example, mass loss through the outer Lagrange points, and direct impact accretion by the CO WD, are consequential angular momentum losses (hereafter CAML) \citep{Webbink1984, KingKolb1995, Schreiberetal2016}, meaning orbital angular momentum losses that are consequences of mass transfer. CAML and the expansion of the degenerate donor due to mass loss, drive the mass transfer instability and ultimately cause the tidal disruption and the merger of the system. Because the accretor is spun up by mass transfer, its spin frequency increases faster than the orbital frequency.

\begin{figure}
    \centering
    \includegraphics[scale=0.36]{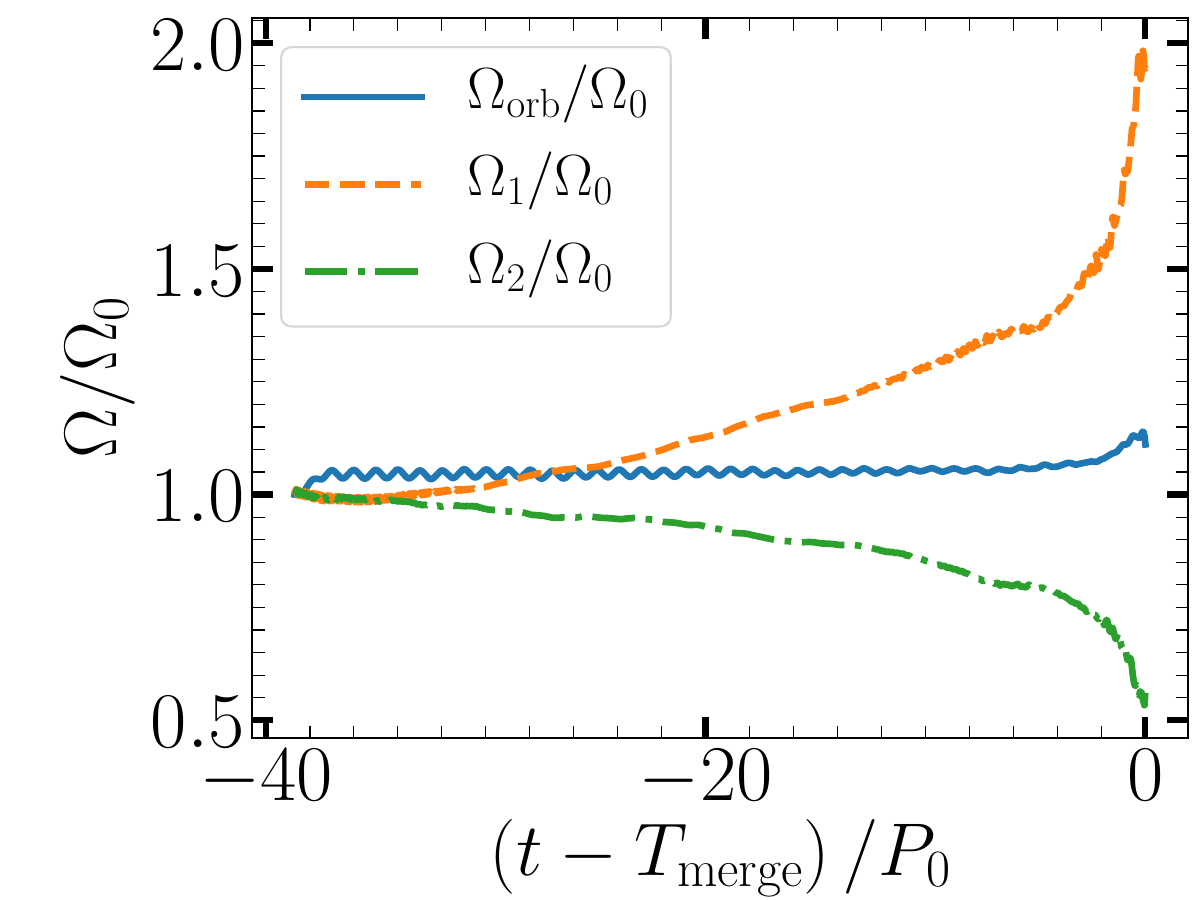}
    \caption{The angular velocity of the orbit ($\Omega_{\rm orb}$), the accretor ($\Omega_1$) and the donor ($\Omega_2$) in the level 12 simulation ({\sc L12} in Table~\ref{tab:past_sims}) divided by the system's initial angular velocity $\Omega_0 = 2 \pi / P_0 = 3.32~{\rm rad/min}$ as a function of time. The system was driven for 1.3 orbits at a rate of one percent of its orbital angular momentum per initial orbit }
\label{fig:l12_bennett}
\end{figure}

In Figure \ref{fig:l12_bennett} we display the angular velocities for the orbit and the spins of the binary components.  
We see in this figure that the orbital angular frequency does not change much while the donor angular frequency decreases somewhat and the accretor spin frequency increases rapidly. This behavior is a characteristic of mass transfer instability and implies that the Darwin instabilities does not play any role in the merging process.
{We further analyze the binary evolution of our highest resolution {\sc L13} run, showing the excellent agreement between our simulations and the analytical expression in Appendix~\ref{app-verification}.}

\subsubsection{The effect of spatial resolution}
\label{ssec:res}

As can be seen in Figure~\ref{fig:q07_level12_rot_cons}, \octo\ conserves mass and energy very accurately to the level of machine precision, regardless of the resolution being used. The source of non-conservation stems from the low values (floor values) of density and energy artificially introduced in a few cells to prevent the occurrence of vacuum conditions. 
In contrast, a finer grid does improve the conservation of angular momentum. For {\sc L10ND} and {\sc L11} the angular momentum deviates by less than 4\% by the time of the merger; however, the strong shearing during the violent disruption 
tends to contribute to conservation errors.
Simulations {\sc L12} and {\sc L13} display a better conservation of angular momentum at the level of less than a percent even 5 orbits after the merger. 

\begin{figure*}
    \centering
    \includegraphics[scale=0.33]{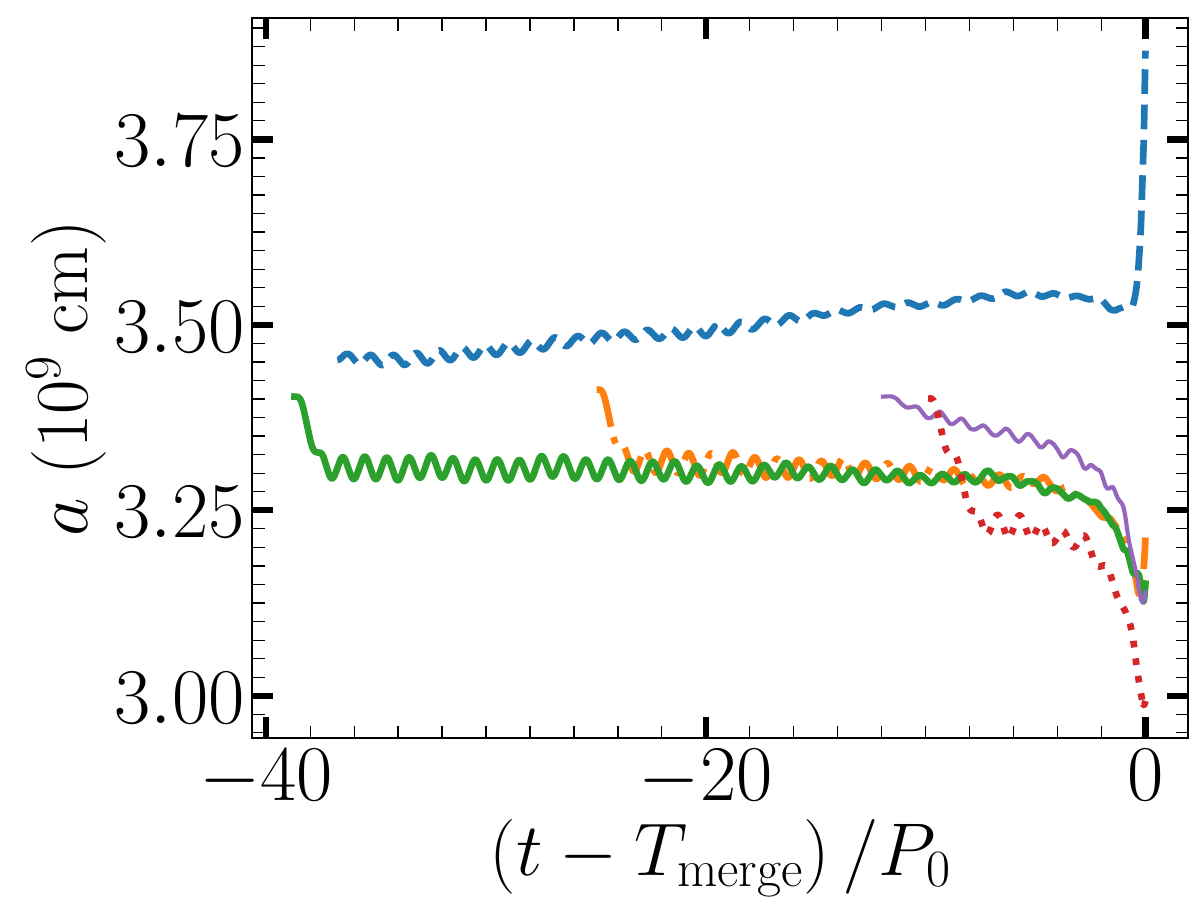}
    \includegraphics[scale=0.33]{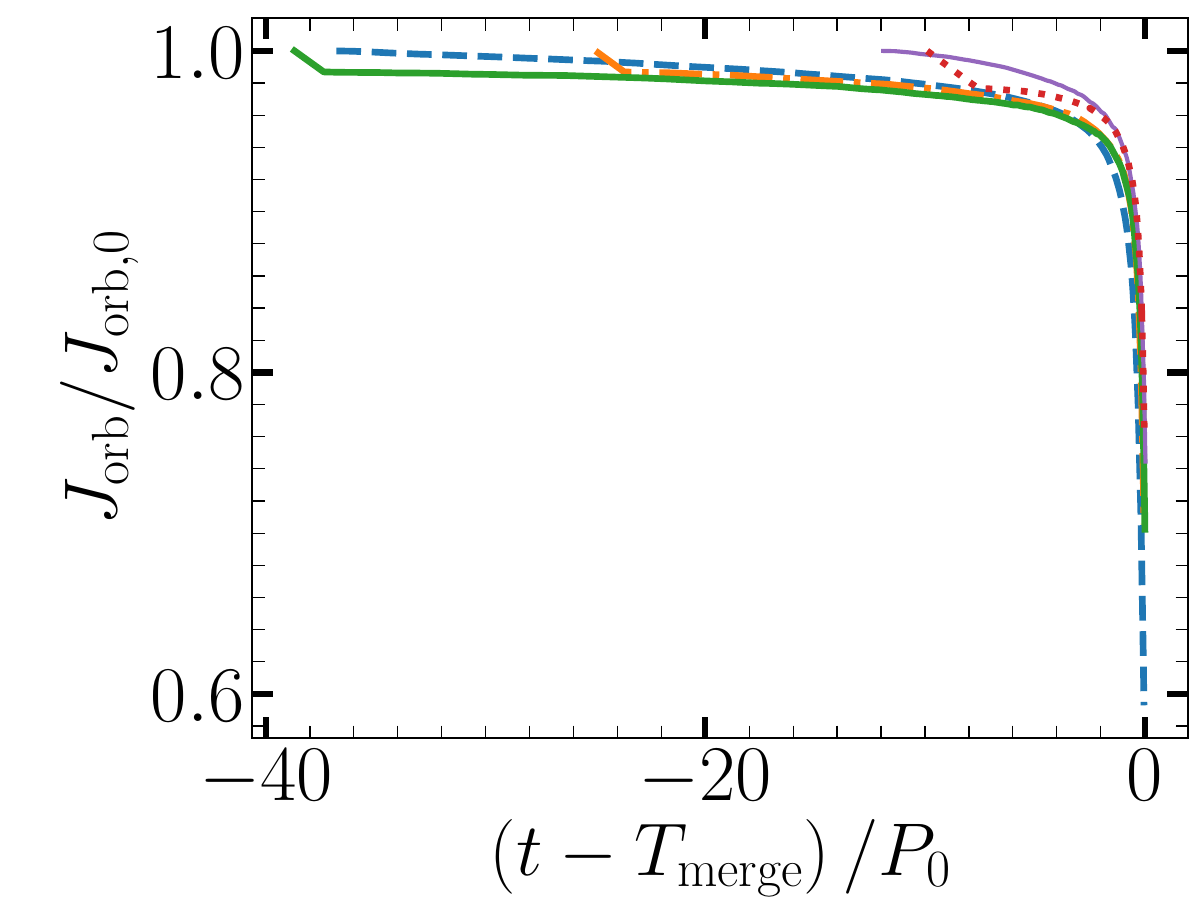}
    \includegraphics[scale=0.33]{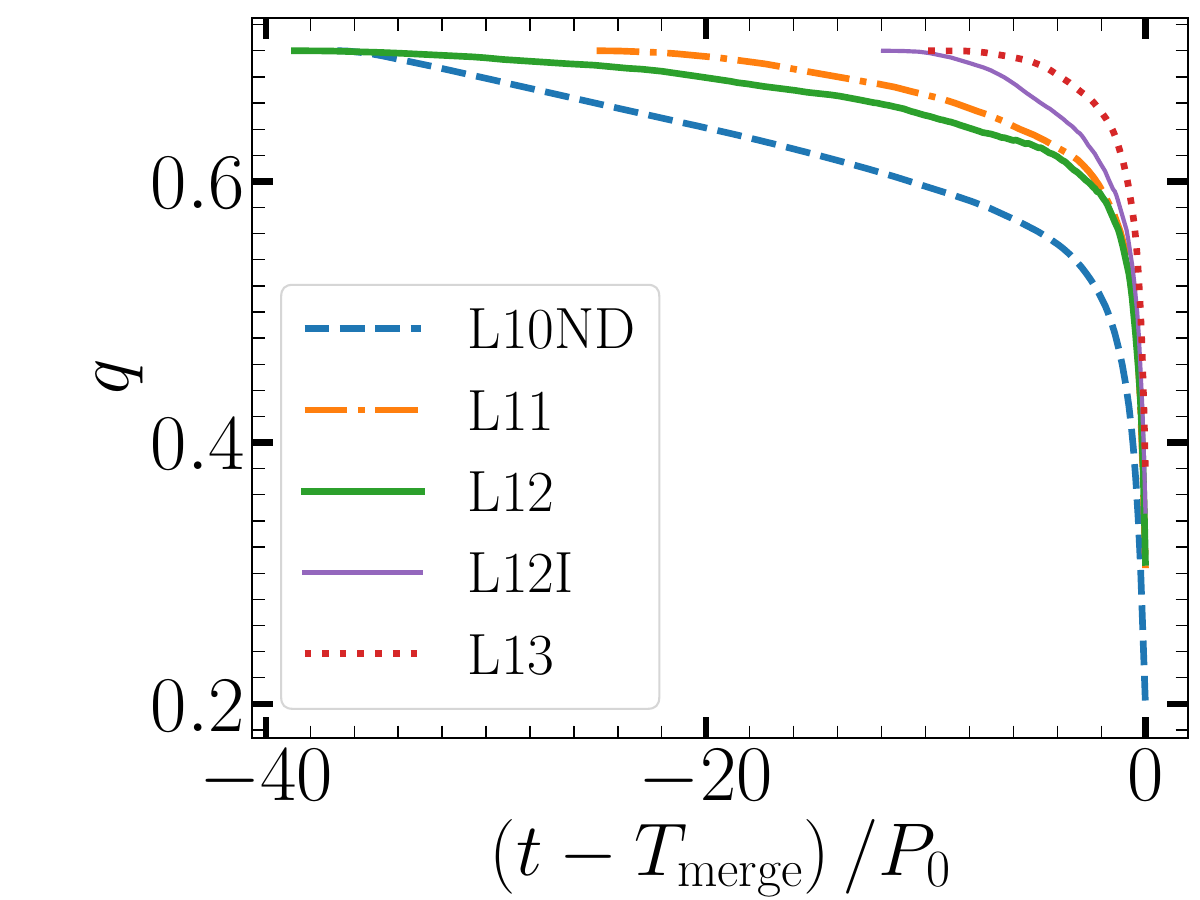}
    \includegraphics[scale=0.33]{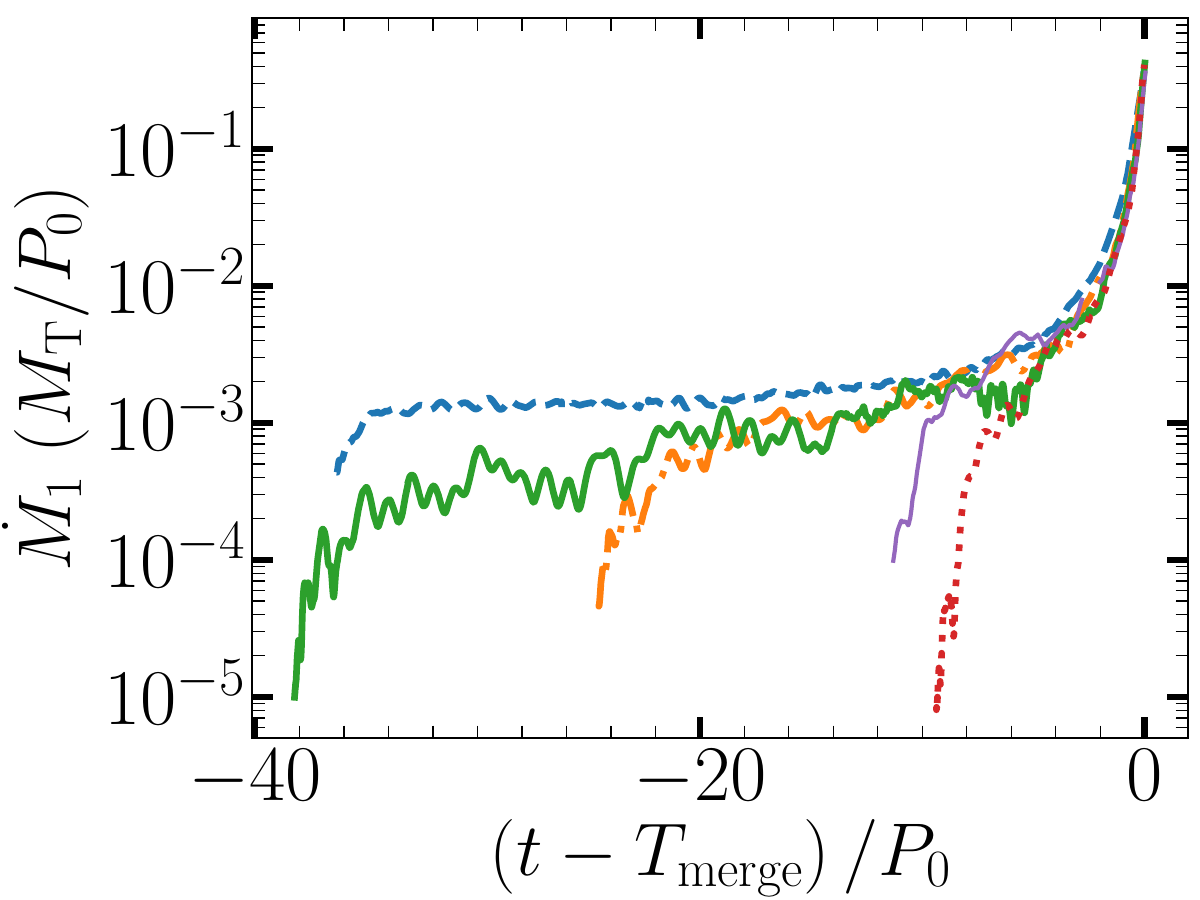}
    \includegraphics[scale=0.33]{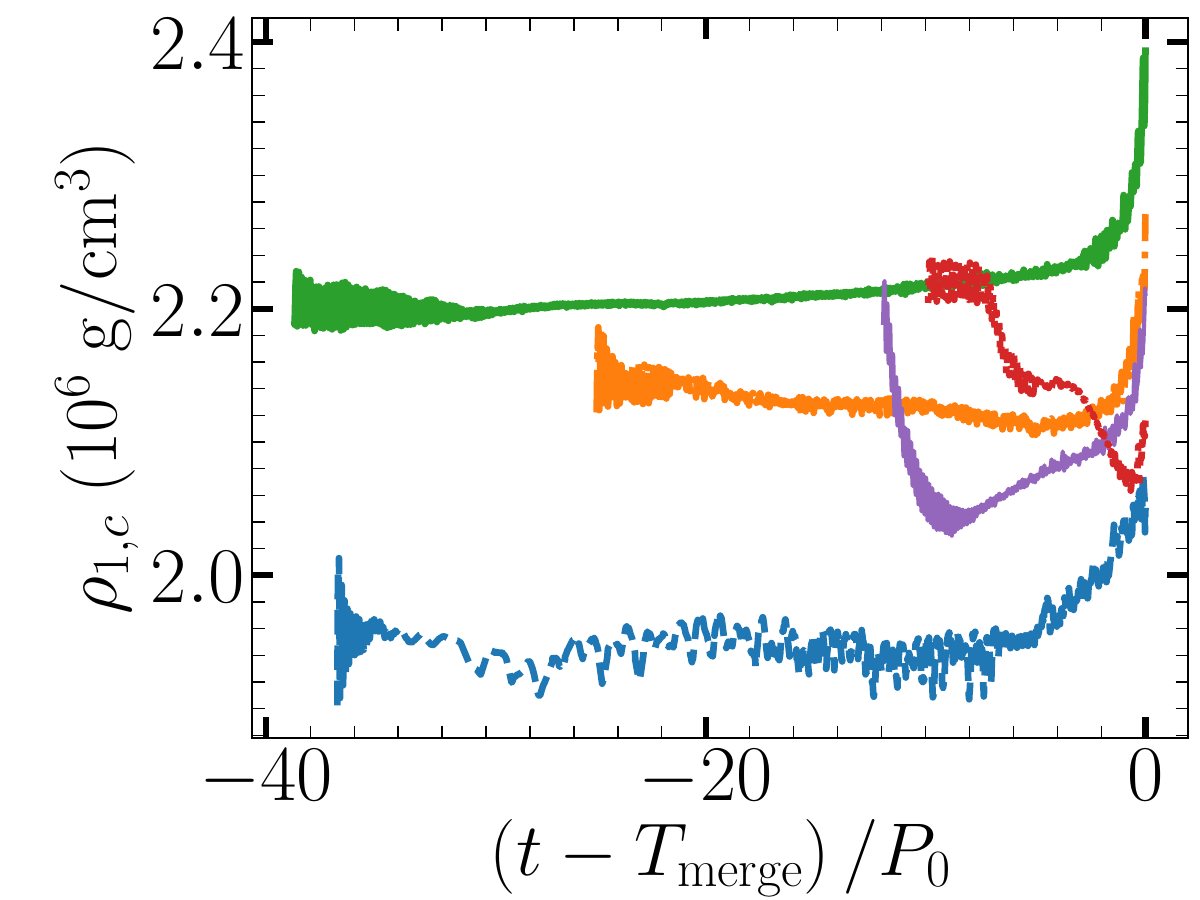}
    \includegraphics[scale=0.33]{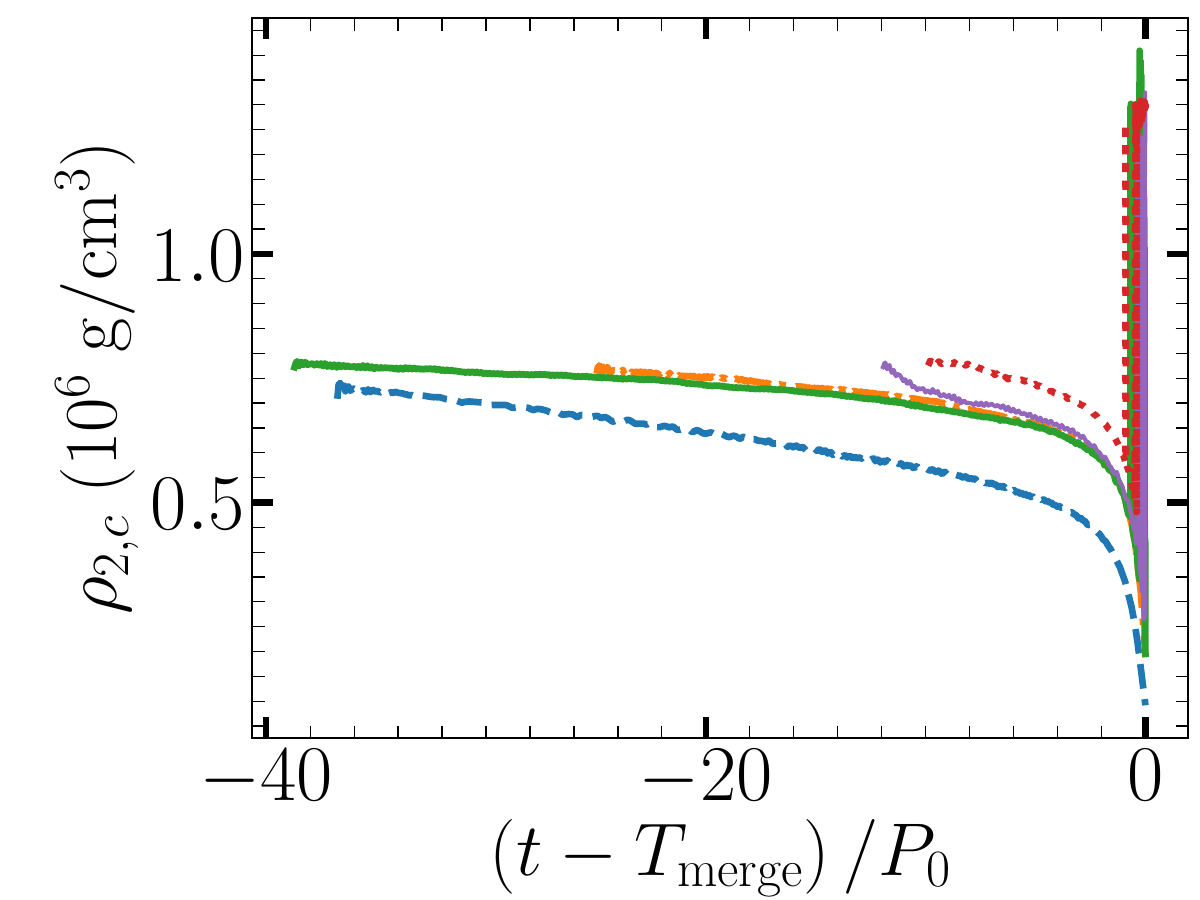}
    \caption{Comparison between different resolutions. Shown are the orbital separation, orbital angular momentum, mass ratio, mass transfer rate, accretor, and donor central densities, all as a function of time. 
    }
    \label{fig:comp_res}
\end{figure*}

In Figure~\ref{fig:comp_res} we present a set of 
diagnostic quantities plotted as a function of time for {\sc L10ND}, {\sc L11}, {\sc L12} (our reference simulation), {\sc L12I} (same as {\sc L12}, but performed in the inertial frame, see a detailed comparison in Section~\ref{ssec:inertial}), and {\sc L13}.
Times have been shifted by $T_{\rm merge}$ so that the diagnostic quantities are shown lined up at the same time before the merger. 

The smallest cell size in simulation {\sc L10ND} is $1.2\times 10^{-3}~{\rm R_{\odot}}$, which lets us resolve the donor diameter into 36 cells. Each additional level of refinement roughly halves the smallest cell size, and doubles the 
number of cells across the donor, so in {\sc L13} the donor diameter is resolved into 290 radial cells. Simulations {\sc L10ND} and {\sc L12I} were not driven, while the other three were driven, {\sc L11} and {\sc L12} for 1.3 orbits and {\sc L13} for 2.3 orbits, as described in Section~\ref{sec:methods}. 
{\sc L12} was driven similarly to {\sc L11} and clearly merges later, as expected of a higher resolution simulation. So as to reduce the computational times, we have driven {\sc L13} by 2.3 orbits, which resulted in a faster merger.

The evolution of the orbital angular momentum is qualitatively the same in all our simulations, {slowly declining at a rate of $10^{-4}-10^{-3}~J_{{\rm orb},0}/P_0$ through most of the evolution (from the moment the driving has stopped until $\approx 5$ orbits before the merging time), and then rapidly dropping at rate of $>10^{-2}~J_{\rm orb}^0/P_0$ just before the merger as the donor is tidally disrupted. } During the
tidal disruption, the orbital angular momentum is transferred to the spin of the accretor as discussed in Section~\ref{ssec:evolution}. 

The mass ratio, initially $0.7$ for all simulations, decreases drastically close to the merger. In fact, the accretor in all of our simulations gains mass throughout the evolution, decreasing the mass ratio monotonically. {\sc L11} (dashed-dotted orange line) and {\sc L12} (solid green line) that were driven at the same rate and the same duration are the most similar, especially as illustrated by the evolution of the mass ratio $q$ and the mass-transfer rate $\dot M_1$. While we expect a similar contact depth in both runs, {\sc L11} overestimates the initial mass-transfer rate because of the larger cell sizes. Simulation {\sc L12} starts at a lower rate and takes around 20 orbital periods to ``catch up" with {\sc L11}, but the rates agree very closely after that. Because the mass transfer rate is a very noisy quantity we smooth it using a Savitzky-Golay filter \citep{savitzky1964smoothing} with a temporal window size equal to half of an orbit. As expected, the initial mass transfer is primarily a function of resolution, and higher resolution simulations yield lower initial mass transfer rates. Nonetheless, the mass transfer rate during the final several orbits before the merger is very similar across all simulations. 

As mass transfer proceeds, the expectation is that the central density of the accretor should increase, while that of the donor should decrease. The behavior of the primary central density shows some variation initially but conforms to expectations as the mass transfer increases.
The secondary's central density behaves in a more homogeneous way: we observe in all simulations a decline just before the merger.

Our conclusions from the resolution study are as follows. First, resolution is important for an accurate angular momentum conservation. In addition, although the merging time itself depends on the initial depth of contact, which is a function of both resolution and amount of driving, the general behavior before the merger is similar regardless of resolution. The WD-WD system merged in all of the simulations, regardless of resolution, supporting that WD-WD systems with mass ratio of 0.7 are unstable once mass transfer begins. This agrees well with previous simulations and analytical expectations (e.g., \citealt{Motl2017}). lastly, in the following section we will show that resolution plays a critical role in determining the properties of the shell of fire (Section~\ref{ssec:SoF}) and the amount of {O$^{16}$} that is being dredged-up during the merger (Section~\ref{ssec:O16}).

\subsubsection{Inertial vs Rotating frame simulations}
\label{ssec:inertial}

To illustrate the advantages of conducting simulations in the co-rotating frame, we ran two simulations at the same resolution: {\sc L12} on the rotating frame, driven for $1.3$ orbits, and {\sc L12I} on the inertial frame, undriven. Simulation {\sc L12I} involves advection of large flows across cells just to represent the orbital motion of the binary components, while mass transfer and internal motions in the binary components would be given by relatively small additional fluxes. Therefore we expect {\sc L12I} to be prone to cumulative errors and larger diffusion. 

Referring to Figure \ref{fig:comp_res} and focusing on the differences between the thick green curves ({\sc L12}) and the thin magenta curves ({\sc L12I}), we note that {\sc L12I} ran to merger much faster than L12, even without driving. The orbital angular momentum decreased from the start, and mass transfer was initially larger and increased rapidly. The central density of the accretor decreased rapidly for a couple of orbits from the beginning due to stronger numerical diffusion, and only began increasing once the mass transfer reached levels comparable to those present in L12, at $\sim 8 P_0$ before the merger. Qualitatively the behavior over the last 4-5 orbits prior to merger was similar for both {\sc L12} and {\sc L12I} as the system rapidly evolved to merger. The main differences can be seen in the binary separation, orbital angular momentum, and binary mass ratio. All of these early differences can be attributed to the accumulation of numerical errors and diffusion. Later in the evolution, when the unstable mass transfer is large and growing exponentially during the few orbits before the merger, the differences between {\sc L12} and {\sc L12I} become less significant since advection dominates in both frames. The main problem with evolutions computed on the inertial frame is that numerical viscosity artificially speeds up the merger. If one is interested in marginally stable/unstable cases, the evolution should be computed in the rotating (comoving) frame.

\subsubsection{Non-driven simulations}
\label{ssec:nondriven_sims}

Most previous merger simulations of WD-WD systems have used some kind of driving mechanism to expedite the merger. Moreover, studies like \cite{Motl2017} and \cite{diehl2021octo} have shown that a shorter driving phase results in a longer evolution to a merger. For practical reasons, the numerical driving rates used in these simulations exceed realistic angular momentum loss rates by factors $\sim 10^9$ or more (from mechanisms like magnetic braking or gravitational waves emission). In this paper, we wanted to investigate the feasibility of simulations closer to realistic driving rates by simply simulating a non-driven system at different resolutions. We know that the lowest mass-transfer rate resolvable by a simulation depends on the level of refinement or spatial resolution. By simulating the evolution of the same binary at increasing resolution we may be able to see convergence to a more realistic evolution. Unfortunately, as discussed below, it turned out that the cumulative effect of numerical errors made these simulation results unreliable.

We begin by discussing the low-resolution non-driven run, {\sc L10ND}. This simulation merged in $37$ orbits, even later than {\sc L11}, which was driven for 1.3 orbits, merging in $25$ orbits, and later than {\sc L13}, which was driven for 2.3 orbits, merging in $10$ orbits. However, {\sc L10ND} merged before {\sc L12}, which was driven for 1.3 orbits, merging in $39$ orbits; see Table~\ref{tab:past_sims}. This is because, with the limited resolution of {\sc L10ND}, the initial mass transfer, even without driving, quickly settles on values of the order of $~\times 10^{-3}$ ($M_{T}/P_0$), higher by a factor of $\sim2-5$ than the mass transfer rate at the start of simulation {\sc L12} (middle left panel of Figure~\ref{fig:comp_res}). Thus, as we anticipated, the resolution plays a critical role in the early evolution of the mass transfer, and {\sc L10ND} is probably a poor approximation to a realistic evolution. 

We have therefore tried to run non-driven simulations of higher resolution, with 11 and 12 levels of refinement. However, these simulations, which start with lower mass transfer rates than their equivalent driven simulations and thus evolve slower, suffer from some numerical issues, happening after the merging time of their equivalent driven simulations. Specifically, we have found out that in these simulations the merger occurs only because the accretor starts expanding for reasons other than the mass transfer. This expansion of the accretor stems from internal, convection-like, instabilities that form inside the accretor and which weren't observed for any of the other simulations (listed in Table~\ref{tab:past_sims}). We could even reproduce those instabilities in single-star simulations and have linked them to the implemented ZTWD EoS (see Appendix~\ref{app-nondriven} for more details). In contrast, the simulations of a polytropic binary WD with the same mass ratio of $0.7$, evolved with an ideal gas EoS (\citealt{diehl2021performance}; see Table~\ref{tab:past_sims}) at four resolutions, all driven with $T_{\rm drv}=2.0$, clearly show convergence with respect to merging time. This further supports that the anomalous behavior observed in some of our non-driven simulations is linked to the EoS. A further investigation of this phenomenon, including the testing of the Helmholtz EoS \citep{Timmes2000} instead of the ZTWD should be considered in order to resolve this issue. We conclude that at this point, with the current implementation of the EoS in \octo,
a convergence study of the evolution of a non-driven system is not feasible.  

\section{The merged object}
\label{sec:post_merger}

We base this discussion on our reference simulation {\sc L12}, which is performed in the rotating frame, 
and has 12 levels of refinement. When necessary, we compare with the results of other simulations. In Section~\ref{ssec:postmerger-structure} we discuss the structure of the merged object, splitting it for convenience into several substructures with characteristic densities and rotational properties, and consider their origins. In Section~\ref{ssec:SoF} we consider nuclear burning and the potential impact of the corresponding energy deposition. Finally, in Section~\ref{ssec:O16} we estimate the dredge-up of core material by extrapolation of results at different resolutions, and discuss the consequences of a small amount of hydrogen in the donor, and its impact on the observed $^{16}$O/$^{18}$O ratio.

\subsection{Structural and rotational properties}
\label{ssec:postmerger-structure}

\begin{figure*}
    \centering
    \includegraphics[scale=0.3,trim={2.05cm 1.55cm 1.95cm 0}, clip]{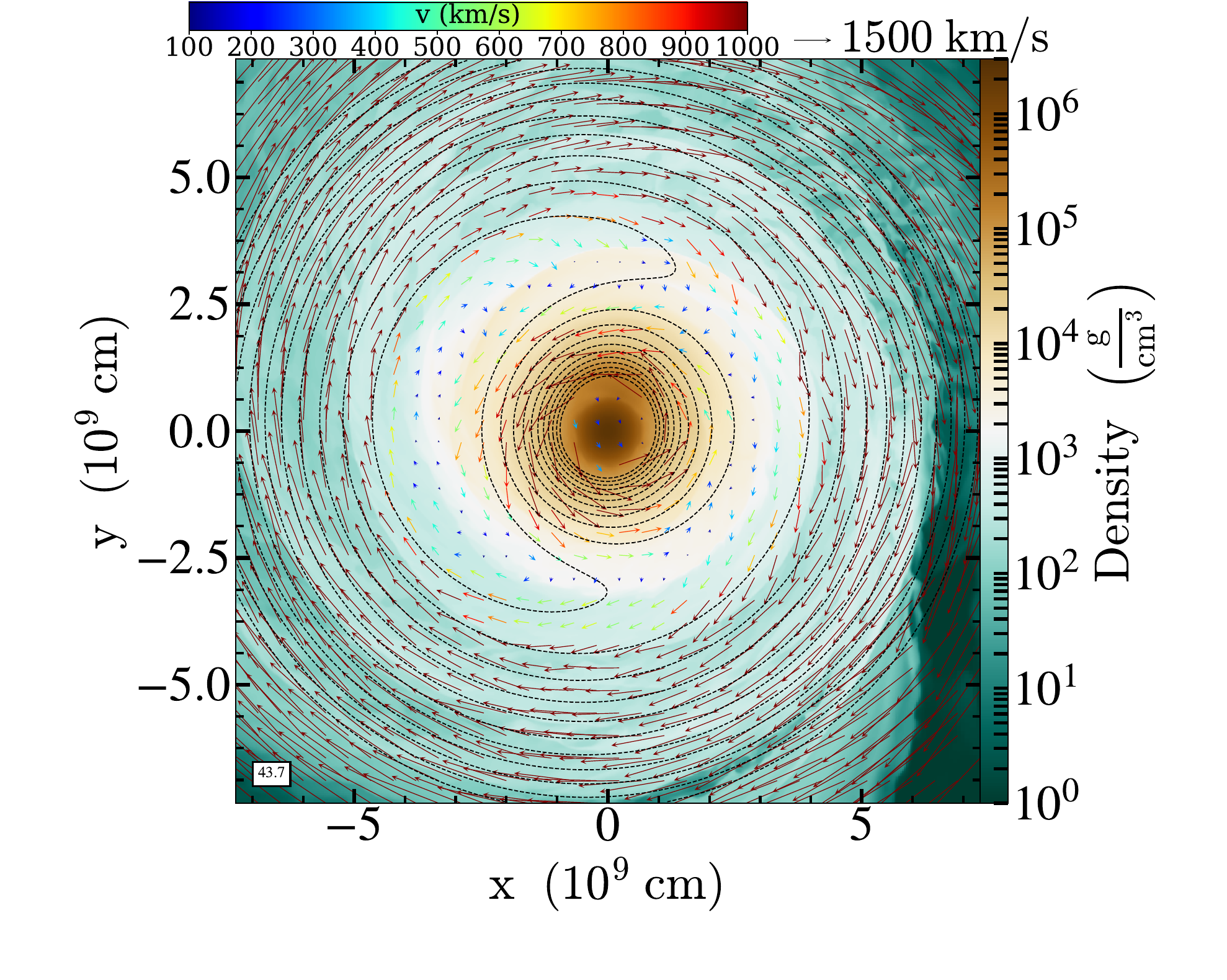} \includegraphics[scale=0.3,trim={2.05cm 1.55cm 1.95cm 0}, clip]{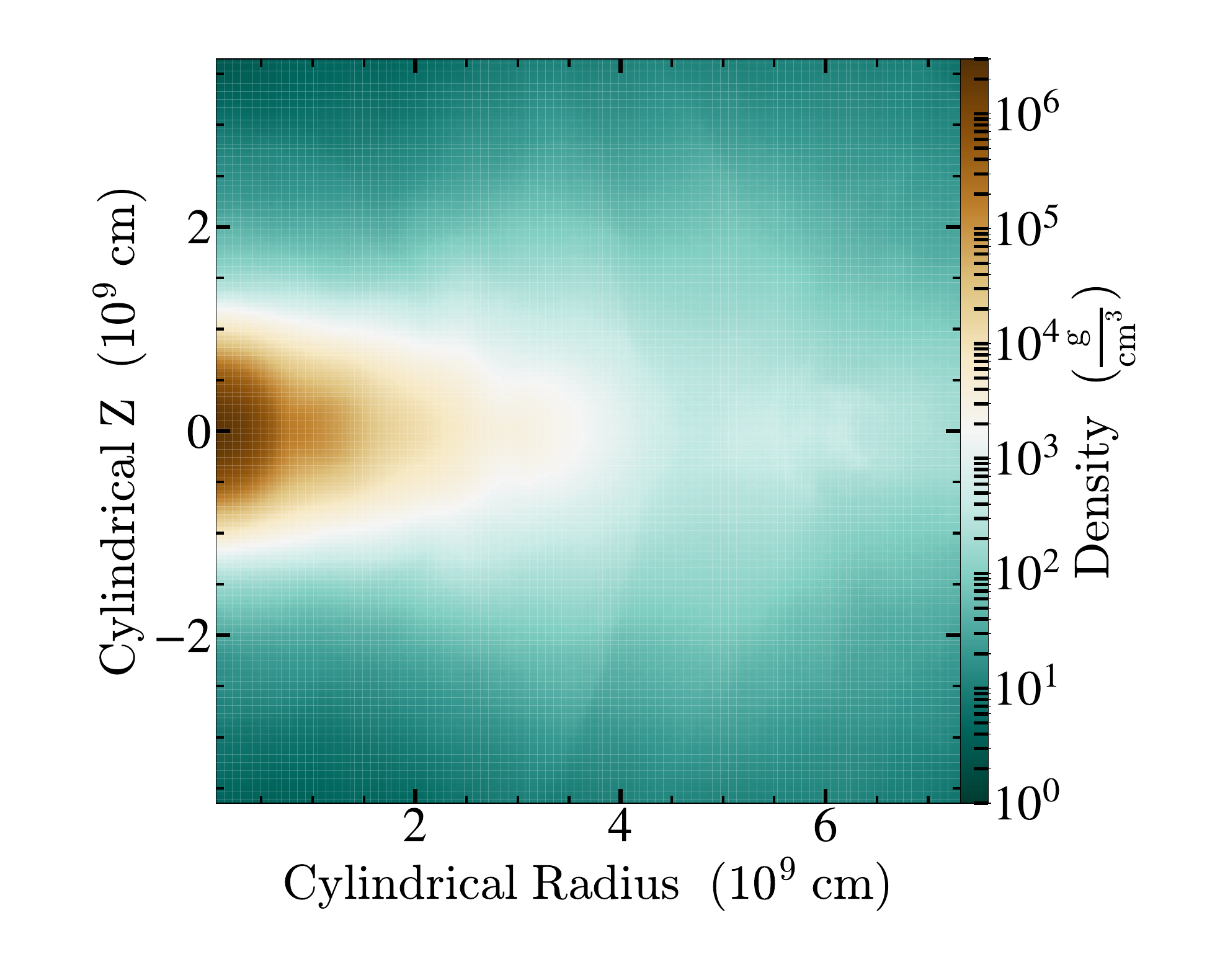}
    \includegraphics[scale=0.4]{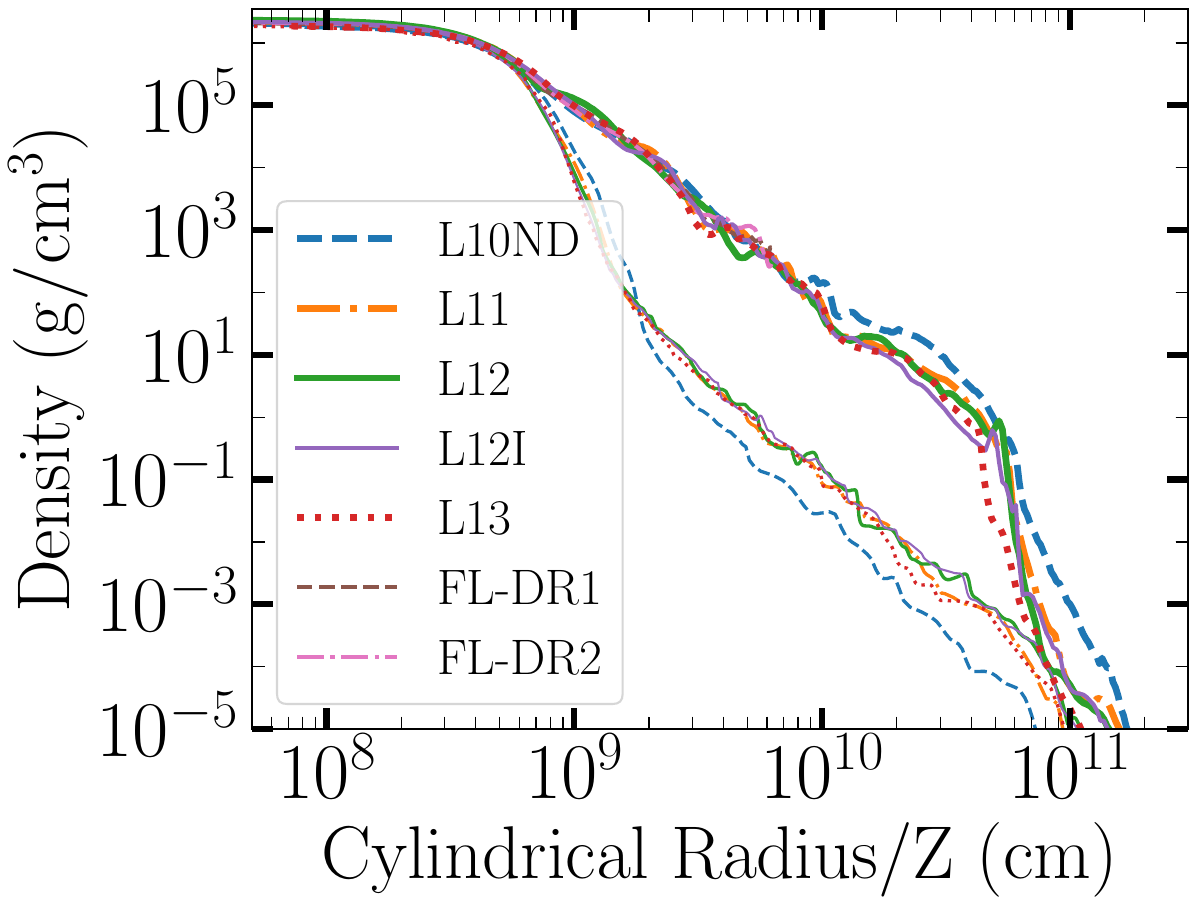}
    \includegraphics[scale=0.4]{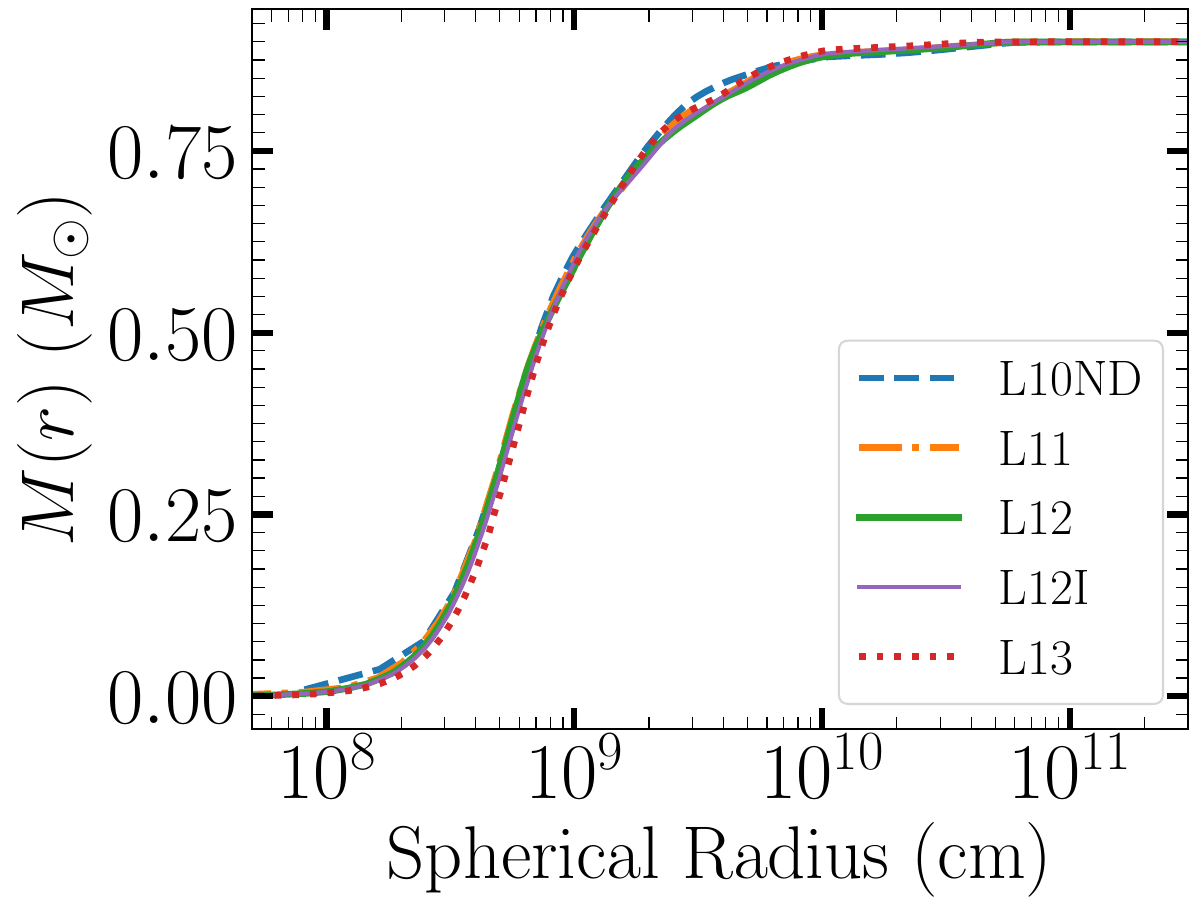}
    \caption{{Mass distribution of the merged object 5 initial orbits past merger}. Top row: a density slice along the equatorial plane (left panel), and a mass-averaged over the azimuthal angle density profile (right panel) for simulation {\sc L12}. Velocity arrows are overlaid with their color indicating the magnitude of the projected velocities in the rotating frame (see text and colorbar values as in Figure~\ref{fig:l12_ev_dens}). Bottom row: the azimuthally-averaged density profile (bottom left, thick lines), the density profile along the z-axis (bottom left, thin lines), and the cumulative mass $M_{\rm in}(r)$ profile of the merged object for all simulations. Colors identify the different resolutions according to the legend.}
    \label{fig:post_merger}
\end{figure*}

\begin{figure*}
    \centering
    \includegraphics[scale=0.3,trim={2.05cm 1.95cm 1.95cm 0}, clip]{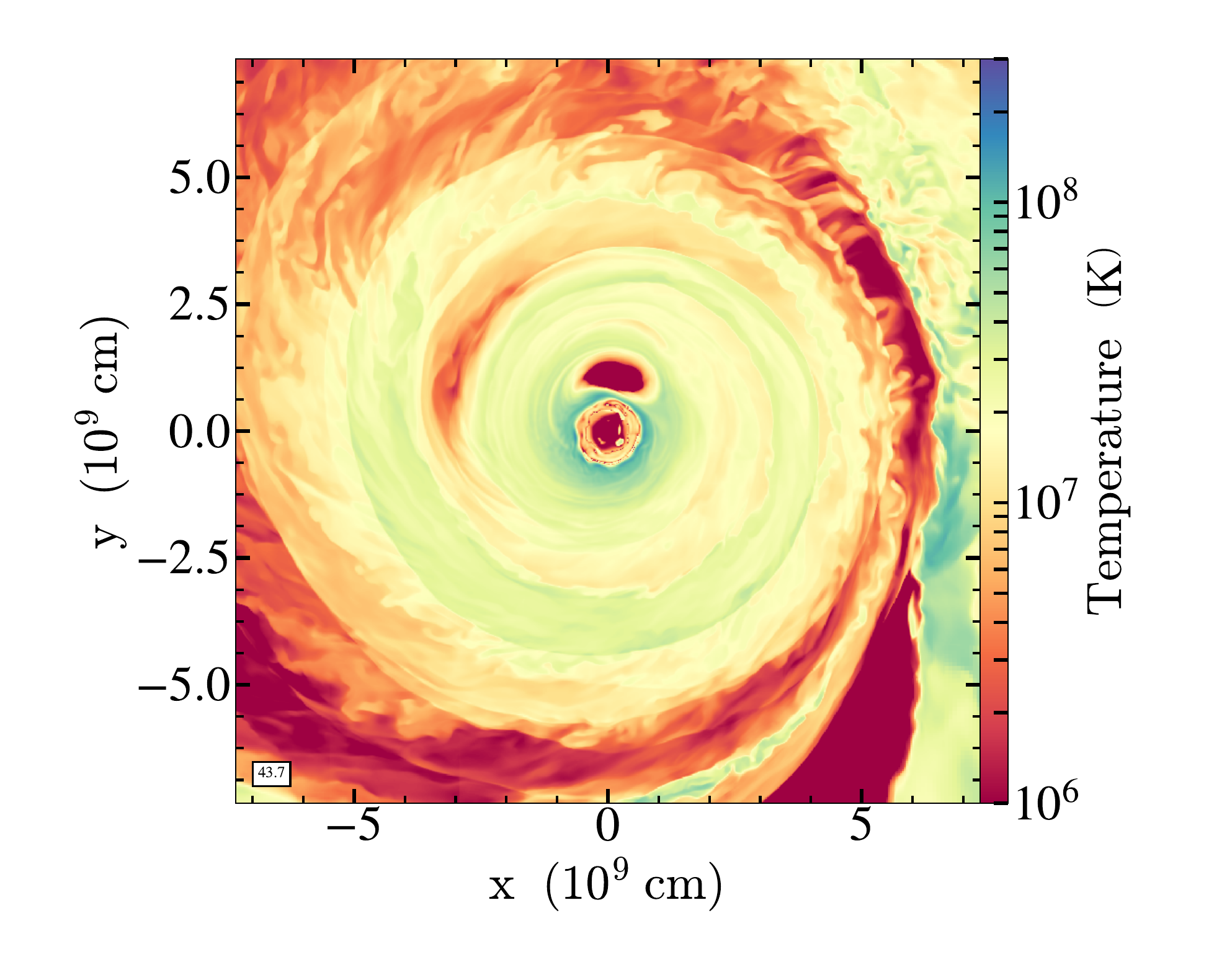} \includegraphics[scale=0.3,trim={2.05cm 1.95cm 1.95cm 0}, clip]{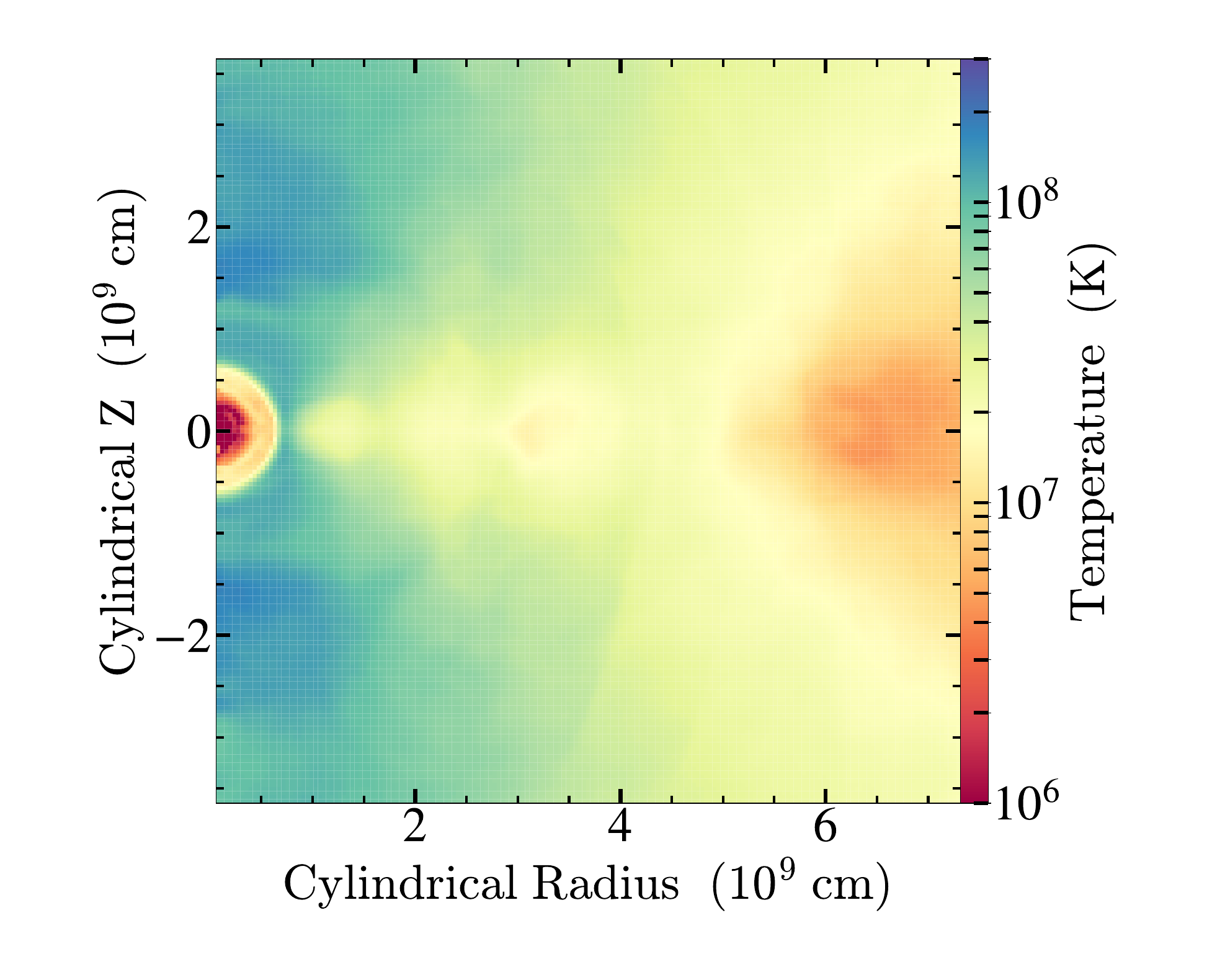}
    \caption{The SoF 5 initial orbits past merger. Equatorial temperature slice (left) and a mass-averaged over the azimuthal angle temperature profile (right) for simulation {\sc L12}}
    \label{fig:SoF}
\end{figure*}

In Figure~\ref{fig:post_merger} we focus on the mass distribution of the merged object, 5 orbits after the merger. In the top row, we show a density slice at the orbital, $xy$, plane (left), and a density profile mass-averaged over the azimuthal angle (right), both for simulation {\sc L12}. Cylindrical radius and cylindrical z are measured from the center of the merged object, \emph{i.e}, the point of maximum density. 
Although difficult to see in the top left panel because of the velocity arrows, there is a higher density ``blob" at the noon position at radii $\sim 1.2\times 10^9$ cm, coincident with the cooler blob, clearly visible in Fig. \ref{fig:SoF}, which will be discussed in the next Section. This is a remnant of the core of the donor, which cannot be tidally disrupted by the accretor before it becomes supported by pressure and rotation, but is expected eventually to be sheared by differential rotation. The effect of this blob is also visible in the top right panel, despite the azimuthal averaging. 
In the bottom left frame, we show the azimuthally mass-averaged density profiles at $z=0$ (thick lines), and the vertical density profile along the $z$-axis (thin lines) for all of our \octo\ simulations. In the bottom right frame we present the cumulative mass profiles inside a sphere of radius $r$ for all of our \octo\ simulations. In addition, we plot the radial density structures on the equatorial plane of the \flower\ simulations (bottom left; thick lines). 

First, we see a remarkably similar structure across all of our simulations, and the following description is valid for each of them. The merged object's radius, defined as the cylindrical radius where the cumulative mass plateaus (bottom right frame), is approximately $R_{\rm out} \approx 3\times10^{10}$~cm. There is some low-density material beyond this radius gradually transitioning to the floor density. A side view (top right frame) and the profiles along the Cylindrical radius and cylindrical $z$-coordinate (bottom-left frame) show that the densest structures consist of a compact oblate spheroid, with an approximate cylindrical radius of $R_{\rm sph} = 10^9$~cm, and a flattened disk with an approximate outer radius of $R_{\rm dsk} = 3\times 10^{9}$~cm around it. We additionally observe that the density contours flare slightly at densities below $10^3$ g/cm$^3$. Gas velocities in the perpendicular plane are typically smaller than 400~km~s$^{-1}$ while the azimuthal velocities on the orbital plane are much larger and as we show later are nearly Keplerian. 

Additional description of the mass distribution in the inner regions, valid for {\em all} the simulations as well, can be gleaned from the bottom left frame of Figure~\ref{fig:post_merger}. The nearly constant density $\rho\sim 10^6$ g/cm$^3$ spherical region $r\le r_1=5\times 10^8$~cm corresponds roughly to the core of the CO WD. There is a gradual transition between this core and $R_{\rm sph}$. For $R_{\rm sph}\le R \le R_{\rm out}$~cm, the equatorial density in the disk-like or extremely oblate extension beyond the spheroid, falls off as $\rho\propto R^{-3}$ and beyond $R_{\rm out}$ it cuts off precipitously as $\rho\propto R^{-8}$. This is consistent with the $M(r)$ behavior shown on the bottom right panel.

Furthermore, we can learn the merged object rotation profile (on the equatorial plane) from the velocity arrows in the top left frame. The plotted velocities are in the frame rotating with the grid at a constant angular frequency of $\Omega_0 = 2 \pi / P_0$. Therefore, a velocity $v'$ in the rotating frame transforms to an inertial frame velocity of $v_{\rm in} = \Omega_0 r+ v'= 550~ {\rm km/s}\cdot r_9 +v'$, where $r_9$ is the cylindrical radius in units of $10^9$~cm. Consequently, the velocity at the transition from the spheroid to the disk just outside $R_{\rm sph}$, on the order of 1000~km~s$^{-1}$ in a counter-clockwise direction in the rotating frame (seen as red arrows), is $\sim 1500$~km/s in the inertial-frame, comparable to the sound speed at the SoF. Beyond $R_{\rm sph}$, $v'$ decreases to near zero at co-rotation, (seen as a white ring) located approximately at $R_{\rm dsk}$, and the direction of rotation in the rotating frame inverts beyond corotation, such that farther out, at a radial distance of 4$\times 10^9$~cm the rotation is of the order of $v'\approx -1000$~km/s in the clockwise direction, still turning counter-clockwise at $\approx 1000$~km/s in the inertial frame. 

\begin{figure*}
    \centering
    \includegraphics[scale=0.4]{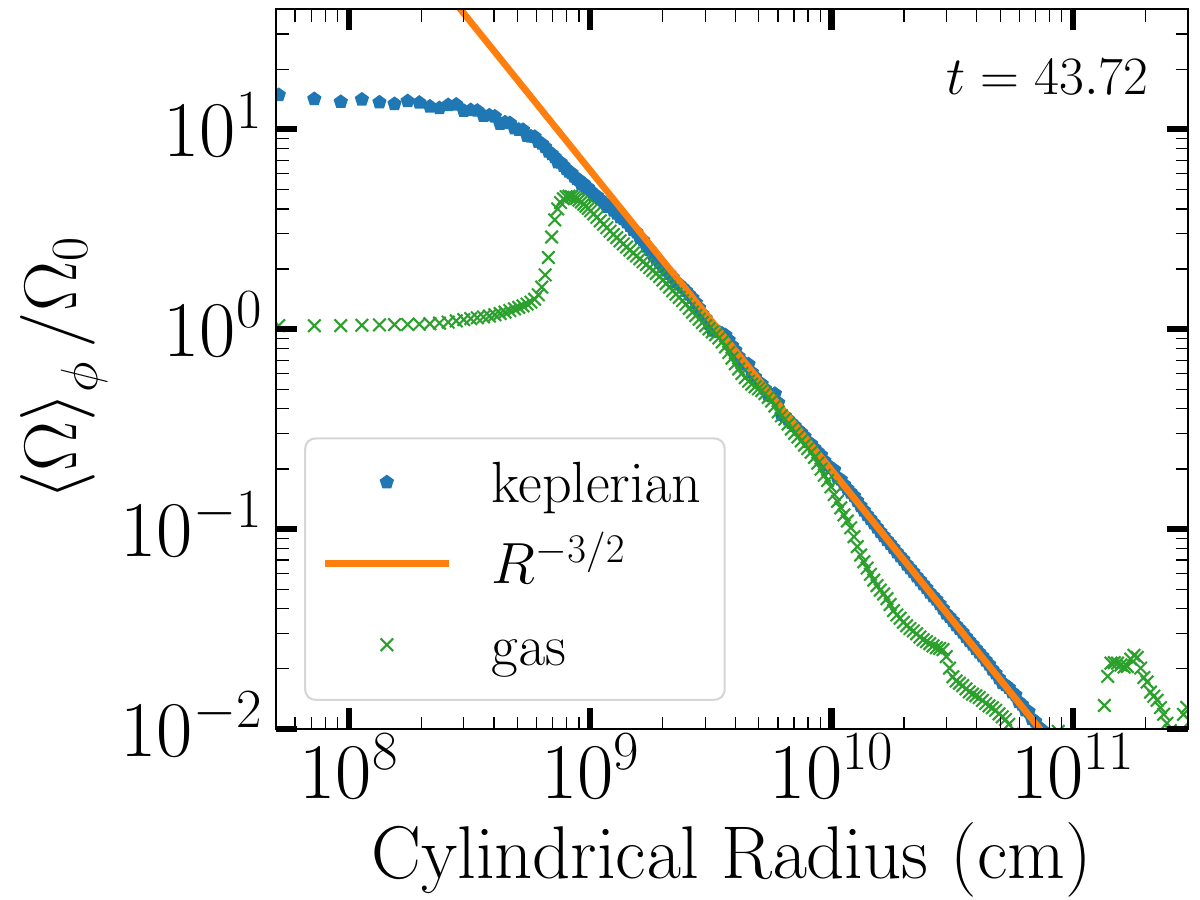}
    \includegraphics[scale=0.4]{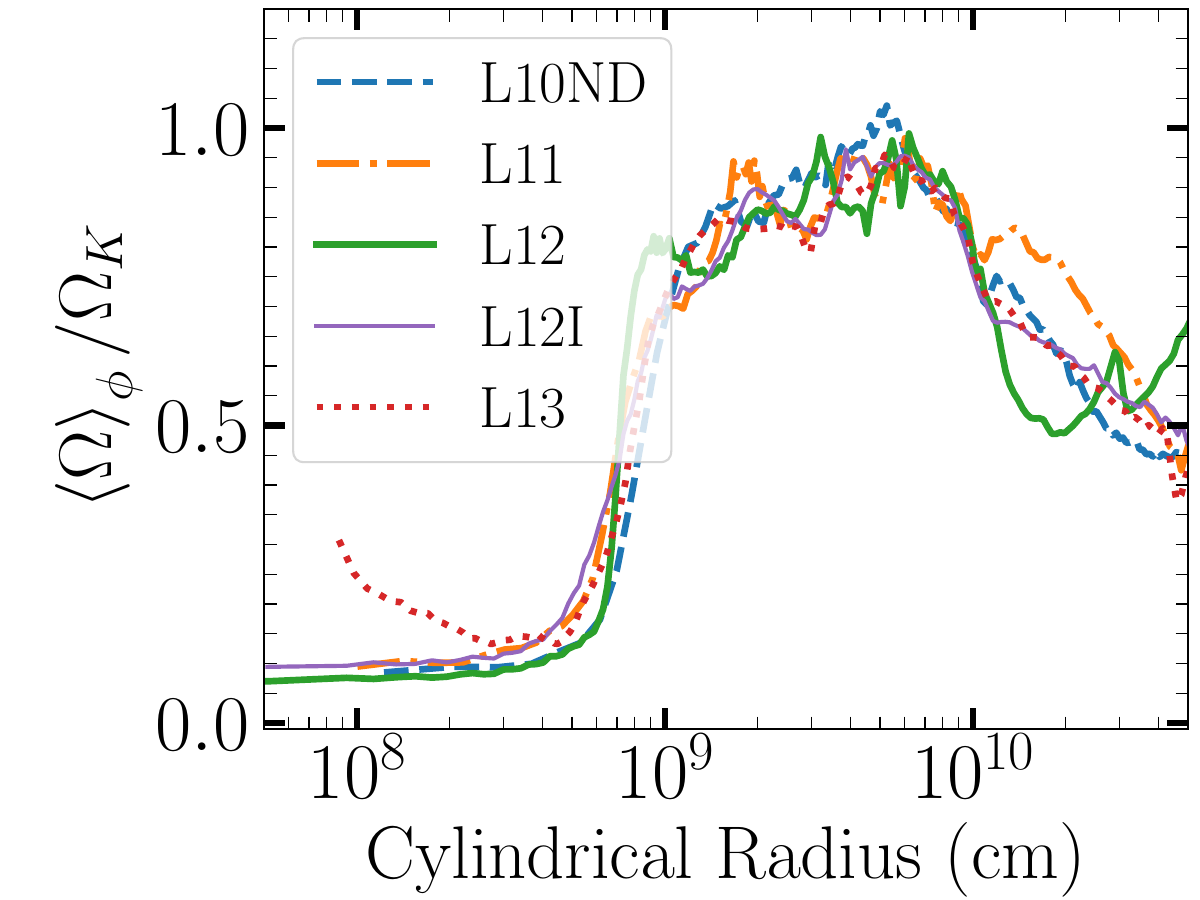}
     \caption{Angular velocity distribution 5 initial orbits past merger. Left: Gas angular velocity (green `X's) and Keplerian angular velocity as computed from the gravitational potential (blue circles) as a function of distance from the merged object center for {\sc L12}. An $\Omega = \sqrt{ G M_{\rm T} /r^{3}}$ relation expected at large distance is over-plotted as a thin orange line. Right panel: The gas angular velocity divided by the Keplerian angular velocity as a function of radius from the center of the merged object of all of our \octo\ simulations.  }
    \label{fig:post_merger2}
\end{figure*}

In Figure~\ref{fig:post_merger2} we show the azimuthally mass-averaged angular velocity profile of the merged object at $z=0$ for simulation {\sc L12}, as a function of cylindrical radius five orbits past merger. The angular velocity here is $\Omega=(xv_y-yv_x)/R^2$, where $x$, $y$, $R$, $v_x$, and $v_y$ are measured with respect to the position and velocity of the merged' object center. 

The angular velocity increases with decreasing radius until just inside $R_{\rm sph}$, where it starts to decrease again and smoothly matches the solid body rotation of the CO WD near $r_1$. The accretor core was initially synchronized to the initial binary frequency, and is still rotating as a solid body at a slightly higher frequency corresponding to the orbital frequency after driving (see Figure~\ref{fig:l12_bennett}). Material inside this radius must be pressure supported since it is well below the local Keplerian velocity, defined as the angular velocity required for pure rotational support,
$\Omega_K(r) = \sqrt{(1/R)d\Phi_g / dR }$, where $\Phi_g$ is the gravitational potential. On large distances the inner mass becomes nearly a constant and $\Omega_K(r) = \sqrt{GM/R^3}$. On the same figure we plot $R^{-3/2}$ relation normalized to the total mass of $M_{\rm T}=0.9~{\rm M_{\odot}}$.

The left panel of Figure~\ref{fig:post_merger2} additionally shows that the strongest shear occurs between $r_1$ and $R_{\rm sph}$. The angular velocity rises rapidly to a maximum value just inside $R_{\rm sph}$ to a sub-Keplerian value consistent with the vertical extent of the spheroid. Beyond the maximum, $\Omega$ decreases, but gradually the relative contribution of pressure to supporting the structure declines. The right panel shows the ratio between the local $\Omega$ and the local Keplerian value. In all runs, the disk is about $0.8$ of the Keplerian value between $r\approx 2R_{\rm sph} = 2\times 10^9$ cm and $R\approx 10^{10}$ cm. Beyond this radius the disk is probably not in equilibrium yet. The material that is still falling back after tidal disruption is unlikely to be supported by pressure, and therefore must have sub-Keplerian azimuthal velocities.

\begin{figure*}
\centering
   \includegraphics[scale=0.5]{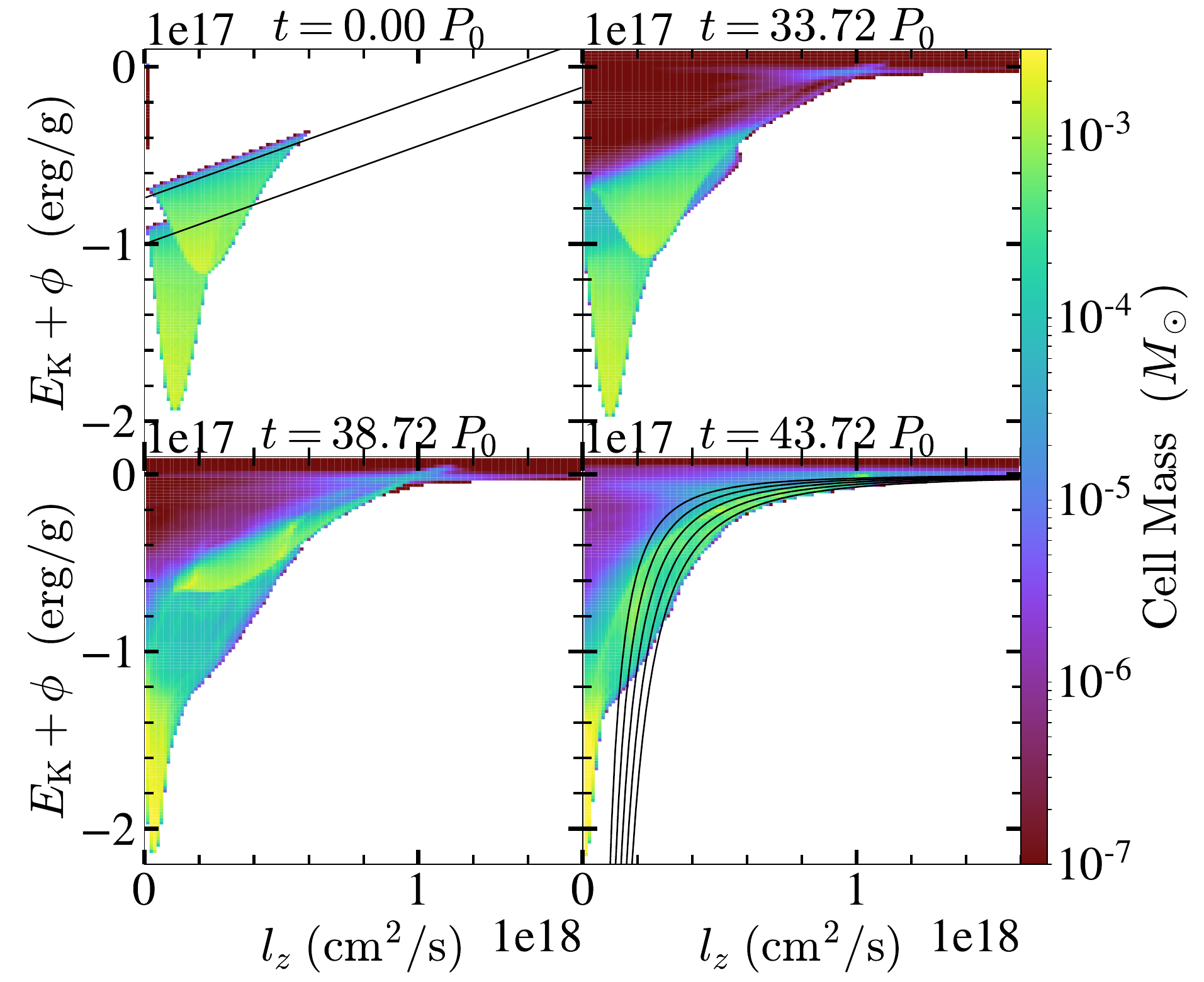}
    \caption{Specific inertial-frame mechanical energy (kinetic plus potential) for all cells for simulation {\sc L12} vs. specific $z$ angular momentum with respect to the center of mass at 4 times: $t=0$, $T_{\rm merge}-5$, $T_{\rm merge}$, and $T_{\rm merge}+5$. The color bar on the right of each frame shows the mass in each bin. See the text for a detailed description.}
    \label{fig:e_lz_diag}
\end{figure*}

Figure~\ref{fig:e_lz_diag} depicts the distribution of specific inertial-frame mechanical energy as a function of the specific $z$ angular momentum with respect to the center of mass of the system in the {\sc L12} simulation at four selected times. We divide the specific energy and the specific angular momentum in the range shown in the figure to 128x128 equally spaced bins. The color indicates the mass in each bin. This provides a picture of the dominant mass structures at different stages of evolution. A similar technique was used by \citet{Hayashi2021} to describe the tidal disruption of a neutron star by a black hole.

At $t=0$, all the mass is in the binary components, with the minima corresponding to the central densest and most bound element in each component. The specific angular momenta of the centers of the components are in the ratio $q^2$, 
with the donor having the largest central $l_z=(M_1/M_{\rm T})^2a_0^2\Omega_0=2.2\times 10^{17}~{\rm cm^2/s}$. The kinetic energy is due just to the synchronous rotation of the system at the initial orbital frequency $\Omega_0$. With $R$ the cylindrical radial distance of any element from the binary axis, we have
 \begin{equation*}
     E = \frac{1}{2}(\Omega_0 R)^2 + \Phi_g = \Omega_0 l_z + \Phi_{\rm eff}\, ,
 \end{equation*}
where $\Phi_{\rm eff} = \Phi_g -\frac{1}{2}(\Omega_0 R)^2$ is the effective potential. The surfaces of both accretor and donor are $\Phi_{\rm eff} =Const$ surfaces, so $E$ is linear on $l_z$ with slope $\Omega_0$. Two lines with that slope are shown for reference. 
 
At $t=T_{\rm merge}-5P_0$, both components are easily distinguishable, but both centers have moved to lower $l_z$ and more negative (bound) values of $E$. The stream material fills the region at low $l_z$ between the two components. The gas leaking out of $L_2$ appears marginally bound/unbound at large $l_z$. At $t=T_{\rm merge}$, the center of the accretor material has moved very close to the center of mass of the system at $l_z=0$, and the donor is being disrupted, so its binding energy is significantly lowered.

Finally, at $t=T_{\rm merge}+5P_0$, the accretor and the center of mass have nearly converged to $l_z=0$, and the binding energy is even greater. 
The material at $l_z\leq 0.5\times 10^{18}~{\rm cm^2/s}$ shown in yellow tones is the most tightly bound accretor matter, rotating nearly as a solid body $l_z=\Omega R^2$, where $\Omega$ is slightly faster than the initial value (see left panel of Fig. \ref{fig:post_merger2}).
For reference, we have drawn lines following $|E|\propto l_z^{-2}$ corresponding to Keplerian orbits around point masses $0.5, 0.6. 0.7, 0.8, {\rm and}, 0.9 M_\odot$. For $l_z\geq 0.4\times 10^{18}~{\rm cm^2/s}$ the highest density structure is consistent with Keplerian or sub-Keplerian orbits around a mass $0.9 M_\odot$.
At intermediate values of $l_z$, we find
structures, seen in green and yellow tones, corresponding to a boundary layer and a vestige of the donor core (the blob) still visible at $E=-0.75 \times 10^{17}~{\rm erg/g},~l_z=0.2\times 10^{18}~{\rm cm^2/s}$. See again Fig. \ref{fig:post_merger2} for comparison.
Note also the green-yellow feature at approximately $0.9\times 10^{18}~{\rm cm^2/s}\le l_z\le 1.1\times 10^{18}~{\rm cm^2/s}$, corresponding to material leaked out of $L_2$, whose highest $l_z$ portion is slightly unbound. All of the above results are consistent with the donor WD being partially disrupted by tides, leaving the core of the donor still surviving, and the envelope being stripped and sheared, with a small amount being leaked out of L2.

\subsection{Nuclear fusion in the merged object}
\label{ssec:SoF}

\begin{figure}
    \centering
    \includegraphics[scale=0.4]{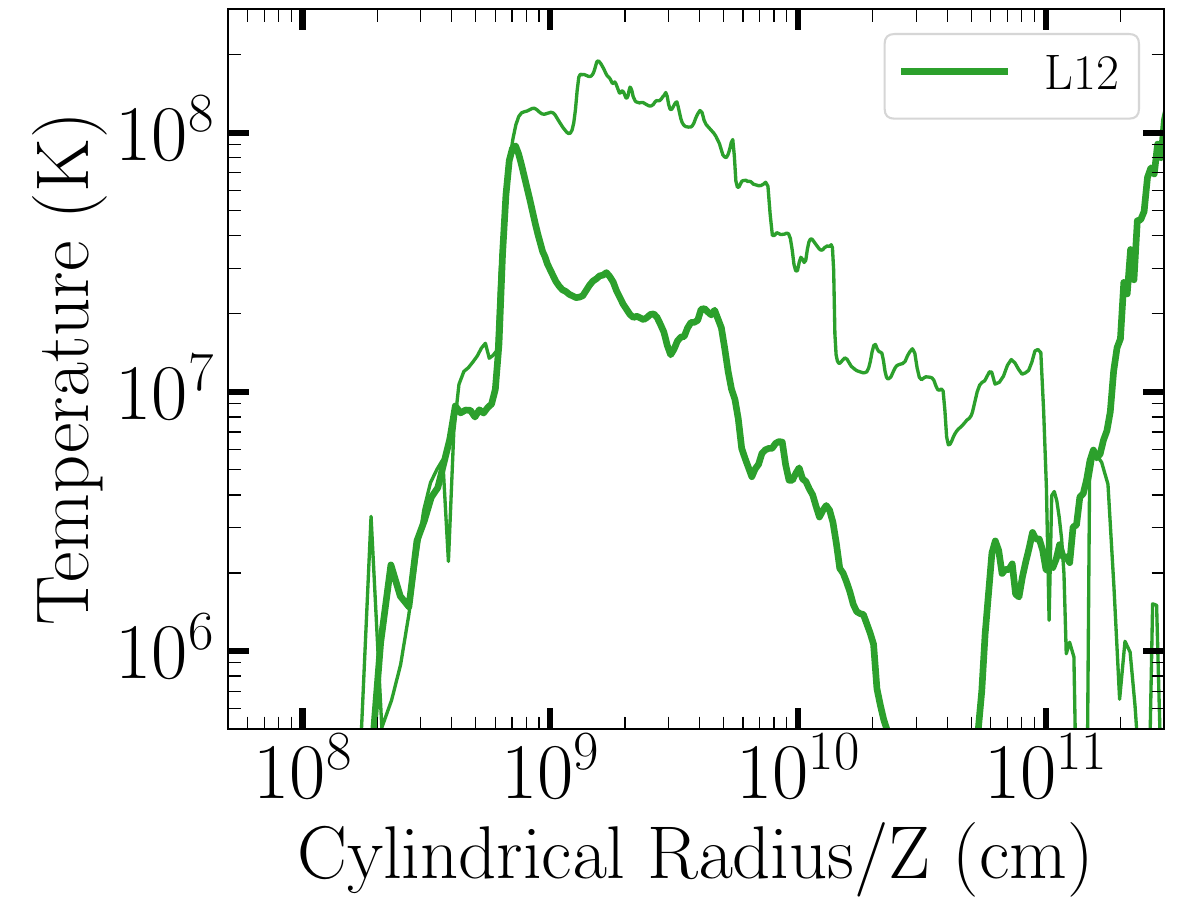}
    \caption{{The azimuthally-averaged radial profile of temperature on the orbital plane (thick lines) and the vertical temperature distribution along the positive $z$-axis for simulation {\sc L12}, 5 initial orbits past merger}}
    \label{fig:temp_prof}
\end{figure}

During a merger of two WDs, temperatures and densities are such that nuclear fusion of helium into carbon and oxygen is taking place at a shell of accreted donor material around the accretor, termed the shell of fire (SoF). 
In Figure~\ref{fig:SoF} we show a temperature slice at the orbital, $xy$, plane (left), and a temperature profile mass-averaged over the azimuthal angle (right), both for simulation {\sc L12}. As in Figure~\ref{fig:post_merger}, Cylindrical radius and Cylindrical z are measured from the center of the merged object, the point of maximum density. The SoF is evident as the hot blue ring around a central yellow circle. In Figure~\ref{fig:temp_prof} we present an azimuthally mass-averaged temperature profile at $z=0$ (thick line), and a temperature profile along the positive $z$-axis (thin line) for {simulation {\sc L12}. All of our \octo\ simulations show a similar temperature profile with a peak temperature between $1-2\times10^8$ and slightly hotter temperatures at the poles. Additionally, due to their lower resolution, in simulations {\sc L10ND} and {\sc L11}, the spherical central core possesses higher temperatures of $2\times10^{7}~{\rm K}$ and $4\times10^{6}~{\rm K}$, respectively.}

{The presence of a cooler remnant of the donor's core, which has not had time to shear and to mix with the surrounding gas, lowers the azimuthal average temperature on the orbital plane (Figure~\ref{fig:temp_prof}, thick line). As a result, the highest temperatures of the SoF occur at higher latitudes.
This cooler remnant can be seen clearly in the equatorial slice of Figure~\ref{fig:SoF} as a red ``blob" adjacent to the SoF at positive $y$ direction, and in the averaged profile (right panel of Figure~\ref{fig:SoF}) as a yellow feature at $R_{\rm sph}<R<R_{\rm dsk}$, which extends symmetrically from $z=-0.4\times 10^9~{\rm cm}$ to $z=0.4\times 10^9~{\rm cm}$ along the vertical axis.}  The presence of this cooler blob had already been noticed in previous simulations \citep{Staff2012, Staff2018}.

\begin{figure*}
\centering
    \includegraphics[scale=0.6,trim={0 0.45cm 0 10.8cm}, clip]{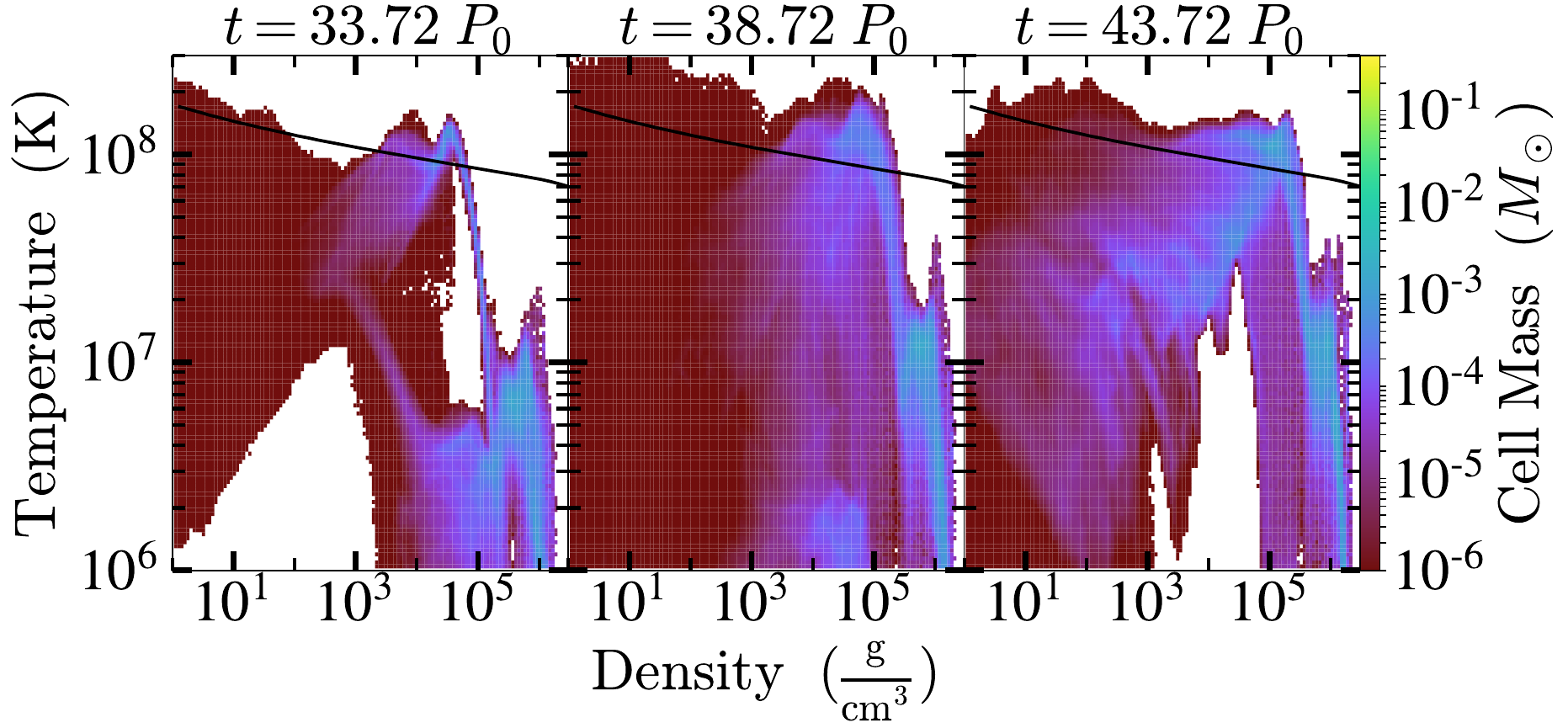}
    \caption{Density-Temperature phase plots for {\sc L12} at three times corresponding (from left to right) to $T_{\rm merge}-5P_0$, $T_{\rm merge}$, and $T_{\rm merge}+5P_0$, respectively. The black line in each panel denotes the threshold for triple alpha burning. The temperature and density are divided each to 128 bins, and the color indicates the mass in each bin}
    \label{fig:rho_T_diag}
\end{figure*}

A more useful display of the He-burning zones is gleaned from Figure~\ref{fig:rho_T_diag}, where we show the time evolution of the mass distribution in density and temperature around the time of the merger. To construct this mapping we divide the domains of density and temperature into evenly spaced 128x128 logarithmic bins, 
and we depict using a color bar how much mass is in each bin. Consequently this display reveals also the main structures of the flow at the three selected times. The structures that extend above the black line meet the conditions for triple alpha burning.
Focusing on the bins with most mass, at $t=33.72P_0$, near the bottom right corner of the left panel, we can see two structures superimposed and roughly shaped as paint-brushed European-style ``1"s, the accretor core as a nearly vertical blue band extending to $\rho\sim 10^6$ g/cc,
with the accretor envelope represented by the serif extending down and to the left.
Similarly, the donor core appears as a vertical band at $\rho\sim 2\times 10^5$ g/cc, with its envlope as the corresponding serif. The ``knee"-like feature above the $3\alpha$ boundary and its downward extension joining the accretor is the accretion belt with the SoF. 
At $t=38.72P_0$ the knee has moved to higher temperatures and densities, and the total mass above the $3\alpha$ boundary has increased. The donor is still discernible. Finally, 5 orbits after the merger, the donor is gone, the hot accretion belt and disk are visible as a broad fanning feature
with an approximate slope of 2/3, which culminates on a slightly cooler knee {(than at time $t=38.72P_0$)} above the $3\alpha$ boundary at even higher densities and extends downward to meet the accretor core. There is now more mass above the $3\alpha$ boundary consistent with Figure \ref{fig:3alpha} (see below).

In Figure~\ref{fig:3alpha} we show, on the top row, the distribution of He-burning power on the equatorial $xy$ plane (left) and on the vertical $xz$ plane (right). Overplotted are four contours of densities (ordered spatially from the exterior to the interior of the merged object) of $10^3, 10^4, 10^5, 10^6$ g cm$^{-3}$, in yellow, and two contours of temperatures, one of $5\times10^7$ K in red, and a second of $10^8$ K in purple.
On the bottom left frame we display the mass of the SoF as a function of time, where the mass of the SoF is determined by summing the mass in the bins above the $3\alpha$-threshold in Figure~\ref{fig:rho_T_diag}. This is an improved method to measure the SoF mass as compared to what was done in earlier simulations \citep{Staff2012, Staff2018} since we find that some helium burning can still take place at densities of $\sim 10^4$ g cm$^{-3}$ (mainly near the poles) or alternatively at temperatures less than $10^8~{\rm K}$ (e.g. at the equator). 
In the bottom right panel of the same figure we plot the total luminosity generated by the triple alpha burning as a function of time in our \octo\ simulations. 
\begin{figure*}
\centering
    \includegraphics[scale=0.55, trim={0 2.2cm 0 7.5cm}, clip]{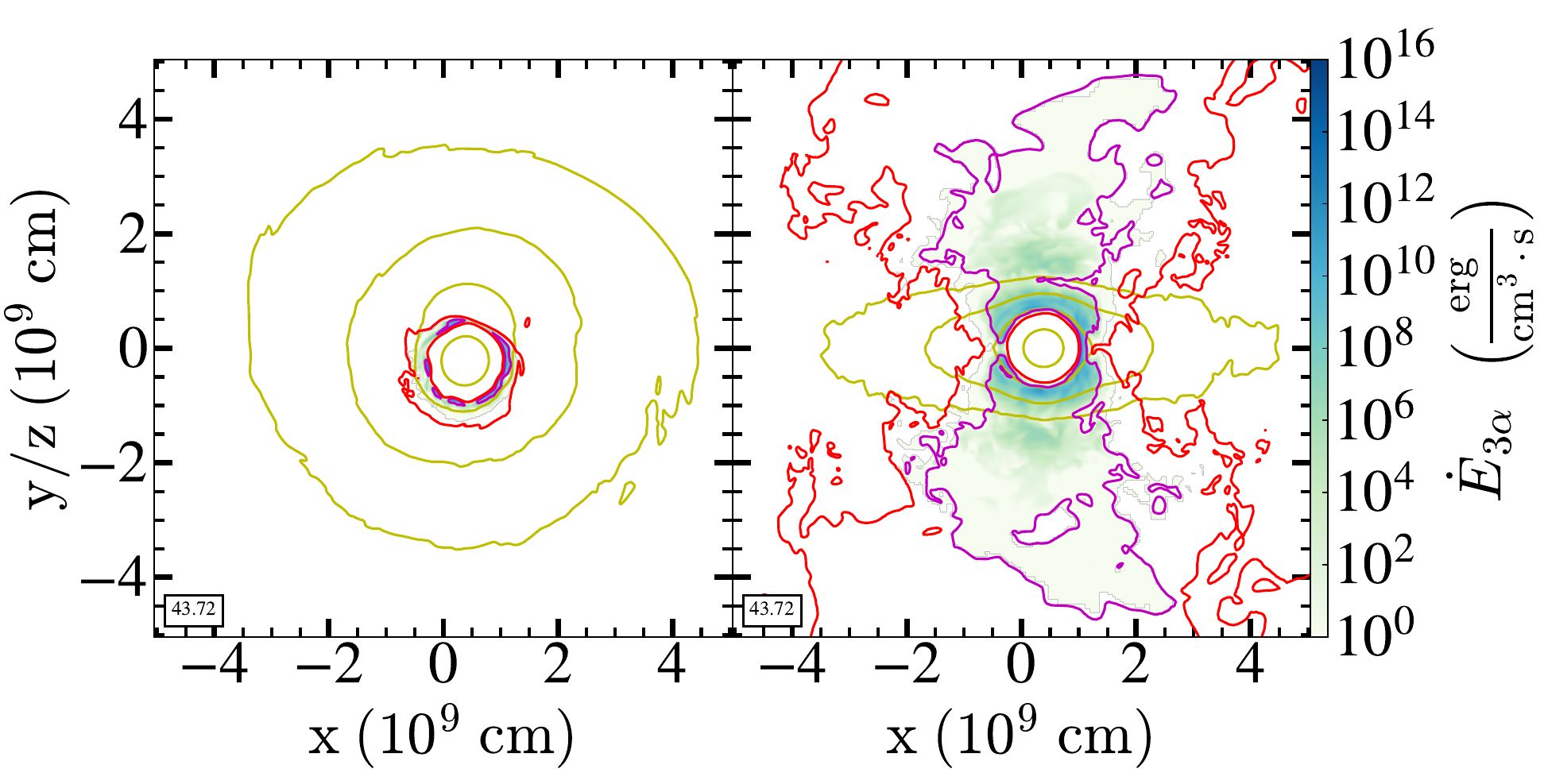}
    \includegraphics[scale=0.4]{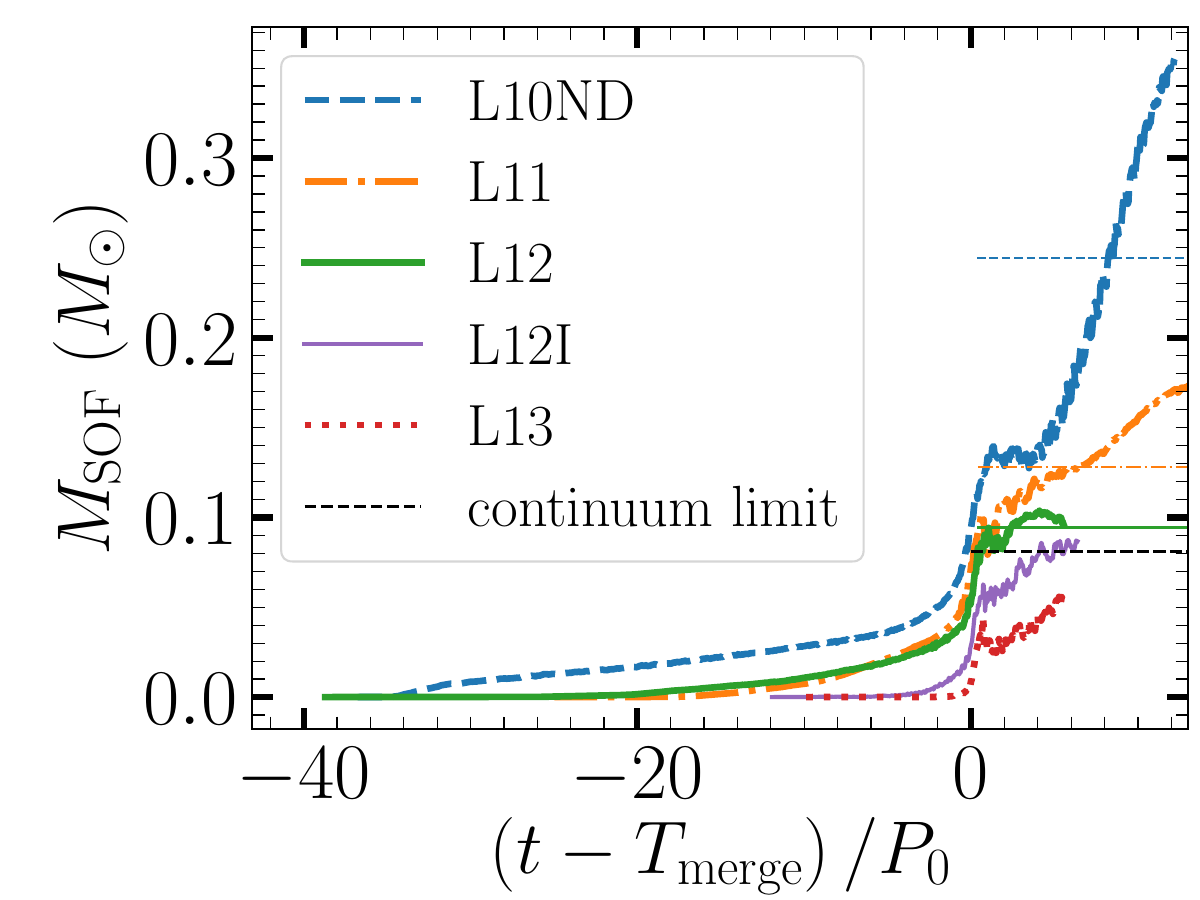}
    \includegraphics[scale=0.4]{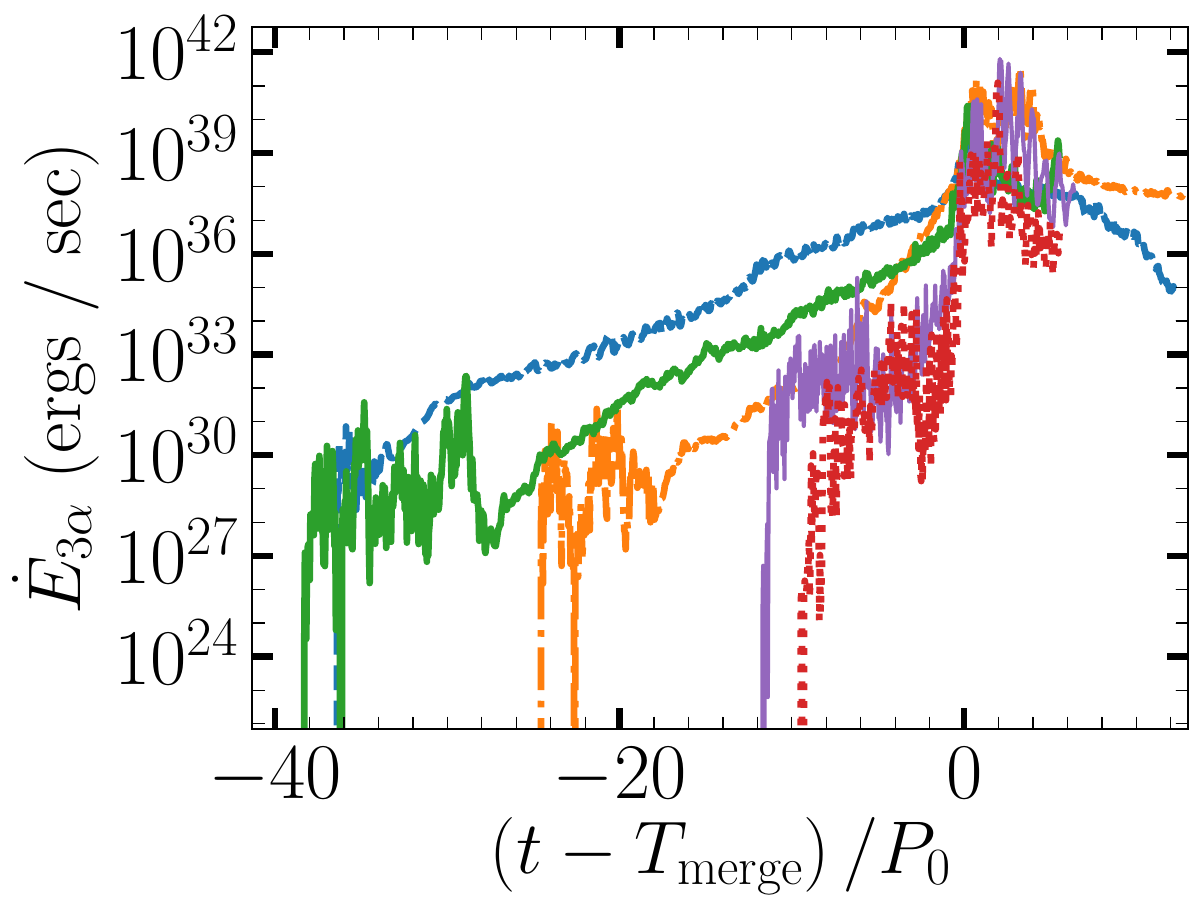}
    \caption{Upper row: Equatorial, $xy$ (left) and meridional, $xz$ (right) slices of helium burning power density five initial orbits past merger for {\sc L12}. Bottom row: The gas mass that has the conditions to burn helium (left) and its burning power (right) as a function of time. \revisionII{The three horizontal lines, dashed blue, dashed-dotted orange, and thick solid green, on the bottom left panel, denote a burning mass averaged between one orbital period after the merger and the latest simulated time, for simulations {\sc L10ND}, {\sc L11}, and {\sc L12}, respectively. These averages are used to derive the continuum limit of our resolution study (plotted as dashed black line; see Table~\ref{tab:postmerger} as well)}}
    \label{fig:3alpha}
\end{figure*}

Comparing the results for different resolutions, it is clear that a higher resolution is needed to get an accurate estimate of the mass of the burning shells and the energy released. In the lower resolution runs, in which the accretor is resolved across fewer cells, the shells interior to the inner boundary of the SoF become hotter with time, so the SoF effectively expands inward to higher densities and the amount of mass in the SoF quickly grows and therefore is probably being overestimated. {For higher resolutions the maximal density that burns increases more slowly \revisionII{with time} and the SoF mass is smaller as a result.} 

{We list in Table~\ref{tab:postmerger} representative values of the SoF for simulations {\sc L10ND}, {\sc L11}, and {\sc L12} with the purpose of examining the convergence of these properties with respect to resolution. \revisionII{We do not include in this analysis simulation {\sc L12I} because unlike the rest of the other simulations it had been performed in the inertial frame of reference and not in the rotating frame. We also do not include simulation {\sc L13} because it shows slight peculiar features in its internal structure like a small decrease in the accretor's central density (bottom left panel of Figure~\ref{fig:comp_res}), and a high central rotation (right panel of Figure~\ref{fig:post_merger2}). In the second column of Table~\ref{tab:postmerger}, we list a representative value for the temperature in the SoF, $T_{\rm SOF}$, that we define as the temperature in which the nuclear power in each simulation is the strongest. This temperature does not vary much with resolution. It oscillates shortly after the merger between $1.3-2.3\times 10^8~{\rm K}$, but then settles down. Averaging over a time period that starts at one initial orbital period after the merger we obtain $T_{\rm SOF}=1.6\times 10^8~{\rm K}$.} 

The SoF density, $\rho_{\rm SOF}$ (third column) is calculated as the maximal density in the SoF averaged over the same time period, which decreases for higher resolution. \revisionII{We then perform a Richardson extrapolation \citep{Richardson27}, by taking results for three increasing resolutions, and extrapolate to the continuum limit to obtain an estimate for the case for cell size going to zero. This yields a density value of $2.9\times10^5~{\rm g~cm^{-3}}$.} 

\revisionII{To check convergence for the mass of the burning shells, we average the values of $M_{\rm SOF}$ (as shown in the bottom left panel of Figure~\ref{fig:3alpha}) at two time points: $t=T_{\rm merge} + P_0$, and the latest simulated time. The obtained values are plotted as three horizontal lines (dashed blue, dashed-dotted orange, and thick solid green, for simulations {\sc L10ND}, {\sc L11}, and {\sc L12}, respectively) in the same figure and are also listed in the fourth column of Table~\ref{tab:postmerger}.} The SoF mass shows a second order convergence to a value of $M_{\rm SOF}\simeq0.08$ with respect to resolution (see the horizontal dashed black line in the same figure and also Table~\ref{tab:postmerger}). }
\begin{table}
 \caption{{Convergence with increasing resolution of SoF properties }}
 \label{tab:postmerger}
 \centering
 \begin{tabular}{clllll}
 \toprule
 Simulation & $T_{\rm SOF}$ & $\rho_{\rm SOF}$ & $M_{\rm SOF}$ & $M_{\rm SOF}^{\rm ^{16}O}$ & $f_{\rm core}$ \\
 &(MK) &(${\rm g~cm^{-3}}$)&($M_{\rm \odot}$)&($M_{\rm \odot}$)&  \\
 \midrule
  {\sc L10ND} & 160 & $7.0\times10^{5}$ & 0.244 & 0.062  & 0.51 \\
  {\sc L11} & 160 & $3.3\times10^{5}$ & 0.128 &  0.032 & 0.50 \\
  {\sc L12} & 160 & $3.0\times10^{5}$ & 0.094 &  0.019 & 0.40 \\
  \bottomrule
  continuum limit & 160 & $2.9\times 10^{5}$ & 0.081 &  0.01 & 0.25 \\
  convergence order & -- & 3.4 & 1.8 & 1.2 & -- \\

  \bottomrule
 \end{tabular}
 \footnotesize{ \revisionII{We perform a Richardson extrapolation \citep{Richardson27} to obtain the continuum limit for $\rho_{\rm SOF}$, $M_{\rm SOF}$, and $M_{\rm SOF}^{\rm ^{16}O}$. $f_{\rm core}=2M_{\rm SOF}^{\rm ^{16}O}/M_{\rm SOF}$}}
\end{table}

\revisionII{The amount of mass in the SoF and its composition (as we will discuss in the next Section) greatly affect the final byproducts of the partial helium burning. These burning products, later on, as the post merger object expands, cools, and evolves to become a carbon star, will get mix with the envelope due to convection. Therefore, they will also determine the surface abundance at the RCB phase. Our aim is to infer from the simulations the estimated conditions expected due to a dynamical merger of a $q=0.7$ system, from which we can ultimately estimate the surface values as the burning products mixed with the envelope.}

Ideally one would like to include the energy deposition by nuclear reactions as the 3D simulation proceeds and to calculate the dynamical reaction to the deposition of this extra energy. \revisionII {However, \octo\ does not yet include a nuclear reaction network and therefore we do not include in the simulation the effects of nuclear energy generation. These reactions probably play a role in reshaping the merged object on a dynamical timescale, and in changing the abundances on a nuclear timescale}. What we can say is that the total energy generated during the simulation can be estimated as the peak luminosity times $\sim 10 P_0$, yielding $E_{3\alpha}\approx 4\times 10^{44}$ erg, which is negligible compared to the various energy components of the system (see Figure \ref{fig:l12_energy}). 

Moreover, we find that the highest power density of the nuclear energy is generated at gas with temperatures $T=1.4-2.0\times 10^{8}~{\rm K}$, and densities $\rho = 1-2.5 \times 10^5~{\rm g~cm^{-3}}$, and equals roughly to $\dot{\epsilon}_{\rm 3 \alpha}^{\rm max} =  5\times 10^{12}~{\rm erg~cm^{-3}~s^{-1}}$. At this range of temperatures and densities, the minimal thermal energy density, assuming a composition of ionized carbon-oxygen, equals $E_{\rm th,~min}\approx 8\times 10^{20}$. Therefore $\approx 10^7$ s are required (roughly $10^5$ orbits) for the nuclear reaction to become important at these regions, which is beyond the scope of the hydrodynamic simulation, and thus can be neglected.

The merged object at the end of our simulations, 5 or even 10 initial orbits after the merger, is not yet spherically (or even fully axially) symmetric. However, following the long-term evolution of the post-merger object by means of 3D hydrodynamic simulations is not feasible as the time step is ultimately limited by the Courant condition. Mapping the 3D object into a 1D implicit code such as {\sc MESA} \citep{Paxton2011, Paxton2013, Paxton2015, Paxton2018, Paxton2019} is a possible way of tackling the long-term evolution provided that we can find an adequate method of averaging the 3D structure of the merged object, which preserves the object's mass and its angular momentum distribution.

In this paper, we average the merged object along different angles based on a similar procedure to that described in section~3.1 of \cite{munson21}. 
{Cells between a polar angle $\theta+d\theta$ and $\theta-d\theta$ are averaged around the azimuthal angle to obtain the quantity at a given stellar radius and the averaging is done for conical shells along both the positive and negative z-axis. The results are consistent with the the temperature distribution shown in Figure~\ref{fig:temp_prof} and the burning power distribution shown in Figure~\ref{fig:3alpha}.
By comparing these results to the models of \cite{munson21} we can evaluate what can occur after the merger event.} Since the models reach He-burning temperatures during the merger, assuming they will follow an RCB-like evolution is reasonable. The input luminosity from the steady He-burning shell should expand the envelope and bring the surface luminosity from 4000 to 10\,000 $L_\odot$ \citep{Menon2013,2019MNRAS.482.2320M,Lauer2019,2020MNRAS.498.2912C}. 

\subsection{\revisionII{The Dredge-up of $^{16}$O and the inferred $^{16}$O/$^{18}$O}}
\label{ssec:O16}

The dredge-up of $^{16}$O, its impact on the surface ratio of $^{16}$O/$^{18}$O (the oxygen ratio), and the discrepancies between the estimated dredge-up by SPH and grid codes were already discussed in \cite{Staff2018}. {Here, we re-examine the amount of dredged-up core material using a similar
approach as in \cite{Staff2018}, at several levels of resolution, using an updated \octo, and an improved definition of the SoF. Using the results obtained at 3 different and increasing resolutions, we
estimate the \revisionII{continuum limit} dredge-up amount using the Richardson extrapolation.} 

We are not aiming here to obtain an accurate estimate of the oxygen ratio since that would require following a nuclear-reaction network simultaneously with the hydrodynamics of the merger and the precise timing of when the outer convective zone connects with the burning products of the SoF \citep{Menon2013,2020MNRAS.498.2912C,munson21}.
\revisionII{Instead, we assume constant conditions, that is constant temperature and density, in the regions of helium burning, and inspect only the most relevant reactions to the production of $^{16}$O and $^{18}$O to infer a minimal possible value of $^{16}$O/$^{18}$O in these regions. We also assume that the reactions proceed to completion which should occur at a timescale dictated by the reaction with the slowest rate. The temperature, density and initial composition of this analysis are derived from our hydrodynamic simulations. A similar single-zone nuclear approach have been performed in \cite{Staff2012}, where in that paper the analysis has been carried-out with a nuclear reaction code. We build on their results in diagnosing the most relevant processes to yield a simple analytic estimate without running a nuclear reaction code.}
Our main conclusion is that the dredge-up of $^{16}$O is accompanied by $^{12}$C, which can help the production of $^{18}$O, if enough hydrogen is present in the donor envelope.

{As we showed in section~\ref{ssec:SoF}, an accurate estimate of the {\em total} mass of the SoF, and regardless of composition, at any time for a given resolution, is obtained by adding the mass above the threshold for triple alpha burning in the $T-\rho$ distribution (e.g. the black line on Figure~\ref{fig:rho_T_diag} for selected times during the L12 run). The results of these estimates for all runs as a function of time are shown in the bottom left panel of Figure~\ref{fig:3alpha}.}

\begin{figure*}
    \centering
    \includegraphics[scale=0.4]{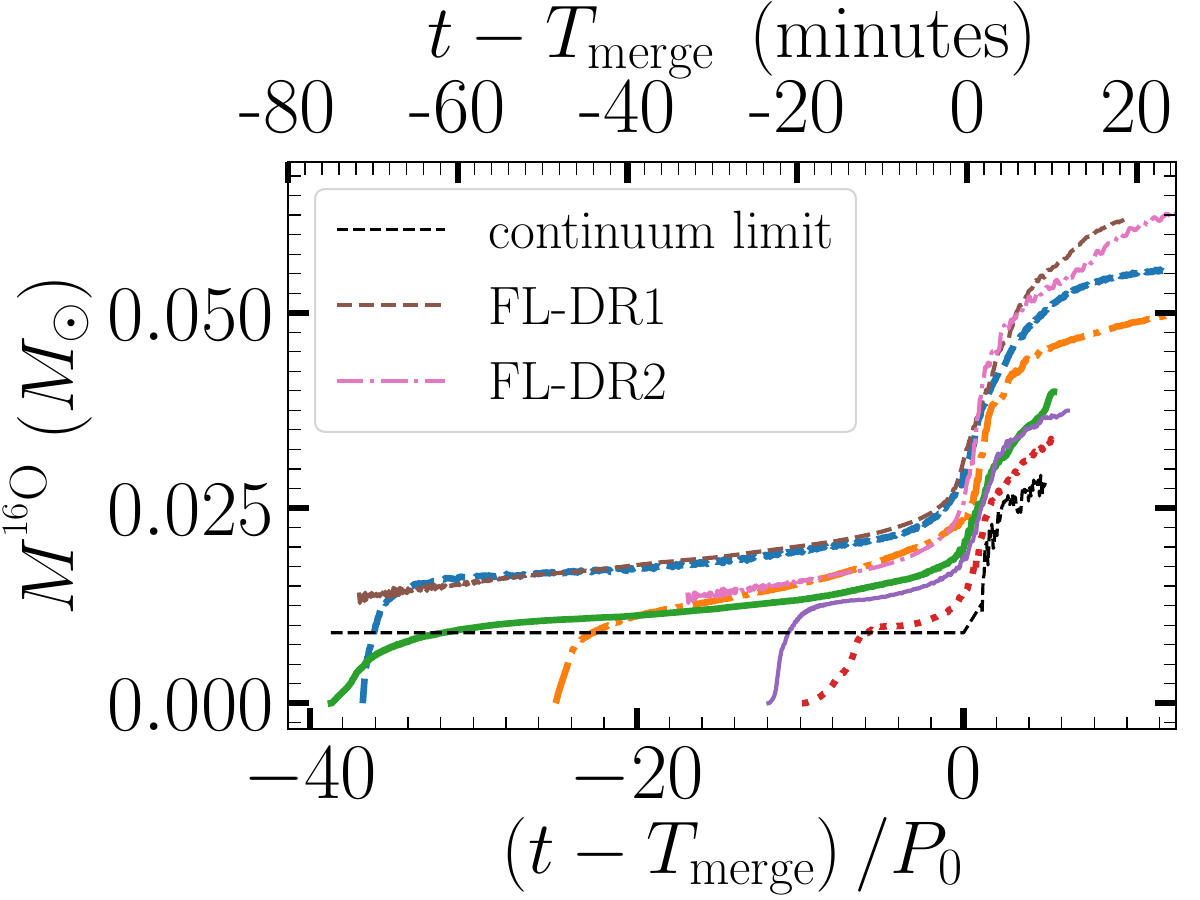}
    \includegraphics[scale=0.4]{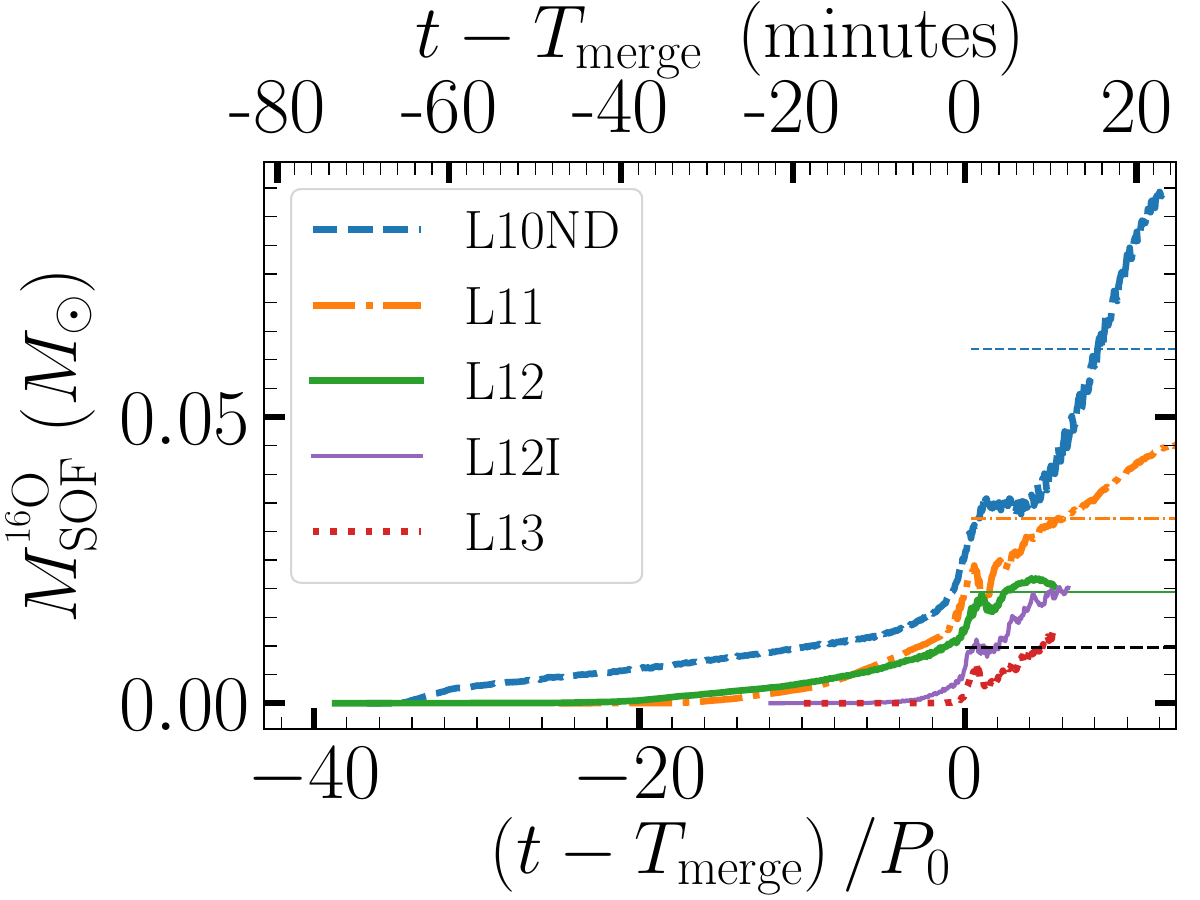}
    \caption{Left: Dredge-up of oxygen-16, defined as mass of oxygen-16 residing at densities lower than $10^5 {\rm \; g\; cm^{-3} }$. The dashed black curve {labeled `Richardson extrap.'} is the Richardson extrapolation for cell size going to zero, derived from simulations L11, L12, and L13. It is composed of `plateau or shelf` value before the merger, and actual values 1-5 orbits after the merger. The `shelf` value is calculated at time $t_{\rm shelf} = - 0.5 (T_{\rm merge} - 4 P_0) $. Right: mass of oxygen-16 in the SoF, where the conditions for helium burning exist. \revisionII{The three horizontal lines, dashed blue, dashed-dotted orange, and thick solid green, denote a value averaged between one orbital period after the merger and the latest simulated time, for simulations {\sc L10ND}, {\sc L11}, and {\sc L12}, respectively. These averages are used to derive the continuum limit value of our resolution study (plotted as dashed black line; see Table~\ref{tab:postmerger} as well).} {The upper axis in both panels shows time in minutes, where the conversion is based on the initial orbital period of simulation {\sc L12}, $P_0=113.6~{\rm s}$}}
    \label{fig:dredge-up}
\end{figure*}

{In Figure~\ref{fig:dredge-up} we show the mass of dredged up $^{16}$O (and also $^{12}$C, assuming equal masses in the dredged-up accretor material) as
a function of time for all of our simulations. Since we evolved three distinct fluids in our simulations: one for the accretor's core (CO), a second for the accretor's envelope (He), and a third for donor material (also He), we could calculate how much mass of the accretor's core material, i.e, the gas that was originally located at the accretor's core, has densities smaller than 10$^5$~g~cm$^{-3}$ and take half for the amount of $^{16}$O  (Figure~\ref{fig:dredge-up} left panel). This is to be compared with Figure~4 from \cite{Staff2018}. }

We also calculated a more precise measurement of how much mass of the accretor's core fluid resides in regions where conditions for helium burning (by the triple alpha reaction) exist for several resolutions, and derive the amount of $^{16}$O (and $^{12}$O), again by taking half. We plot this quantity in the right panel of Figure~\ref{fig:dredge-up}. The main difference between these quantities is that on the left panel, mass that resides outside the SoF and at lower densities, is still counted, while the right panel informs us the composition in the regions where triple-alpha burning happens. \revisionII{Similar to the bottom left panel of Figure~\ref{fig:3alpha}, we additionally plot on the right panel an average $M^{^{16}{\rm O}}_{\rm SOF}$ between two time points, $t=T_{\rm merge} + P_0$, and the latest simulated time, for simulations {\sc L10ND}, {\sc L11}, and {\sc L12} as horizontal lines (in dashed blue, dashed-dotted orange, and thick solid green, respectively). These averages are used to derive the continuum limit dredge-up value of our resolution study (plotted as dashed black line; see Table~\ref{tab:postmerger} as well)}

The general behavior of the dredge-up curves on the left panel of Figure~\ref{fig:dredge-up} consists of three phases: an initial fast rise, perhaps dominated by numerical diffusion, a monotonic slow increase until the merger, and a steep increase during the merger, most likely a combination of real dredge-up and numerical diffusion. It makes sense therefore to take values of the dredge-up at the same time after the merger at different resolutions and to apply the Richardson extrapolation to estimate the corresponding value \revisionII{in the continuum limit}. {The curve thus obtained is shown as a black dashed line in the left panel of Figure~\ref{fig:dredge-up} and is composed of $\sim$100 data points with an average convergence order of $\sim$1 (and a standard deviation of $\sim0.2$)}. Based on this analysis we conclude that the final mass of dredged-up $^{16}$O, after subtracting the fast rise ``shelf" value shown as horizontal black line at pre-merger times, is $\sim 0.02 M_\odot$. However, not all of the dredged-up $^{16}$O 
ends up in the SoF, see the discussion below.
The reason that analogous SPH simulations essentially found no dredge-up 
may be attributed to the known inability of standard implementations of 
SPH to correctly resolve mixing at fluid boundaries unless certain modifications are adopted \citep{Read2010, RuizBonilla2022}.

The observations suggest that in most RCB stars the ratio $^{16}{\rm O}/^{18}{\rm O} \leq 10$. Some RCB stars have higher values for this ratio but still much less than the solar value 
$\sim 500$. Most of the $^{18}$O comes from the reaction 
$^{14}{\rm N}(\alpha, \gamma)^{18}{\rm F}(\beta^+)^{18}{\rm O}$. Thus 
the amount of $^{14}$N present in the SoF limits the amount of $^{18}$O produced, since some $^{18}$O may burn to $^{22}$Ne. 
The amount of $^{14}$N in the SoF comes from two possible 
sources: what was originally present in the He WD, plus what
is synthesized in the SoF during burning through proton captures 
$^{12}{\rm C}(p, \gamma)^{13}{\rm N}(\beta^+)^{13}{\rm C}(p, \gamma)^{14}{\rm N}$, which requires the presence of { sufficient H (see below)}.

{Considering first the simpler case in which no H is available in the SoF, the mass of $^{18}{\rm O}$ would be limited by the small amount of $^{14}{\rm N}$ originally present in the donor. The amount of $^{16}$O would mainly depend on the dredge-up from the accretor's core. Assuming an initial $^{14}{\rm N}$ mass fraction of ${\rm X}_{\rm ^{14}N}^0$ in the SoF, and \revisionII{$0.5f_{\rm core}$ mass fractions of $^{12}{\rm C}$ and $^{16}{\rm O}$ while the rest is helium, the minimal mass-fractions ratio of $^{16}{\rm O}/^{18}{\rm O}$ in the SoF would be}} 

\begin{equation}
\revisionII{\left(^{16}{\rm O}/^{18}{\rm O}\right)_{\rm SOF}=(7/18)f_{\rm core}/{\rm X}_{\rm ^{14}N}^0,\;{\rm with\; no\; hydrogen}.    }
\label{eq:o_ratio_no_h}
\end{equation}
\revisionII{Here, 
$f_{\rm core}=2M_{\rm SOF}^{\rm ^{16}O} /M_{\rm SOF}$ is the mass fraction of the accretor's core fluid in the SoF, which can be inferred from  our simulations.} In our low resolution simulations, \revisionII{{\sc L10ND} and {\sc L11},} we obtain $f_{\rm core}\simeq 0.5$, while simulation {\sc L12} yields $f_{\rm core}\simeq 0.4$, and the extrapolation to infinite resolution yields a lower fraction of $f_{\rm core}=0.25$ (a first 
order convergence; see horizontal lines in the right panel of Figure~\ref{fig:dredge-up} and Table~\ref{tab:postmerger}). With no hydrogen and this amount of dredge-up, values of $^{16}{\rm O}/^{18}{\rm O}$ less than 10 are impossible unless ${\rm X}_{\rm ^{14}N}^0\ge4\%f_{\rm core}$. However, the amount of $^{14}{\rm N}$ is diluted by the accretor material that has been dredged-up to the SoF, so the mass fraction of $^{14}{\rm N}$ in the He WD donor star, ${\rm X}_{\rm ^{14}N}^{\rm He-WD}$, should be ${\rm X}_{\rm ^{14}N}^{\rm He-WD}\ge4\%\left(f_{\rm core}/(1 - f_{\rm core})\right)$ for $^{16}{\rm O}/^{18}{\rm O}\le 10$. This imposes a substantial restriction, specifically because observations of most RCB stars show mostly subsolar metallicities (e.g., \citealt{2000A&A...353..287A}). Obtaining a ratio of the order of unity without the presence of hydrogen in these cases is unlikely. 

With H in the SoF, on the other hand, we can create more $^{14}$N by {\em two} proton captures for every $^{12}$C, as seen in the reaction chain  $^{12}{\rm C}(p, \gamma)^{13}{\rm N}(\beta^+)^{13}{\rm C}(p, \gamma)^{14}{\rm N}$. Because the half-life of $^{13}$N is 9.9 minutes and the proton capture rate of $^{12}{\rm C}(p, \gamma)$ in the SoF is $0.6$ seconds, it is fair to assume that the available $^{12}$C will capture protons before $^{13}{\rm C}(p, \gamma)^{14}{\rm N}$ takes place. Furthermore, because the average temperature in the SoF is only 150MK, there will be a negligible amount of additional $^{12}$C generated via triple-alpha processes. Therefore, only the protons remaining after $^{12}$C is depleted in the SoF are available for $^{13}{\rm C}(p, \gamma)^{14}{\rm N}$, \revisionII{and only they will contribute to the formation of additional $^{18}$O, which will reduce the $^{16}{\rm O}/^{18}{\rm O}$ ratio. When there are enough protons to convert all the $^{12}$C to $^{14}{\rm N}$, any additional protons do not contribute to the production of additional $^{18}$O and will not change this ratio. The $^{16}{\rm O}/^{18}{\rm O}$ in the SoF can therfore be written as:
\begin{equation}
        \begin{array}{lll}
        \left(^{16}{\rm O}/^{18}{\rm O}\right)_{\rm SOF}& = (7/18)f_{\rm core}/{\rm X}_{\rm ^{14}N}^0 \times& \\
        & \left\{ \begin{array}{ll}
            1 & {\rm X}\leq f_{\rm core}/24, \\
            \left(1 + f_1 - f_2\right)^{-1} & f_{\rm core}/24\leq {\rm X}\leq f_{\rm core}/12, \\
            \left(1 + f_2\right)^{-1} & f_{\rm core}/12\leq {\rm X},
        \end{array} \right. &
        \end{array}
\label{eq:o_ratio}    
\end{equation}
where $f_1=14{\rm X}/{\rm X}_{\rm ^{14}N}^0$, $f_2=7f_{\rm core}/\left(12{\rm X}_{\rm ^{14}N}^0\right)$, and ${\rm X}$ is the hydrogen mass fraction in the SoF. }

\begin{figure*}
    \centering
    \includegraphics[scale=0.4]{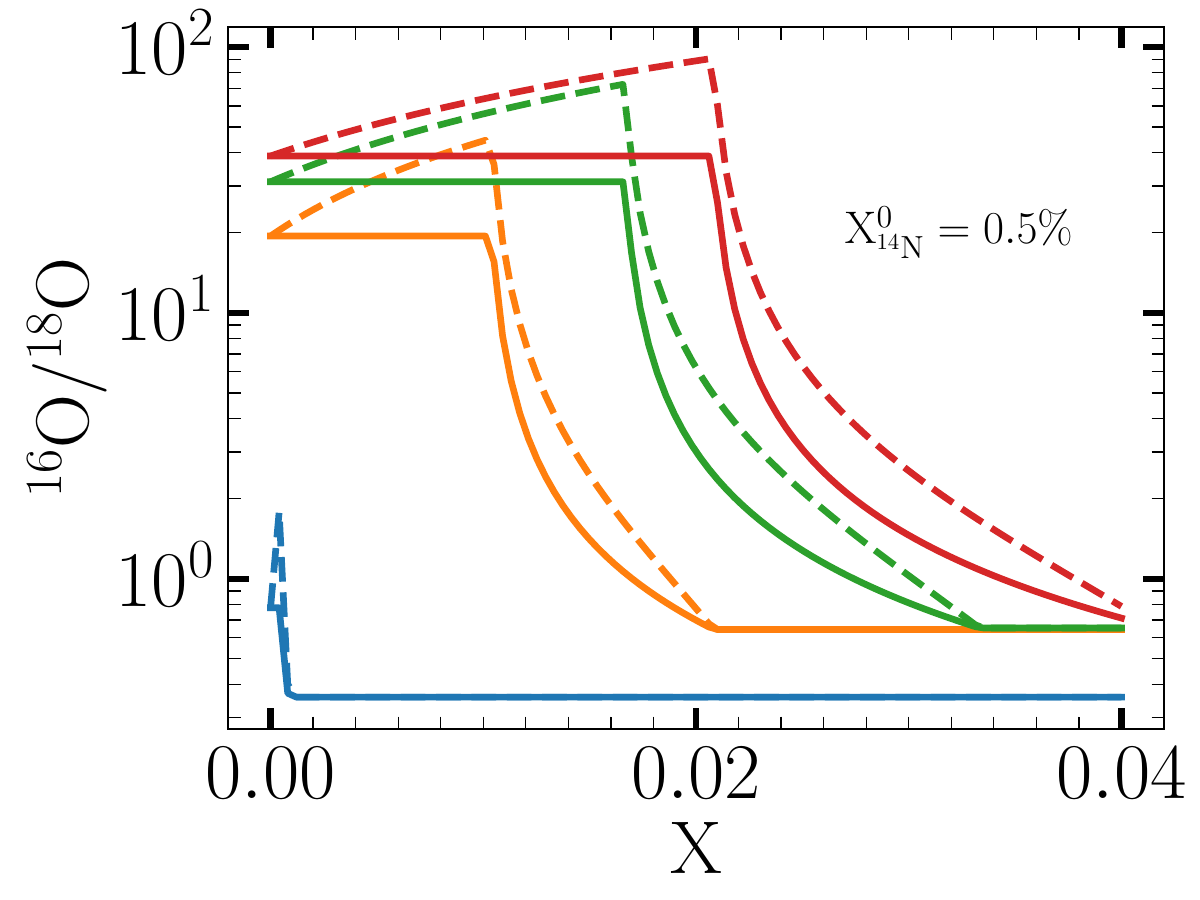}
    \includegraphics[scale=0.4]{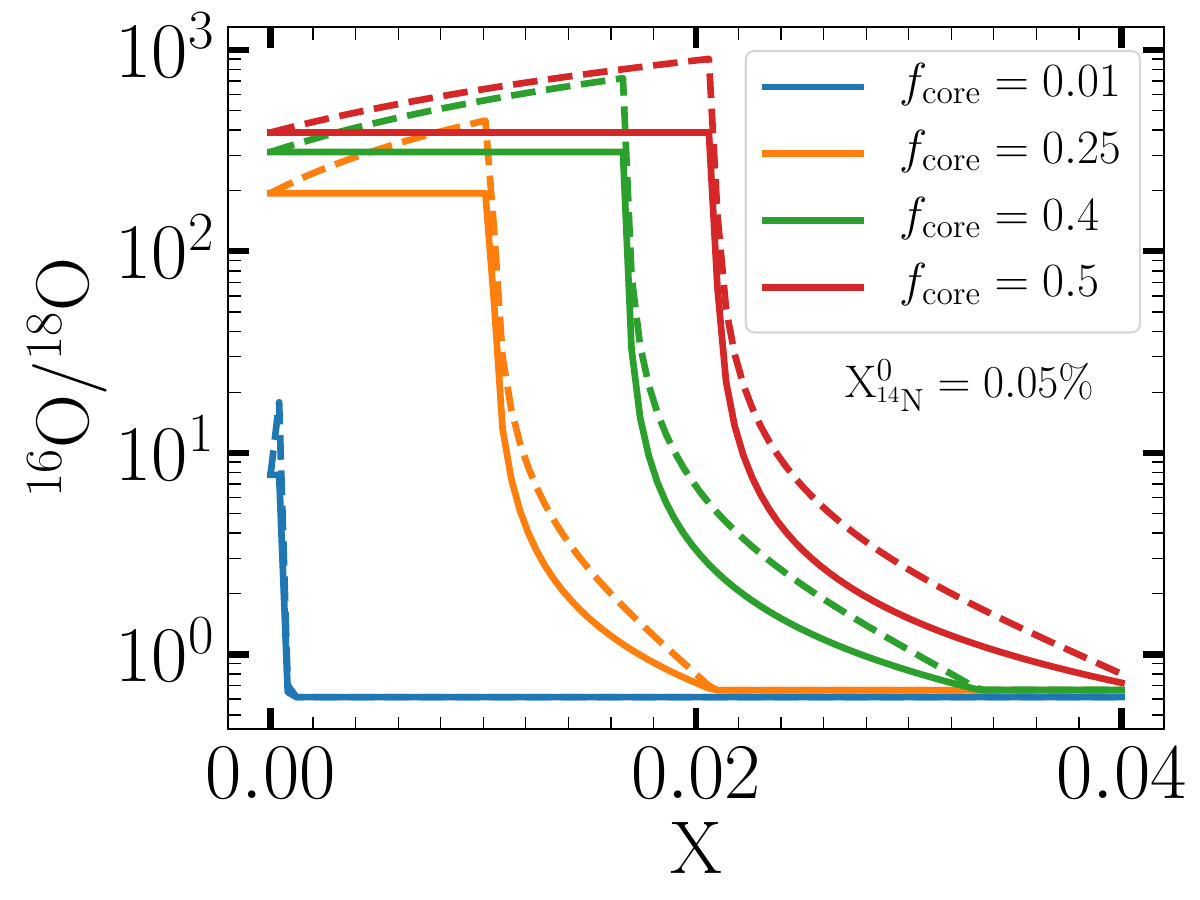}
    \caption{{The $^{16}{\rm O}/^{18}{\rm O}$ ratio in the SoF as a function of H mass fraction, X, \revisionII{assuming constant density and temperature}, for two representative initial abundances of $^{14}{\rm N}$. Left: X$^{0}_{^{14}{\rm N}} = 0.5\%$. Right: X$^{0}_{^{14}{\rm N}} = 0.05\%$. \revisionII{Solid lines are according to eq.~\ref{eq:o_ratio}.} Dashed lines also include alpha captures on $^{13}$C, which creates more $^{16}$O and therefore increase the ratio. \revisionII{ We additionally assume that all the involved reaction chains $^{12}{\rm C}(p, \gamma)^{13}{\rm N}(\beta^+)^{13}{\rm C}(p, \gamma)^{14}{\rm N}$, $^{14}{\rm N}(\alpha, \gamma)^{18}{\rm F}(\beta^+)^{18}$ proceed to completion, and that $^{12}$C captures all the protons before $^{13}{\rm C}(p, \gamma)^{14}{\rm N}$ takes place.}}}
    \label{fig:dredge-up-theory}
\end{figure*}

\revisionII{We plot this function in Figure~\ref{fig:dredge-up-theory} for two representative values of ${\rm X}_{\rm ^{14}N}^0$ of $5\times 10^{-3}$ (left panel) and $5\times 10^{-4}$ (right panel; corresponding to subsolar metallicities) in solid lines and for four values of $f_{\rm core}$: 0.01 (blue), 0.25 (orange), 0.4 (green), and 0.5 (red).}
Later on, as the burning proceeds, some $^{16}$O can be created by the remaining $^{13}{\rm C}$ via $^{13}{\rm C}(\alpha, n)^{16}{\rm O}$, which would be contributing to increasing the ratio \revisionII{. The ratio including this reaction is plotted in the same figure as dashed lines.}

{As can be seen, there are two regimes with a sharp transition between them. For low X values, the ratio is mostly dependent of the initial nitrogen abundance (comparing the left and right panels with $X_{\rm ^{14}N}^0$ of 0.5\%, and 0.05\%, respectively). For higher X the ratio decreases to a value $\le 1$. The location of the transition is a function of the accretor mass fraction in the SoF, $f_{\rm core}$. It can be generally written as ${\rm X} > f_{\rm core}/24$, where at twice this mass fraction at ${\rm X} > f_{\rm core}/12$, the ratio falls down to $<1$. For our infinite resolution extrapolation, $f_{\rm core}=0.25$, a minimal amount of ${\rm X}>1$\% is required to obtain the observed values in RCBs.}

{To estimate the H mass fraction in the SoF we assume that any amount of hydrogen resident in the He WD envelope (outside the hydrogen free core) gets mixed completely into the SoF, i.e., ${\rm X} = M_{\rm He-env}^{\rm H}/M_{\rm SOF}$, where $M_{\rm He-env}^{\rm H}$ is the hydrogen mass in the envelope of the He WD. Under this assumption an amount of $M_{\rm He-env}^{\rm H}=
(f_{\rm core}/24)M_{\rm SOF}
\approx 10^{-3}M_{\rm \odot}$ 
is required. For the case considered here, where the He WD is the result of a previous complex binary interaction, which may include several episodes of mass transfer and one or two full common envelope interactions, the amount of hydrogen in the He WD envelope can only be poorly constrained. Alternatively, if a merger takes place in an environment that is richer in H, like a merger inside a common envelope or a merger with a sub-giant star as proposed by \cite{Shen2023}, they could also produce the right ratios. In the latter cases the rest of the hydrogen envelope must becomes unbound in the process, so by the time the post-merger object evolves to its RCB phase (on the order of $\sim 10^{2}-10^{3}~{\rm yr}$), no hydrogen is seen.}

{Finally, two other factors can increase the $^{16}{\rm O}/^{18}{\rm O}$ ratio, although we claim here that they play only minor roles. One is the amount of $^{16}$O that is dredged-up from the accretor {\em outside} the SoF. If we assume a total mixing of the mass in the SoF to the surface, the surface ratio will be 
\begin{equation}
    \begin{split}
    \left(^{16}{\rm O}/^{18}{\rm O}\right)_{\rm surf}=\left(M_{\rm SOF}^{\rm ^{16} O}+M_{\rm out}^{\rm ^{16}O}\right)/M^{\rm ^{18}O} \\
    = \left(^{16}{\rm O}/^{18}{\rm O}\right)_{\rm SOF}\left(1+M_{\rm out}^{\rm ^{16}O}/M_{\rm SOF}^{\rm ^{16} O}\right).
    \end{split}
    \label{eq:sof_surf}
\end{equation} 
Where $M_{\rm out}^{\rm ^{16}O}$, is the $^{16}$O mass outside the SoF as a result of the dredge-up from the accretor. We find in our infinite resolution extrapolation $0.02~M_{\rm \odot}$ of ${\rm ^{16}O}$ at low densities (Figure~\ref{fig:dredge-up} left panel) and that $0.01~M_{\rm \odot}$ of it resides in the SoF (Figure~\ref{fig:dredge-up} right panel), while the remaining $0.01~M_{\rm \odot}$ must reside outside the SoF. Consequently $M_{\rm out}^{\rm ^{16}O}\simeq M_{\rm SOF}^{\rm ^{16} O}$, and this dredge-up to outer layers only doubles the ratio. Additionally, the $^{18}$O will eventually burn to $^{22}$Ne through the reaction $^{18}{\rm O}(\alpha, \gamma)^{22}{\rm Ne}$. However, \cite{Staff2012} found that the relatively low temperature in the SoF will delay this reaction to $10^{12}~s\approx 30,000~{\rm yr}$ (see the green line in their figure~16, for example). At this point, the convection deep into the SoF has already stopped (see e.g., \citealt{Menon2013,2020MNRAS.498.2912C,munson21}), and therefore the $^{16}{\rm O}/^{18}{\rm O}$ ratio in the SoF will no longer affects the surface ratio. In \cite{munson21} the convection to the SoF persists up until $100~{\rm yr}$ and then stops. This gives a window of time for the production of $^{18}{\rm O}$ without being processed further to $^{22}{\rm Ne}$.}

\section{Summary and Conclusions}
\label{sec:concl}

In this paper, we have carried out a set of simulations of WD mergers to further our understanding of the origin and phenomenology of RCB stars. All our simulations have total masses in the narrow range $0.85-0.9 M_\odot$ because of observational constraints, and a mass ratio  $q=M_2/M_1=M_{\rm donor}/M_{\rm accretor}$=$0.7$ because previous simulations have indicated that this ratio yields the optimal conditions for incomplete helium burning. Our simulations extend previous studies by using a new and improved code \octo\ and
comparing the outcomes for a range of resolutions, at  10, 11, 12, and 13 levels of refinement. The finest grid cell for {\sc L10ND} was $1.2\times 10^{-3} R_\odot$, while the finest cell for {\sc L13} was $1.43\times 10^{-4} R_\odot$ (Table~\ref{tab:past_sims}). The outcomes of our resolution study show that a finer grid results in better angular momentum conservation (Figure~\ref{fig:q07_level12_rot_cons}) and allows us to follow the evolution with lower mass transfer rates (Figure~{\ref{fig:comp_res}}). However, this also translates to a longer and costlier evolution to merger. We find excellent agreement between our {\sc L13} evolution and the analytic predictions right up to the tidal disruption and merger, where the analytic treatment fails (Appendix~\ref{app-verification}). This comparison serves both to verify our method and illustrates the limitations of the simple analytic approach. 

Another comparison was with respect to the choice of frame (Section~\ref{ssec:inertial}). Inertial frame simulations suffer from numerical diffusion due to the advection of the binary stars across the grid cells, which results in higher initial mass transfer rates and a rapid evolution through a merger (Figure~\ref{fig:comp_res}). Rotating frame simulations avoid these problems and thus they are much more suitable for the simulation of marginally stable or unstable binaries.

The comparison of our simulations with equivalent ones carried out with a different grid code, \flower, have yielded remarkably similar results (Appendix~\ref{app-flower}). However, \octo\ has the advantage of using adaptive mesh techniques, allowing us to simulate a larger domain than in \flower. In \octo\ simulations we therefore could follow the merged object up to large distances and catch that the density profile outside the disk falls like $\rho \propto R^{-8}$ (Figure~\ref{fig:post_merger}).

We then analyzed the merged object in Section~\ref{sec:post_merger}. Its structure is an inner pressure-supported, sphere, slightly oblate, rotating as a solid body, with approximate radius $R_{\rm sph} = 1 \times 10^9$~cm surrounded by a disk (Figure~\ref{fig:post_merger}). The inner part of the disk is in Keplerian rotation, while the outer parts rotate slower than the Keplerian speed. This could be due to partial pressure support or that these outer regions would eventually fall back (Figure~\ref{fig:post_merger2}). 

We have analyzed the temperature and density conditions of the gas after the merger to determine how much of the gas is likely to undergo nuclear fusion (Section~\ref{ssec:SoF}). As in previous studies, we have found a region of hot temperatures around the accretor with the right conditions for helium burning. However, due to the presence of a cold blob of gas, a remnant of the disrupted donor, regions of higher latitudes are hotter than the equatorial, and the SoF does not consist of complete shell, but has some hole(s) in it (Figures~\ref{fig:SoF}, \ref{fig:temp_prof}, and \ref{fig:3alpha}). {The amount of fusing gas is a function of resolution, but it converges to $\sim 0.08$~\msun (see Figure~\ref{fig:3alpha} and Table~\ref{tab:postmerger}). The energy generated by the nuclear reactions over the time span of $\sim 20~{\rm min}$ is $\sim 10^{44}$~erg still negligible compared to any other energy component in the simulation, thus justifying the fact that we do not include nuclear effects on the evolution of the simulation}.

The most important goal of our research was to re-examine the dredge-up of material from the accretor at several resolutions and to obtain an improved estimate of the dredged-up mass by extrapolating to an ``infinite" resolution (Figure~\ref{fig:dredge-up}). {Our extrapolated value of $M^{\rm ^{16}O} = 0.02~M_{\rm \odot}$ at lower densities than $10^5~{\rm g~cm^{-3}}$ (Figure~\ref{fig:dredge-up} left panel) bridges the gap in the amount of dredged-up $^{16}$O mass found in previous $q=0.7$ simulations between SPH vs finite-volume codes (see Table~1 of \citealt{Staff2018}), and our result sits right at the middle. In addition, the extrapolated amount of $0.02~M_{\rm \odot}$ of accretor mass in the SoF (Figure~\ref{fig:dredge-up} right panel and Table~\ref{tab:postmerger}), and outside the SoF, is lower than previously found (see \citealt{Staff2012} Table~3), which consequently results in two times smaller accretor mass fraction in the SoF of $f_{\rm core}=0.25$ (see Table~\ref{tab:postmerger} vs. \citealt{Staff2012} Table~2). However, our analysis based on the most relevant nuclear reactions (Figure~\ref{fig:dredge-up-theory}) indicates that even with this relatively mild dredge-up of $^{16}$O from the accretor, creating $^{18}$O only from the available $^{14}$N in the SoF might not be enough. In order to obtain the observed ratio of ${\rm ^{16}O}/{\rm ^{18}O}\lesssim 10$, H burning on the $^{12}$C that accompanied the dredge-up of $^{16}$O should take place, forming additional $^{14}$N that will be further processed to $^{18}$O.}

Our attempts to simulate evolution with no driving yielded problematic results (see Section \ref{ssec:nondriven_sims}), which were examined in some detail using both \octo\ and \flower\ (see Appendix \ref{app-nondriven}).
We conclude that with the current implementation of the ZTWD EoS, non-driven simulations from which we can infer further physical insights are impossible. The driven simulations, on the other hand, merged before any instability started to grow, and we are confident that the merger occurred because of the unstable mass transfer (see Section~\ref{ssec:mass-transfer} as well as Appendix~\ref{app-verification}). We do note that the implication of the adapted EoS on the nuclear reactions and the dredge-up is unclear and should be investigated in detail in the future. We tested how a slight change in the derivation of the thermal energy (with or without the dual-energy formalism) could affect the simulation results and found only a very minor difference (Appendix \ref{app-def}). 

Finally, to accelerate \octo\ simulations we have developed a GPU-accelerated version over the last few years. We outlined our progress in Appendix~\ref{app-scaling}, where we showed runtimes on the GPU partition of Perlmutter using our CUDA kernels, and a noticeable GPU speedup over the CPU runs. Our aim is to expand on our results shown in this work by adding more GPU optimizations, as well as a Kokkos version, allowing us to significantly speed-up future simulations and to target a wider range of supercomputers.

\section*{Acknowledgments}
{We thank K. J. Shen for insightful comments, and the anonymous referee for helpful remarks.}
This research used resources from the National Energy Research Scientific Computing Center, a U.S. Department
of Energy Office of Science User Facility operated under
Contract No. DE-AC02-05CH11231.
Portions of this research were conducted with high performance computational resources provided by the Louisiana Optical Network Infrastructure (http://www.loni.org).
This research was supported in part by Lilly Endowment, Inc., through its support for the Indiana University Pervasive Technology Institute.
This material is based upon work supported by the National Science Foundation under Grant No. CNS-0521433. This work was also supported by National Science Foundation Award 1814967.
Any opinions, findings and conclusions, or recommendations expressed in this material are those of the authors, and do not necessarily reflect the views of the National Science Foundation (NSF).
This work was supported in part by Shared University Research grants from IBM, Inc., to Indiana University.
Octo-Tiger's GPU development and testing was partially supported by a Swiss National Supercomputing Centre (CSCS) grant under project ID s1078.
This work also required using and integrating a Python package for astronomy, yt (http://yt-project.org, \citealt{Turk2011}).
JF thanks the Lorraine and Leon August Endowment to LSU for support.
This work was supported by the U.S. Department of Energy through the Los Alamos National Laboratory. Los Alamos National Laboratory is operated by Triad National Security, LLC, for the National Nuclear Security Administration of U.S. Department of Energy (Contract No. 89233218CNA000001). LA-UR-24-28926.

\section*{Data Availability}
\octo\ is available on GitHub\footnote{\url{https://github.com/STEllAR-GROUP/octotiger}} and was built using the following build chain\footnote{\url{https://github.com/STEllAR-GROUP/OctoTigerBuildChain}}. On Queen-Bee and BigRed \octo\ version \cite{dominic_marcello_2021_4432574} was used and on NERSC's Perlmutter the pre-release state of v0.9.0 was used. We had to use different versions due to some fixes for the A100 GPUs Perlmutter (see Appendix~\ref{app-scaling}). 

\bibliographystyle{mnras}
\bibliography{bibliography}
\bsp

\clearpage
\appendix

\section{Zero-Temperature White Dwarf Equation of State}
\label{app-eos}
When using the Zero-Temperature White Dwarf (ZTWD) equation of state (EoS) in Octo-Tiger, the fluid is modeled using a combination of a zero-temperature Fermi fluid and an ideal gas. The total pressure, $P$, is the sum of pressure from the zero-temperature fluid, $P_{\rm deg}$, and the ideal gas internal (thermal) pressure, $P_{\rm th}$. The gas energy, the quantity \octo\ evolves, is the sum of internal (thermal), degenerate, and bulk kinetic energy densities.
\begin{equation}
    E = E_\mathrm{th} + E_\mathrm{deg} + E_\mathrm{kin}.
\end{equation}

The energy density of the degenerate electron gas, $E_{\rm deg}$, is given by the relation,
\begin{equation}
E_{\rm deg} = H_{\rm deg}\rho  - P_{\rm deg},
    \label{edeg}
\end{equation}
where 
\begin{equation}
H_{\rm deg} = \left( 8A /B \right) \left[ \left( x^2 + 1 \right)^{1/2} - 1\right]
    \label{hdeg}
\end{equation}
is the specific enthalpy of the degenerate electron gas, \begin{equation}
P_{\rm deg} = A \left[ x \left( 2x^2 - 3 \right) \left( x^2 - 1\right)^{1/2} -3 \sinh^{-1}{x}\right],
\label{pdeg}
\end{equation}
and
\begin{equation}
x \equiv  \left( \rho / B \right)^{1/3}.
\label{x}
\end{equation}
The constants $A$ and $B$ are
\begin{equation}
\begin{split}
    A  & = \pi m_e^4 c^5 / (3h^3) = 6.00228 \times 10^{22} {\rm\; dynes\;cm^{-2}}, \\
    B  & = \mu_e \times 8 \pi m_p m_e^3 c^3 / (3 h^3) = \mu_e \times 9.81011 \times 10^5 {\rm \; g \; cm^{-3}},
\end{split}
    \label{AB}
\end{equation}
where $\mu_e$ is the average ratio of nucleons to electrons. We assume that throughout each white dwarf (and basically throughout the whole domain) 
$\mu _e = 2$, hence, $B=1.96202\times 10^6 {\rm \; g \; cm^{-3}}$.

For numerical reasons, at low density regions \mbox{($x<0.001$)}, we approximate $E_{\rm deg}$ by a Taylor expansion
\begin{equation}
E_{\rm deg}\left(x<0.001\right)=2.4 \times A x^5.
\label{small_edeg}
\end{equation}

As described in \cite{Marcello2021}, \octo\ evolves a second variable for the energy, the ``entropy tracer", $\tau$ \citep{Motl2002}. The inclusion of this additional variable allows for the proper evolution of shocks while still retaining a precise calculation of the thermal energy in regions where the kinetic energy dominates over the non-degenerate energy components. 
The thermal energy density therefore is computed according to
\begin{equation}
    E_{\rm th} = 
\begin{cases}
    E - E_{\rm kin} - E_{\rm deg}, & \text{if } E - E_{\rm kin} - E_{\rm deg}  \geq \epsilon_1 E\\
    \tau^{\gamma},             & \text{otherwise}
\end{cases},
\label{eq:def_th}
\end{equation}
where $\gamma = \frac{5}{3}$ is the ratio of specific heats.
In addition, the entropy tracer is reset using $\tau = \left(E_{\rm th}\right)^{\frac{1}{\gamma}}$,
 in computational cells which satisfy 
\begin{equation}
    E - E_{\rm kin} - E_{\rm deg} > \epsilon_2 E
\label{eq:dual_energy2}
\end{equation}
for at least one of the adjacent cells or the cell itself. Otherwise it is left alone. Note the additional degenerate energy terms in \ref{eq:def_th} and \ref{eq:dual_energy2} compared to what is described in \cite{Marcello2021}. Throughout this paper, we use the default values $\epsilon_1=0.001$ and $\epsilon_2=0.1$ as was chosen by ~\cite{Bryanetal1995}. 

The combination of using both a ZTWD EoS and a dual energy formalism implies that in the WD interiors, where the degenerate energy dominates, the entropy tracer is not directly updated from the gas energy and is only affected by diffusion. Initially, the entropy tracer is set to a very low value inside the stars, which corresponds to a very low, practically zero, temperature. However, as the simulation evolves in time the entropy tracer begins to increase inside the stars, which effectively slightly heats the stars. To test whether this heating is significant, we have run a simulation where we disable the dual energy formalism and found that this has only a very little effect on the results (see Appendix~\ref{app-def}). 

Finally, the temperature is calculated as
\begin{equation}
 T = \frac{\left(\gamma - 1 \right)E_{\rm th}}{n K_{\rm B}},
 \label{eq:T_octo}    
 \end{equation}
where
\begin{equation}
 n = \sum_i \frac{\rho_i}{\mu_i m_{\rm H}},
 \label{eq:n_octo}    
 \end{equation}
 $\mu_i$ is the molecular weight of the $i$th specie, $m_{\rm H}$ is the mass of an hydrogen atom, and $K_{\rm B}$ is the Boltzmann constant. The total pressure is the sum of the degenerate and thermal pressure components
\begin{equation}
 P = P_{\rm deg} + \left(\gamma - 1 \right)E_{\rm th},
 \label{eq:P_octo}    
 \end{equation}

\section{Verification of the early mass transfer with analytical considerations}
\label{app-verification}

In what follows we analyze the results of our numerical simulations focusing on the results of the highest resolution {\sc L13} run, with two goals in mind: a check of the accuracy of the simulations, and a better understanding of the evolution of the binary system. We start from an approximate analytical expression for the total angular momentum of the binary as the sum of the two-body orbital angular momentum for point-mass components plus the contribution from the spins, where $I_1$ and $I_2$ are the moments of inertia of the components with respect to their respective centers of mass: 

\begin{equation}
J = M_1 M_2 \sqrt{Ga \over M } (1 - e^2)^{1\over 2} + I_1 \Omega_1 + I_2 \Omega_2    
\end{equation}
 This expression is approximate in that we are using the simple Keplerian two-body orbital frequency for point masses, valid if both components remain spherically symmetrical. This approximation gets gradually worse as the binary separation is reduced and tidal distortions increase.
We construct from the above the following logarithmic derivatives:
\begin{multline}
    {\dot {J} \over J_{\rm orb} } = {\dot{M}_1 \over M_1} + {\dot{M}_2 \over M_2} +
    {1\over 2}{\dot{a} \over a} - {1 \over 2} {\dot{M} \over M} - {e \dot{e} \over (1 - e^2)} +\\
    {\dot{I}_1 \Omega_1 + I_1 \dot{\Omega}_1 + \dot{I}_2 \Omega_2 + \Omega_2 \dot{\Omega}_2 \over J_{\rm orb}},
\label{eq:logdJdot}
\end{multline}
where the part of the right-hand-side that is on the first line is the $\dot{J}_{\rm orb} / J_{\rm orb}$, while the term that is on the second line is $\dot{J}_{\rm spin} / J_{\rm orb}$:

\begin{multline}
    {\dot{J}_{\rm spin} \over J_{\rm orb}} = {J_1 \over J_{\rm orb}} \left({\dot{I}_1 \over I_1} + {\dot{\Omega}_1 \over \Omega_1}   \right) + {J_2 \over J_{\rm orb}} \left( {\dot{I}_2\over I_2} + {\dot{\Omega}_2 \over \Omega_2} \right).
\end{multline}
Finally, we set an explicit expression for the left-hand-side of Equation~\ref{eq:logdJdot}:
\begin{equation}
    \dot{J} = \dot{J}_{\rm sys} + \dot{M} {j}_{\rm L2}
\end{equation}
where $\dot{J}_{\rm sys}$ represents systemic angular momentum losses, either numerical (see below) or real, such as gravitational radiation and magnetic braking, and $j_{\rm L2}$ is the effective specific angular momentum of gas at L2.
In our simulations the initial driving is set to $\dot{J}_{\rm sys} = - J_{\rm orb} \times 0.01 / P_0$ for a time $0\le t\le T_{\rm drv}$, and zero for 
$t>T_{\rm drv}$. Mass loss from the binary occurs mostly through L2, but also through L3, and gravitational coupling between the binary and the 
mass outflow will affect the average specific angular momentum of the outflow. Thus ${j}_{\rm L2}$ should be regarded as an effective specific angular momentum of the outflow.

While the above equations are in principle valid for the instantaneous values of the masses and orbital parameters, in practice the determination of these quantities and their derivatives from the numerical simulation, requires averaging over data points covering a time $\sim P_0$, where $P_0$ is the initial orbital period. Thus these equations should be considered effectively orbit-averaged equations. To construct the various terms in Eqn.~\ref{eq:logdJdot}, we obtain at every instant $t$ the best-fitting elliptical orbit, based on the data points for the interval $(t-P_0, t)$. This fit yields the orbit-averaged values for the separation $a$, the eccentricity $e$, and their derivatives, and \octo\ provides the rest of the quantities.

\begin{figure*}
    \centering
    \includegraphics[scale=0.75]{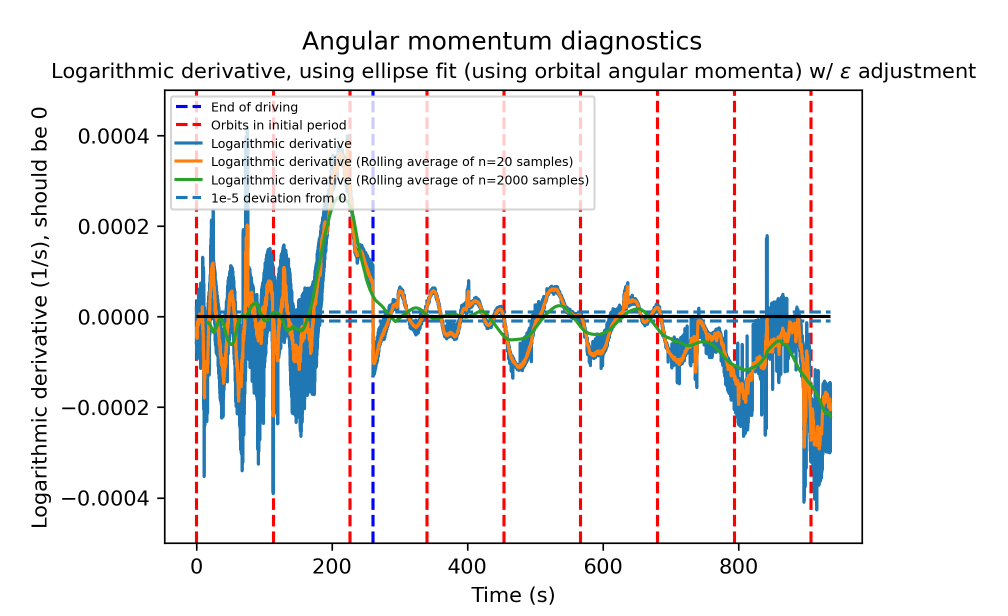}
    \caption{The graph shows the instantaneous differences between logarithmic derivatives on the left- and right-hand sides of the equation (\ref{eq:logdJdot}), which should be zero at all times.
    Rolling averages of this value are also plotted. The red vertical lines are each placed one orbital period apart, and the blue vertical line indicates the time at which driving ceased. 
    }
\label{fig:l13_bennett}
\end{figure*}

Fig. \ref{fig:l13_bennett} shows the instantaneous (blue) difference between the logarithmic derivatives on the left- and right-hand sides  of Equation~\ref{eq:logdJdot} based on the results of simulation {\sc L13}. The orange and green curves represent the above differences averaged over 20 and 2000 diagnostic samples, where a sample is approximately 5 timesteps. Perhaps the most remarkable message conveyed by 
this figure is how well the orbit-averaged equations describe the numerical evolution of the binary
which is free of the simplifying assumptions of the analytic model. Significant deviations from the simple model only occur during the initial driving and the final dynamical merger. Tidal distortions of both components, mass loss through the outer Lagrange points, and ultimately the inability to define unequivocally what constitutes the donor during its disruption, ultimately make the simple model inapplicable. The fact that the numerical simulation matches the expectations from the simple model precisely when the eccentricity is small ($e\approx 0.01$, see top left panel in Fig. \ref{fig:l12_diags}) and the components are nearly spherical strengthens our confidence in the validity of our results. 

\section{{Binary evolution comparison between \flower\ and \octo}}
\label{app-flower}

We find that the \flower\ simulations are very comparable to the high resolution \octo\ simulations with respect to resolution and to the amount of driving. The {\sc L12} simulation has a similar number of cells across the donor to simulations FL-1 and FL-2, while the {\sc L13} simulation has a similar total number of cells in the domain, but higher spatial resolution of the donor (Table~\ref{tab:past_sims}). The only difference between {\sc FL-1} and {\sc FL-2} is that the former is driven for one orbit, while the latter is driven for two, which is comparable to {\sc L12}, which is driven for 1.3 orbits, and to {\sc L13} which is driven for 2.3 orbits.

In Figure~\ref{fig:comp_flower} we show a set of diagnostic quantities as a function of time for the four simulations, similar to what we show in Figure~{\ref{fig:comp_res}. The merging times compare as one would expect: the simulations driven for $\sim$2 orbit merge at 10 and 17 orbits, for the {\sc L13} and {\sc FL-1}, respectively, while those driven for $\sim$1 orbits merge at 37 and 39 orbits, for the {\sc L12} and {\sc FL-2}, respectively.
\begin{figure*}
    \centering
    \includegraphics[scale=0.33]{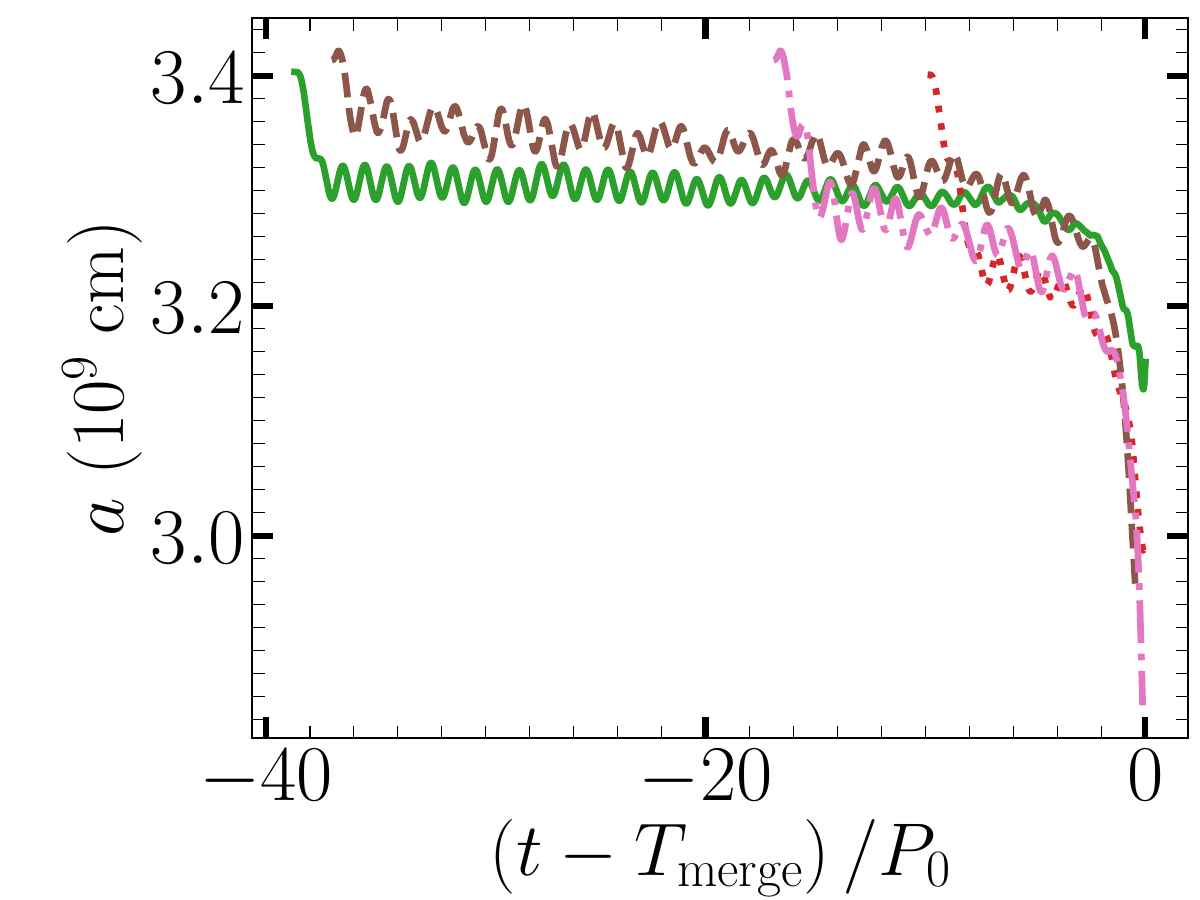}
    \includegraphics[scale=0.33]{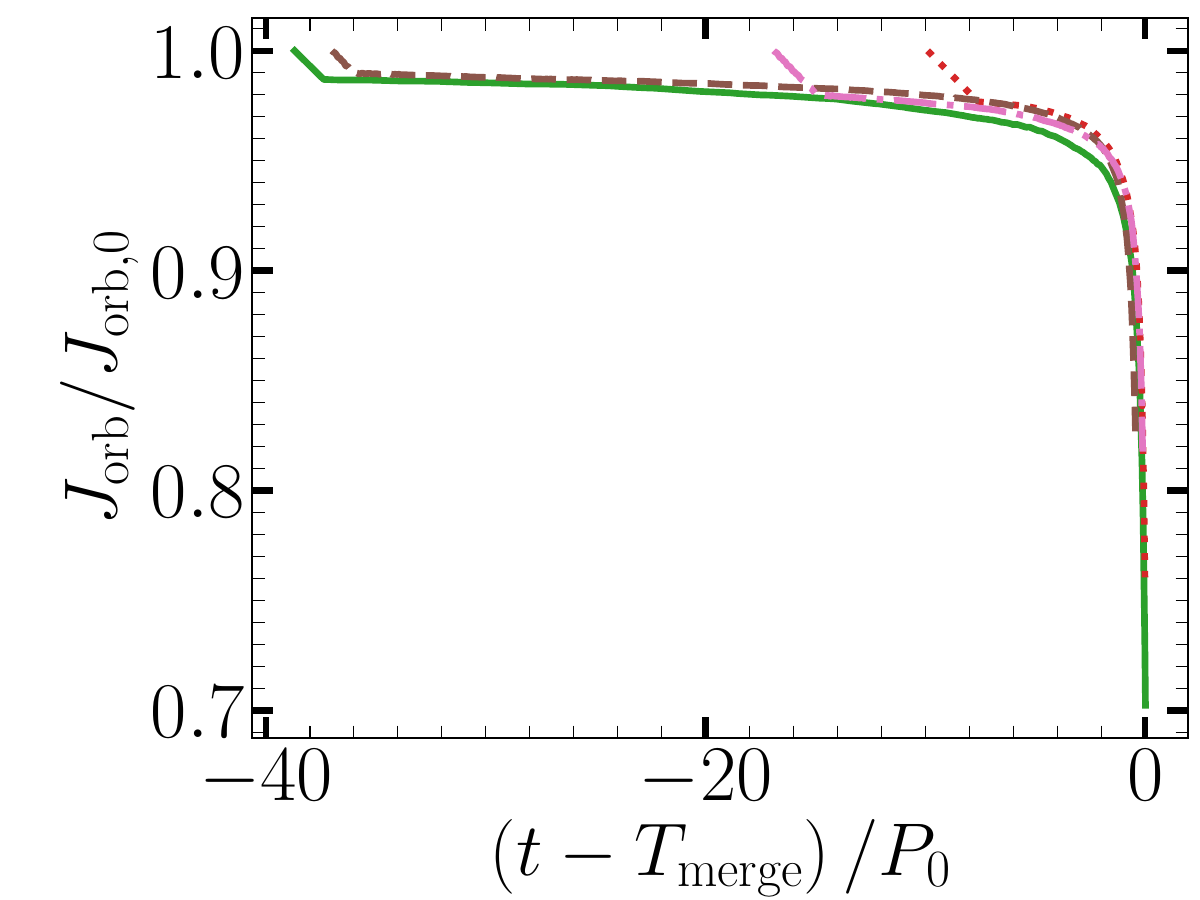}
    \includegraphics[scale=0.33]{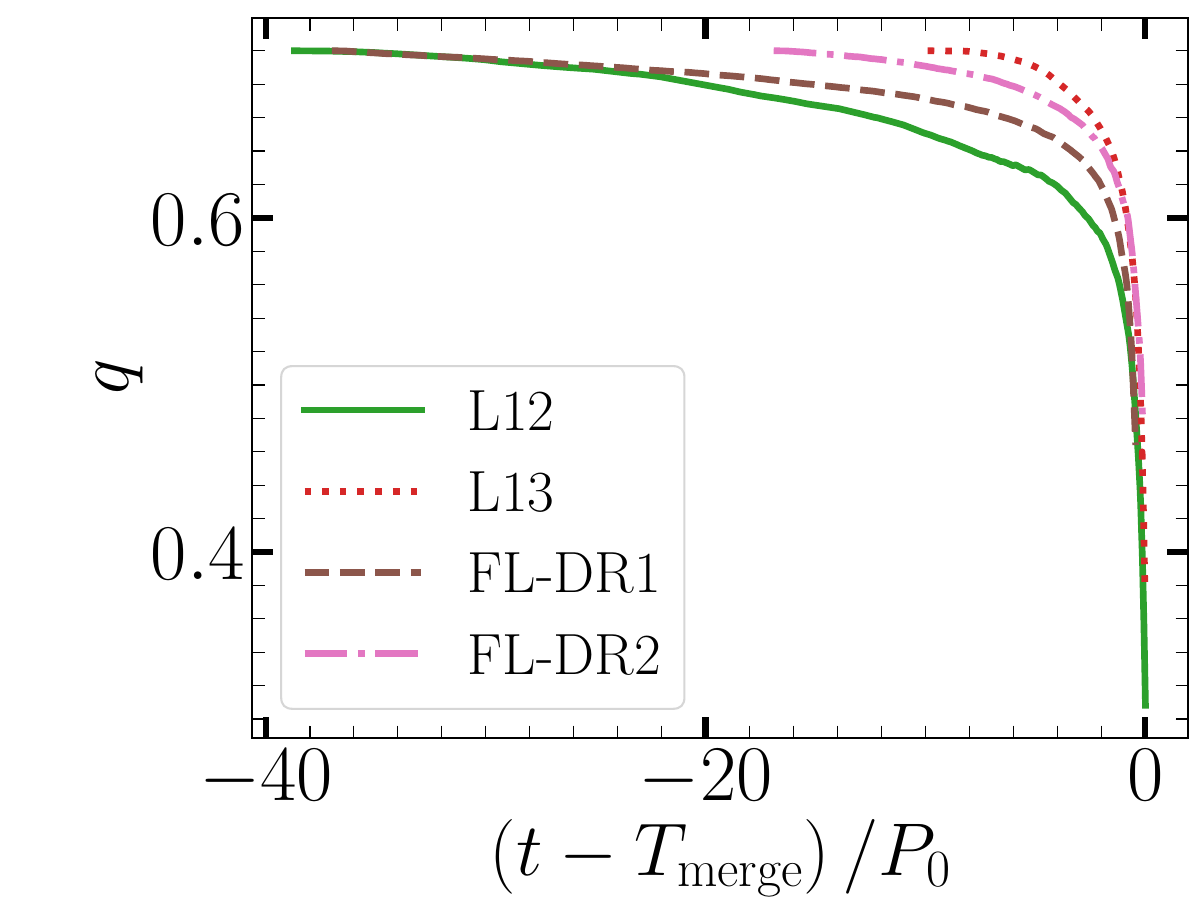}
    \includegraphics[scale=0.33]{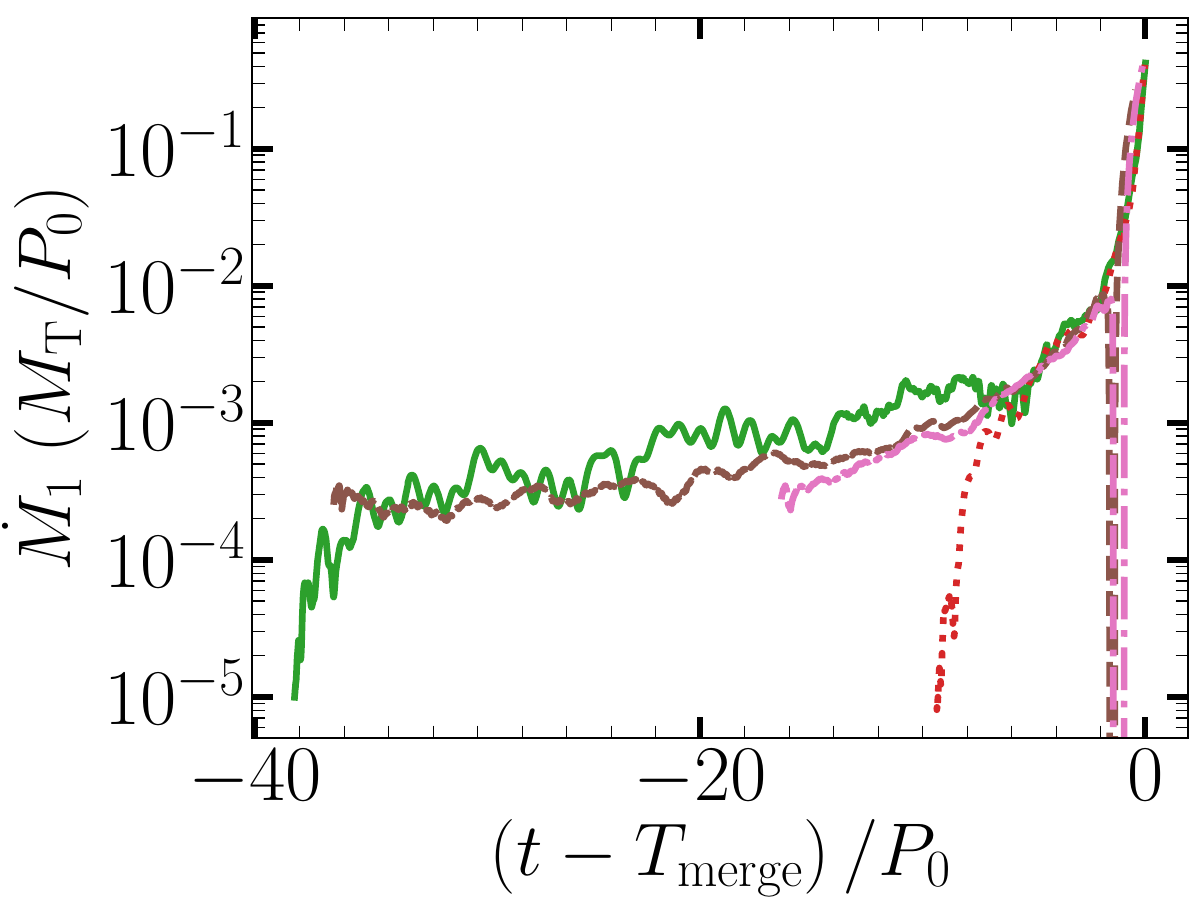}
    \includegraphics[scale=0.33]{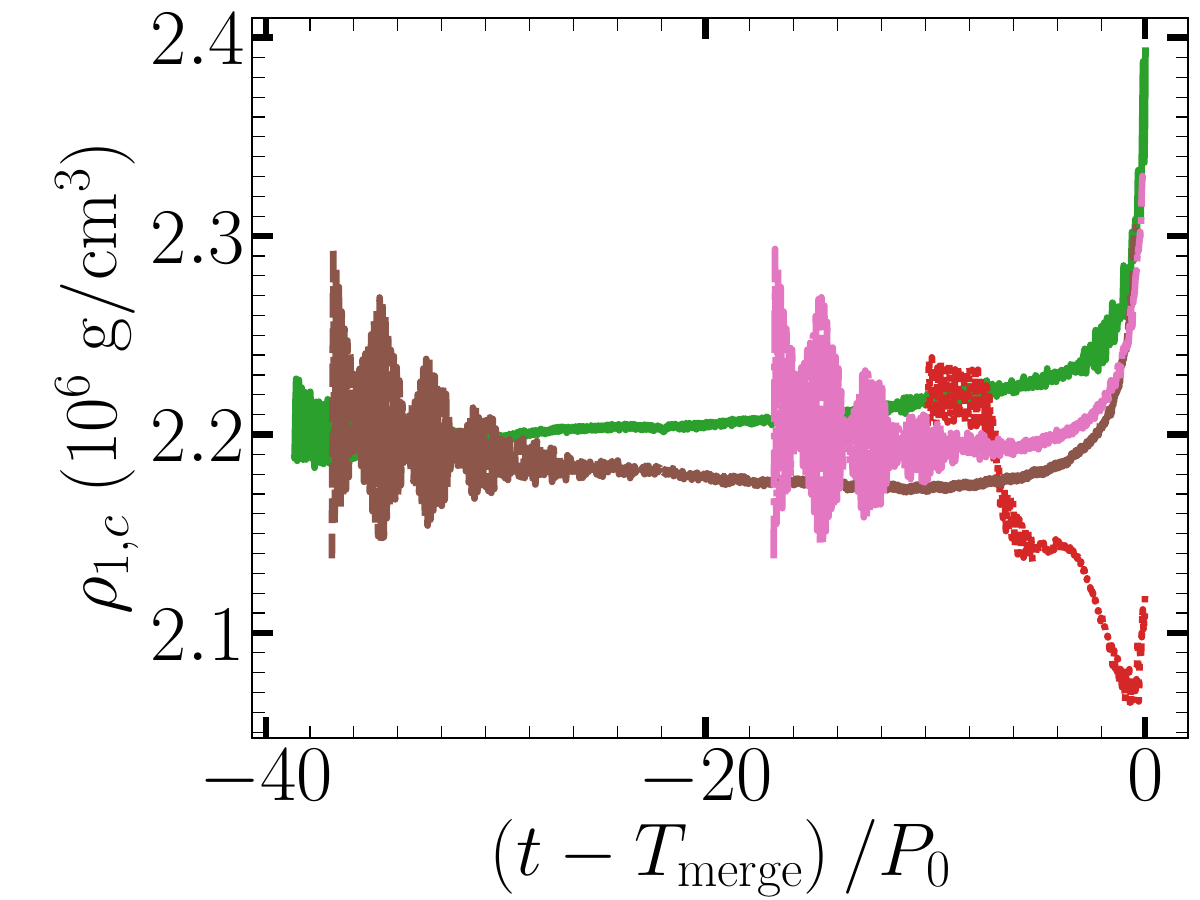}
    \includegraphics[scale=0.33]{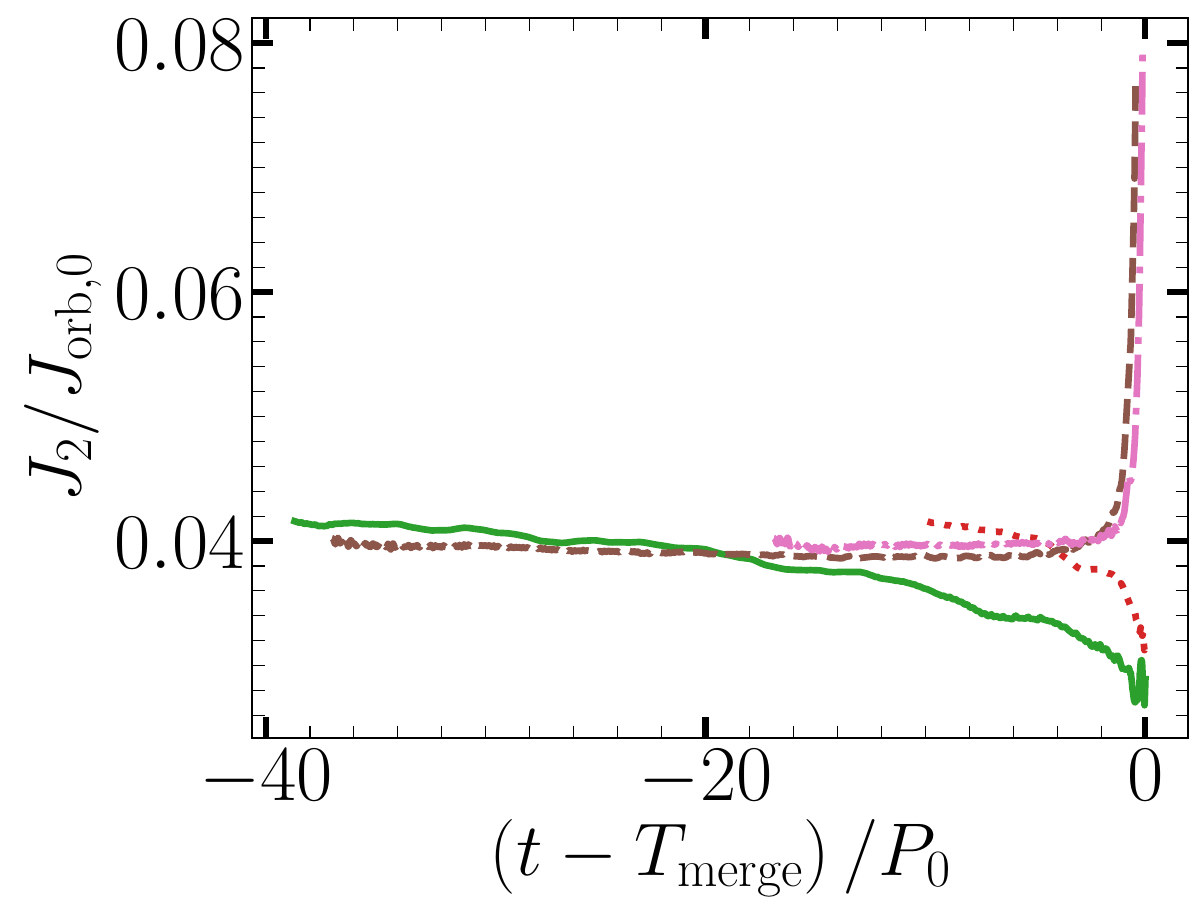}
    \caption{Comparison between {\sc L12} and {\sc L13} and the two \flower\ simulations. Orbital separation and orbital angular momentum (first row), mass ratio and mass transfer rate (second row), maximum accretor density and donor angular momenta (third row)}
    \label{fig:comp_flower}
\end{figure*}

The only anomaly in behavior is for {\sc L13}, which exhibits a decrease in the central density of the accretor before the merger, not exhibited by any other simulation in this comparison. Most noticeable is the extremely consistent mass transfer rate onto the accretor for all four simulations. 

Below are some links to the movies showing density slices of the \flower\ simulations:

\href{https://www.youtube.com/watch?v=0hiB-20QnbE&t=27s}{Equatorial Slice of Q0.7 driven for 1 orbit}

\href{https://www.youtube.com/watch?v=Gn7_YiAJvzA&t=43s}{Meridional Slice of Q0.7 driven for 1 orbit}

\href{https://www.youtube.com/watch?v=aR-2HuPwqSc&t=7s}{Equatorial Slice of Q0.7 driven for 2 orbits}

\href{https://www.youtube.com/watch?v=vGqrPORO5W4&t=138s}{Meridional Slice of Q0.7 driven for 2 orbits}

\section{Non-driven simulations - hitting the numerical barrier}
\label{app-nondriven}

When trying to simulate a non-driven `more realistic` evolution one should expect a long (in wall-clock time) and expensive (in computer resources) simulation. Moreover, on hundreds (or thousands) of orbits numerical errors that accumulate can come into effect. For example, instead of the actual mass transfer (which can be very slow), deviations from angular momentum conservation can start dictating the evolution. We wanted to take advantage of the excellent angular momentum conservation in \octo\ (see Figure~\ref{fig:q07_level12_rot_cons}, for example), and the fact that \octo\ can scale very well to a high number of cores to try and run non-driven simulations in a reasonable wall clock time. Unfortunately, as we will describe below, we have hit another numerical barrier, which has never been reached before. This time, the issues stem from the specific usage of the ZTWD EoS.

{In Table~\ref{tab:nondriven_sims}} we present the non-driven simulations {\sc L11ND}, {\sc L12ND} that are otherwise identical to the previously discussed {\sc L11}, {\sc L12}. Diagnostic plots for these four simulations are shown in Figure~\ref{fig:comp_nd}.}
\begin{table*}
 \caption{Our non-driven simulations}
 \label{tab:nondriven_sims}
 \centering
 \begin{tabular}{cllllllllll}
 \toprule
 Source & Code & $M_{\rm tot}$ & EoS & $T_{\rm drv}$ & $T_{\rm merge}$ & Resolution; $\Delta x_{\rm min}$ ($R_{\rm \odot}$)& donor res & $a_0$ & $P_0$ & Hybrid \\
 &&(\msun) &&($P_0$)&($P_0$)&(cells/particles)&(radial cells)& ($10^9$ cm) & (sec) & \\
 \midrule
  {\sc L11ND} & Octo (Bd) & 0.9 & ZTWD &  0 & 51.4 & 2.5 M; $5.73\times10^{-4}$ & 73 &3.413&114.0& Yes\\
  {\sc L12ND} & Octo (Bd) & 0.9 & ZTWD &  0 & -- & 5.3 M; $2.86\times10^{-4}$& 145 &3.403&113.6& Yes\\
  \bottomrule
 \end{tabular}
 \footnotesize{ Notations are as in Table~\ref{tab:past_sims}}
\end{table*}

\begin{figure*}
    \centering
    \includegraphics[scale=0.33]{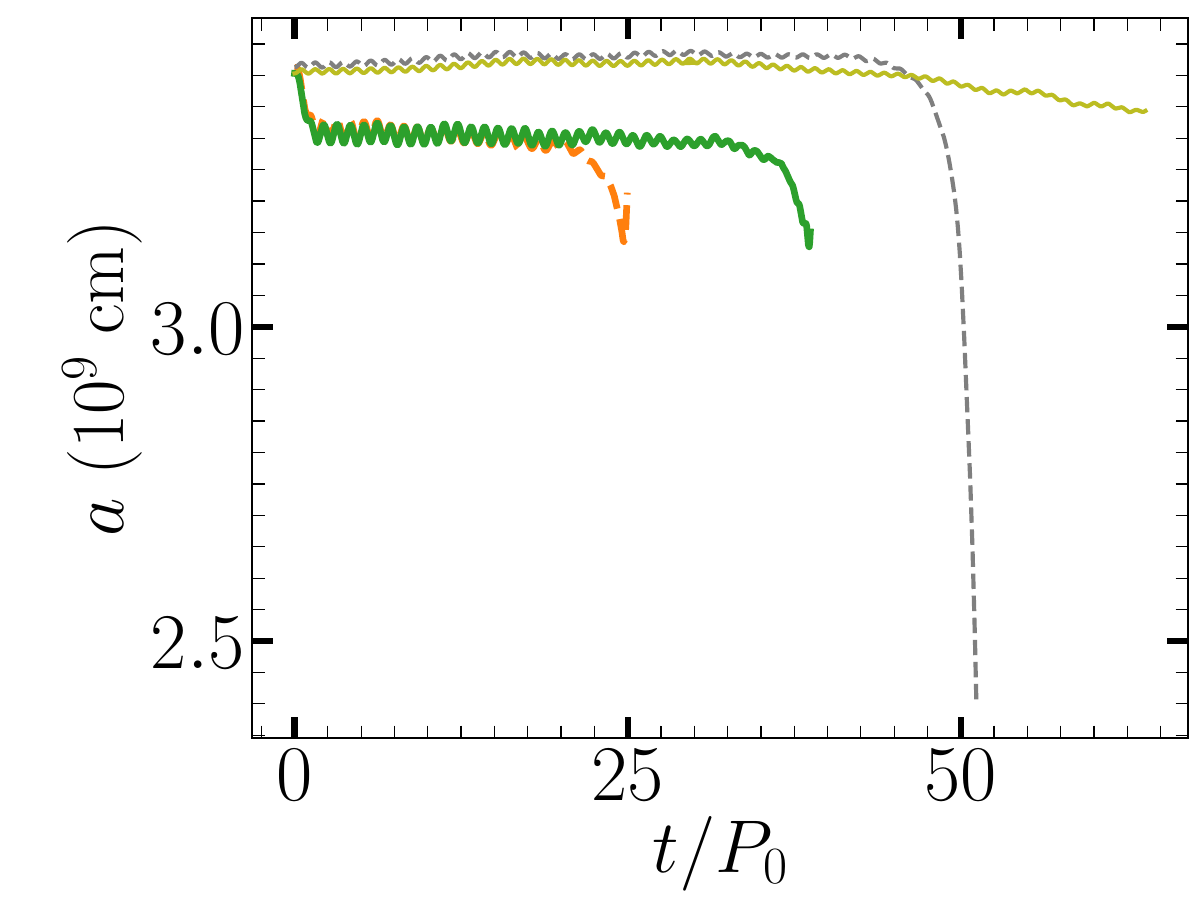}
    \includegraphics[scale=0.33]{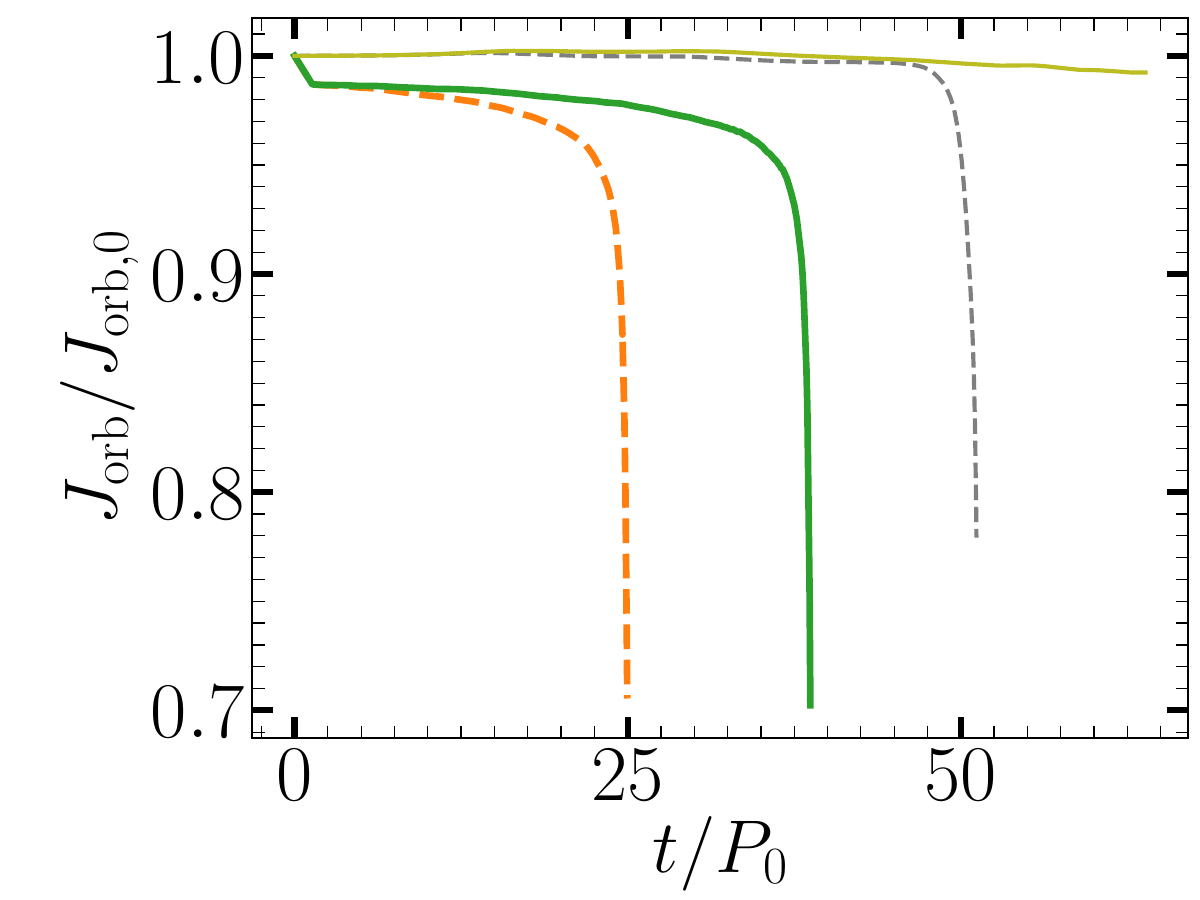}
    \includegraphics[scale=0.33]{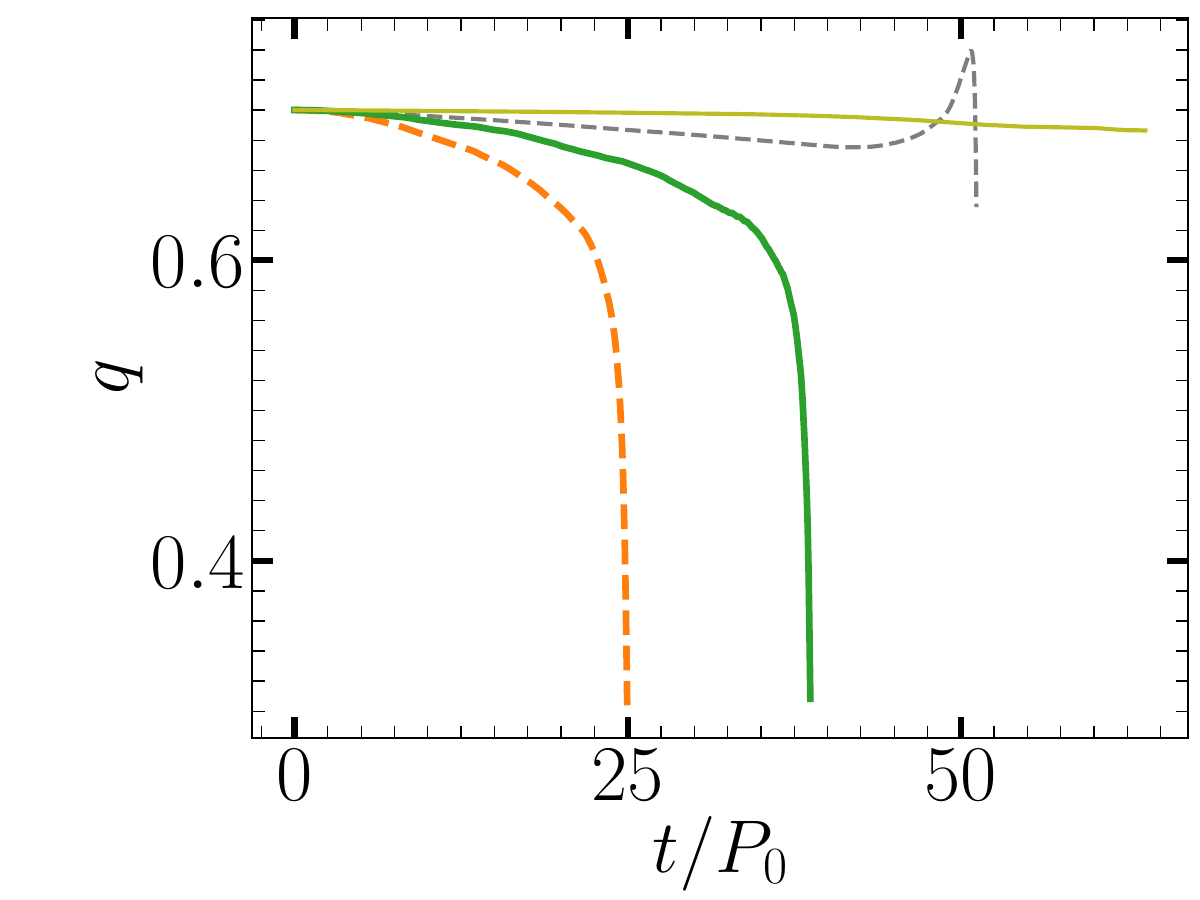}
\includegraphics[scale=0.33]{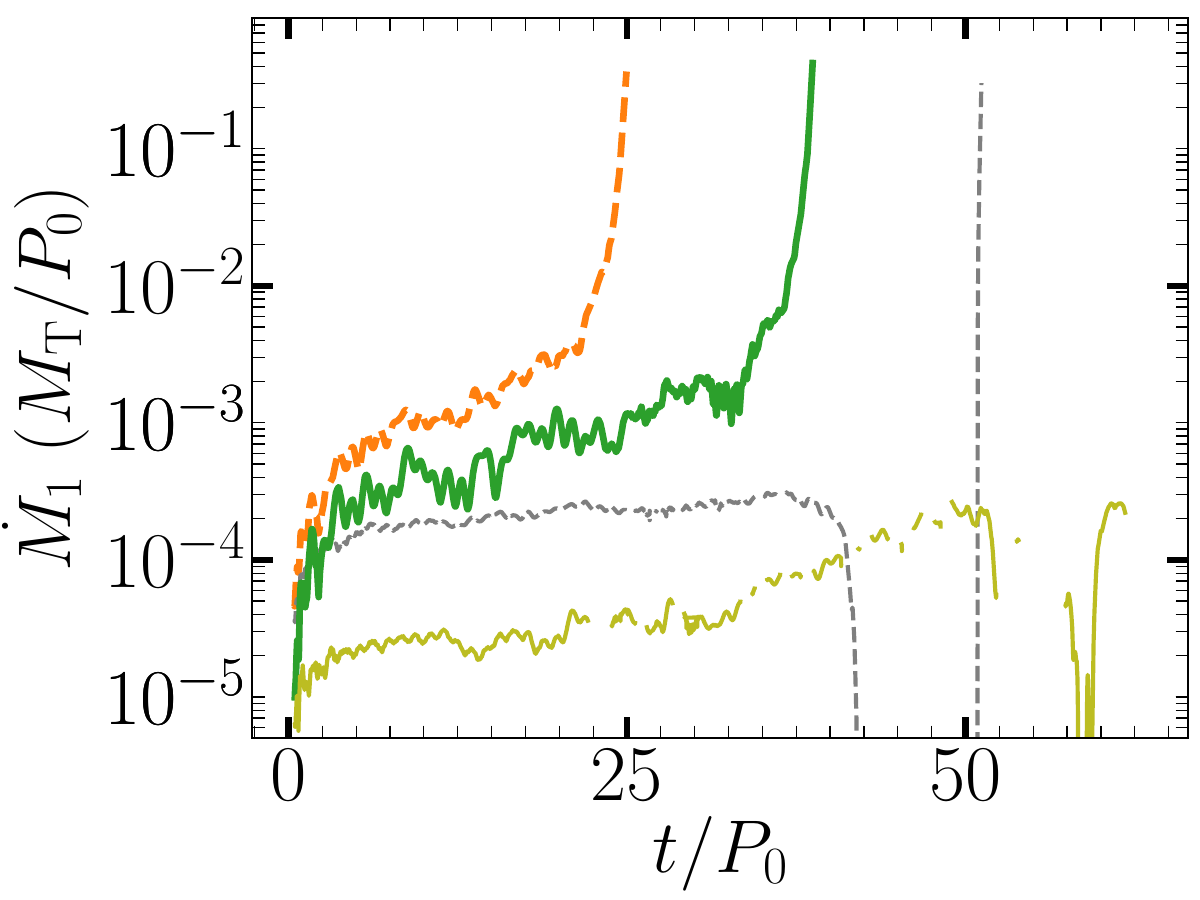}
    \includegraphics[scale=0.33]{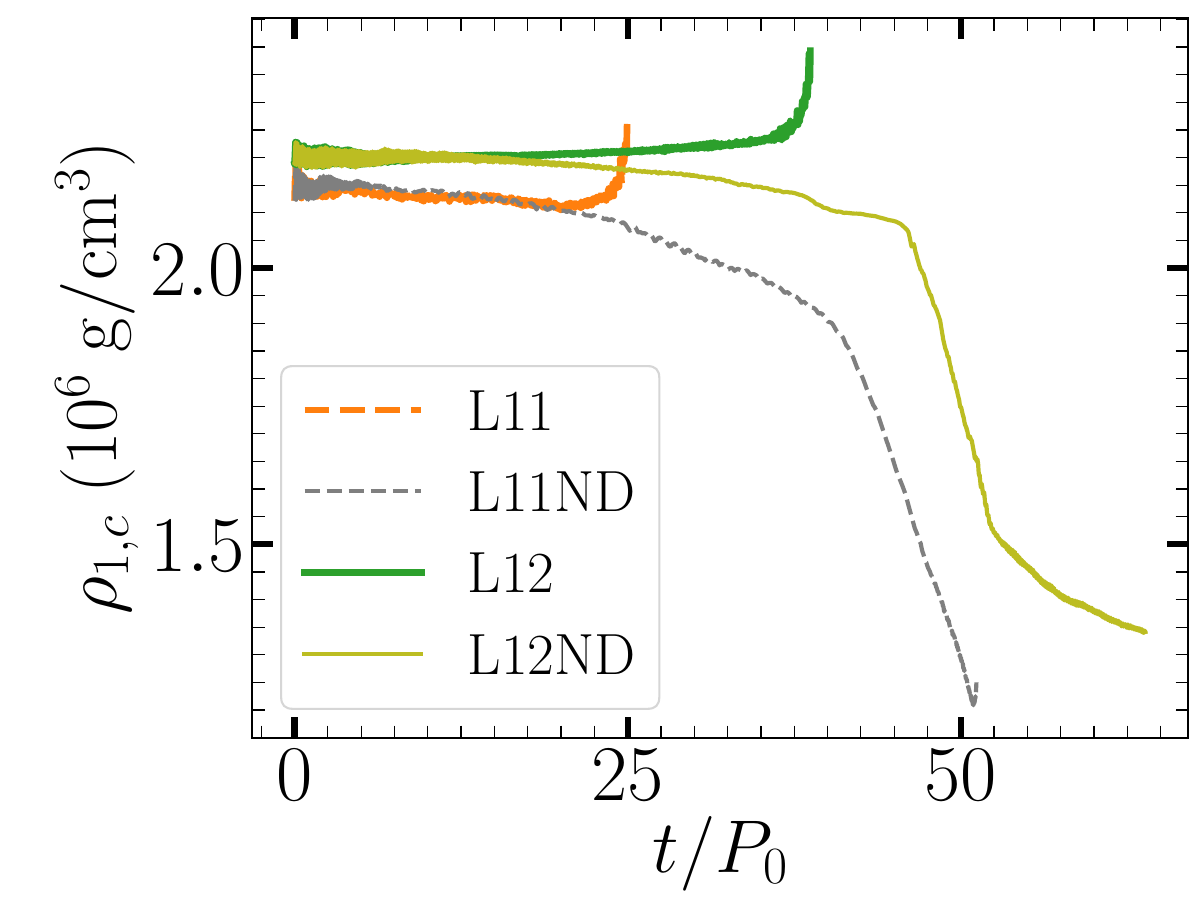}
    \includegraphics[scale=0.33]{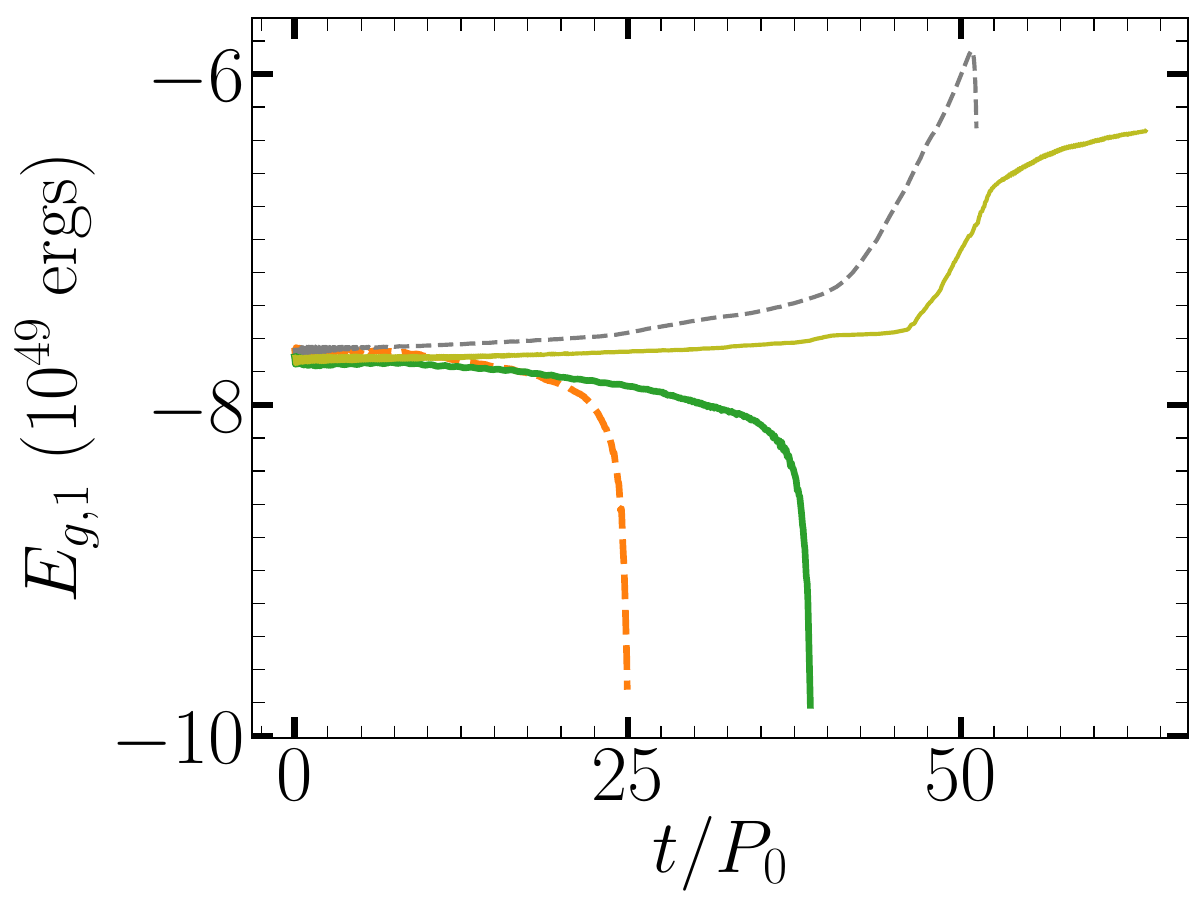}
    \caption{Comparison between the non-driven simulations {\sc L11ND}, and {\sc L12ND} and the driven simulations {\sc L11}, and {\sc L12} 
    }
    \label{fig:comp_nd}
\end{figure*}

At first glance, the evolution of the orbital angular momentum agrees very well and again shows remarkable resemblance, slowly declining through most of the evolution but then rapidly dropping just before the merger for simulations {\sc L11}, {\sc L11ND}, and {\sc L12} (we stopped {\sc L12ND} before it merged). Also, with no surprise, the non-driven {\sc L11ND} merged much later than the driven {\sc L11}.

However, in simulation {\sc L11ND}, the primary (a more massive star that should be the accretor) is donating mass to the secondary (less massive star) and the mass ratio grows above $0.7$ before rapidly decreasing right before merger. Likewise, the mass transfer rate between donor and accretor is positive for driven simulations with the mass accretion rate in {\sc L11ND}, {\sc L12ND} becoming negative. 

Furthermore, {\sc L11ND} and {\sc L12ND} are clearly anomalous with the central density of the primary declining (and the potential energy increasing) throughout the simulation. The accretor expands in these non-driven simulations, overflows its Roche lobe, and transfers mass to the secondary star, which ultimately triggers the merger. The expansion itself is caused by an instability, convection-like, that is forming inside the accretor. We stopped {\sc L12ND} the moment we witnessed the expansion to save computer time.

To isolate this problem we have tested whether we can reproduce this expansion in a single star simulation. In this test, we created a hydrostatic model of a white dwarf using the ZTWD EoS. We have run two resolutions, which correspond to the resolutions of the {\sc L10ND} and {\sc L11} binary simulations, with respect to the number of cells across the star's diameter. We fixed the star mass to the mass of the accretor and thus the star structure closely follows the initial accretor structure in our binary simulations. Like the stationary star test in \cite{Marcello2021}, the grid size was twice the size of the star. On low resolution, the star remains stationary on thousands of dynamical times (the initial orbital period, $P_0$, shown in the plots is $114~{\rm sec}$ and is equal to 33 dynamical times). This gives us further confidence in our {\sc L10ND} simulation, where such an expansion was not observed. However, on the {\sc L11} equivalent resolution (64 cells across), the same instabilities that developed in the binary simulations {\sc L11ND}, and {\sc L12ND} are observed.

\begin{figure*}
    \centering
    \includegraphics[scale=0.2]{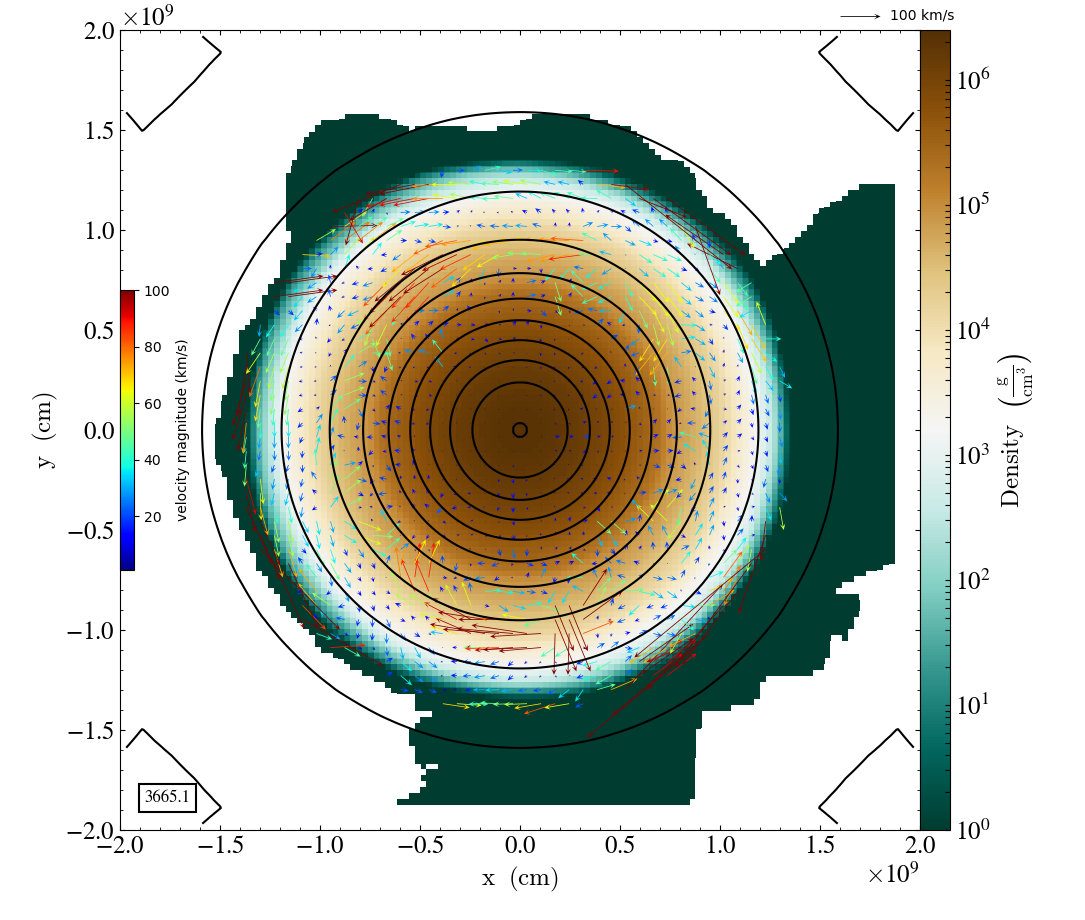}
    \includegraphics[scale=0.2]{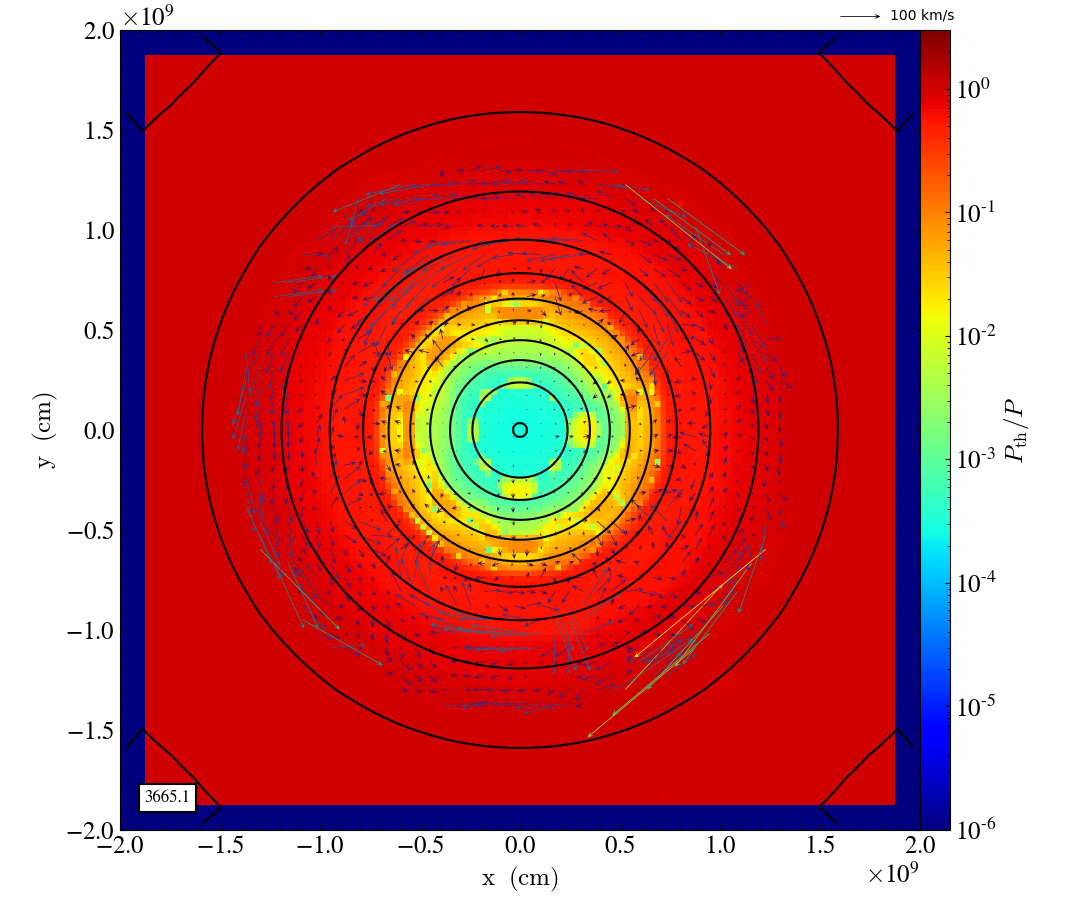}
    \includegraphics[scale=0.2]{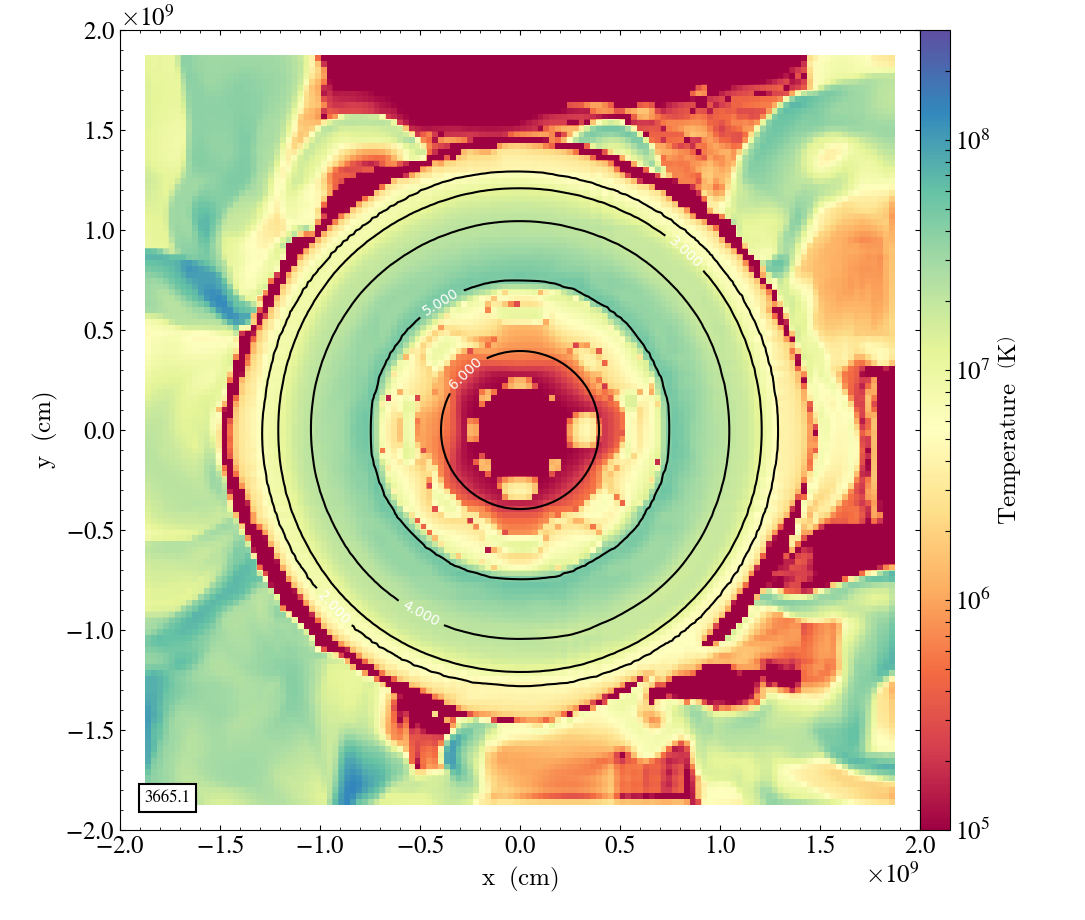}
    \caption{Internal convection-like instabilities in a single star simulation. Density (left), thermal pressure divided by total pressure (center), and temperature (right) slices along the equatorial plane are plotted. We show equipotential contours and velocity arrows on the left and center panels, while on the right panel, we show density contours. Hot bubbles that form inside the stars buoy outward, disrupting the stability of the star}
    \label{fig:single_ztwd_bubbles}
\end{figure*}

We illustrate these internal convection-like instabilities in Figure~\ref{fig:single_ztwd_bubbles}. We plot density (left), thermal pressure divided by total pressure (center), and temperature (right) at the time they begin to form, at approximately $32P_0$. Each panel shows a slice through the equatorial plane from the single star simulation where we resolve 64 cells across the star's diameter and where the thermal energy is computed according to the dual-energy formalism (DEF; Eq.~\ref{eq:def_th}). On the most left and central panels, we also plot equi-potential contours and velocity arrows, while on the right panel, we plot density contours.
Bubbles of hotter temperature than their surroundings are clearly seen in the interiors of the star at radii of $\sim 0.2-0.5\times 10^{9}~{\rm cm}$. In these bubbles, the thermal pressure is also higher compared to the thermal pressure of their surroundings (the {\it total pressure} is not that different though). Those bubbles tend to rise outward in a convection like motion, disrupting the structure of the star. As a consequence, instabilities are formed along the star surface, which results in a complex flow. We found out that this behavior precedes the sudden rapid expansion of the star.

\begin{figure*}
    \centering
    \includegraphics[scale=0.28]{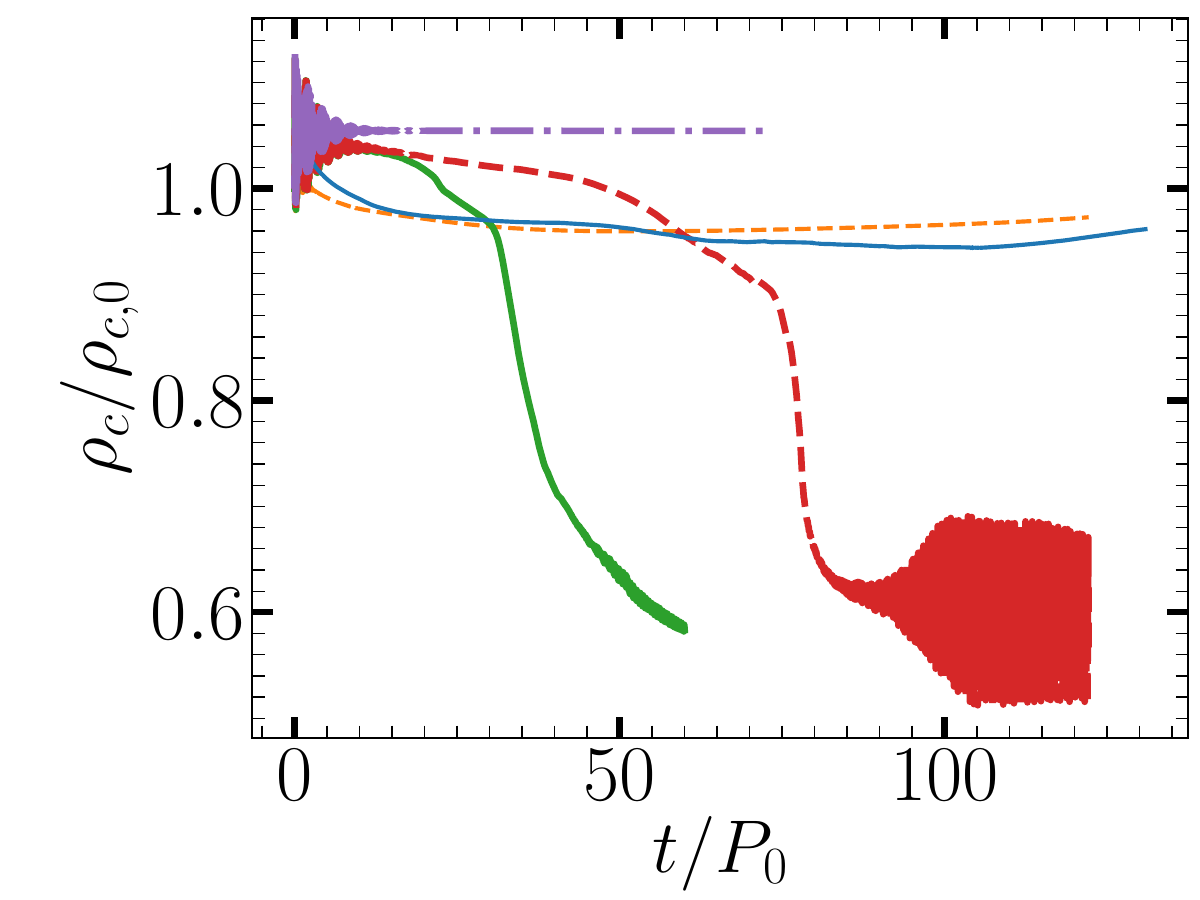}
    \includegraphics[scale=0.28]{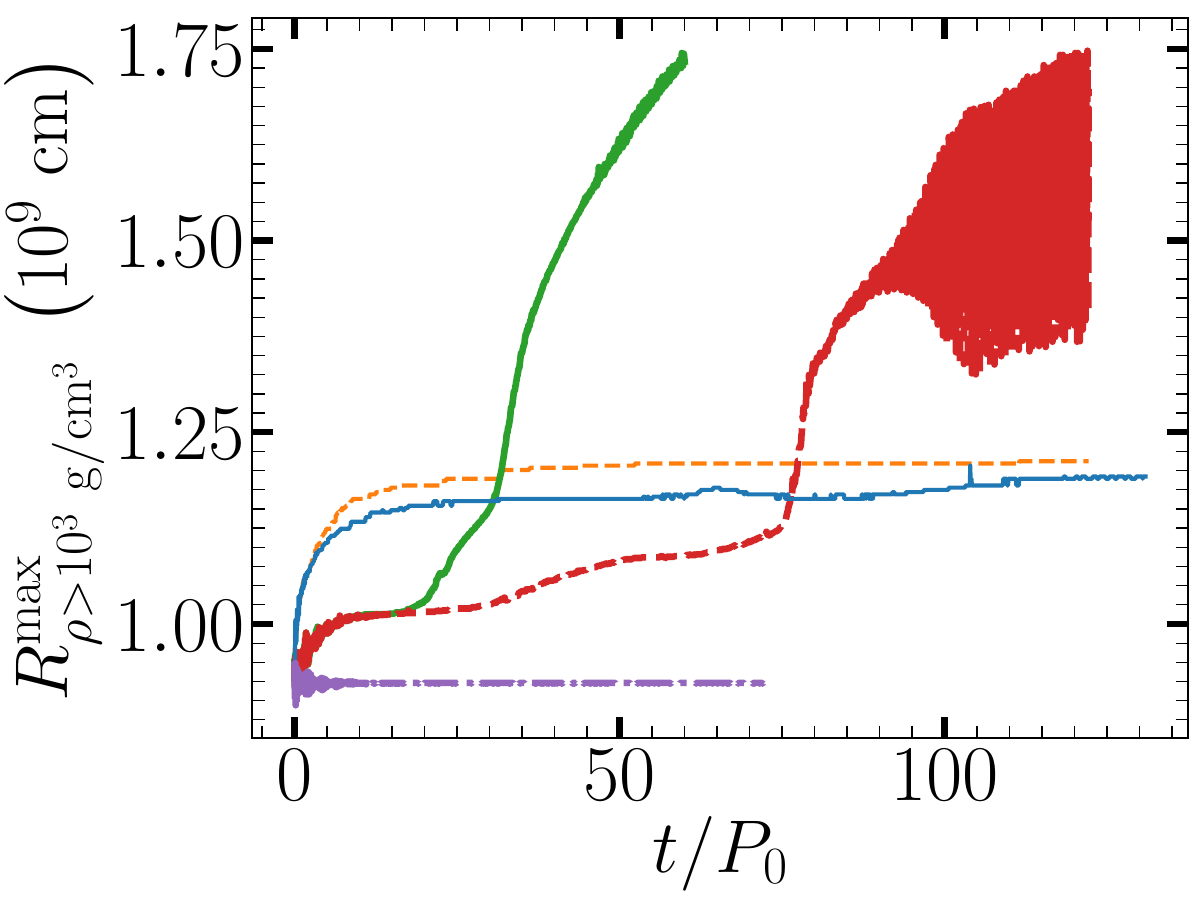}
    \includegraphics[scale=0.28]{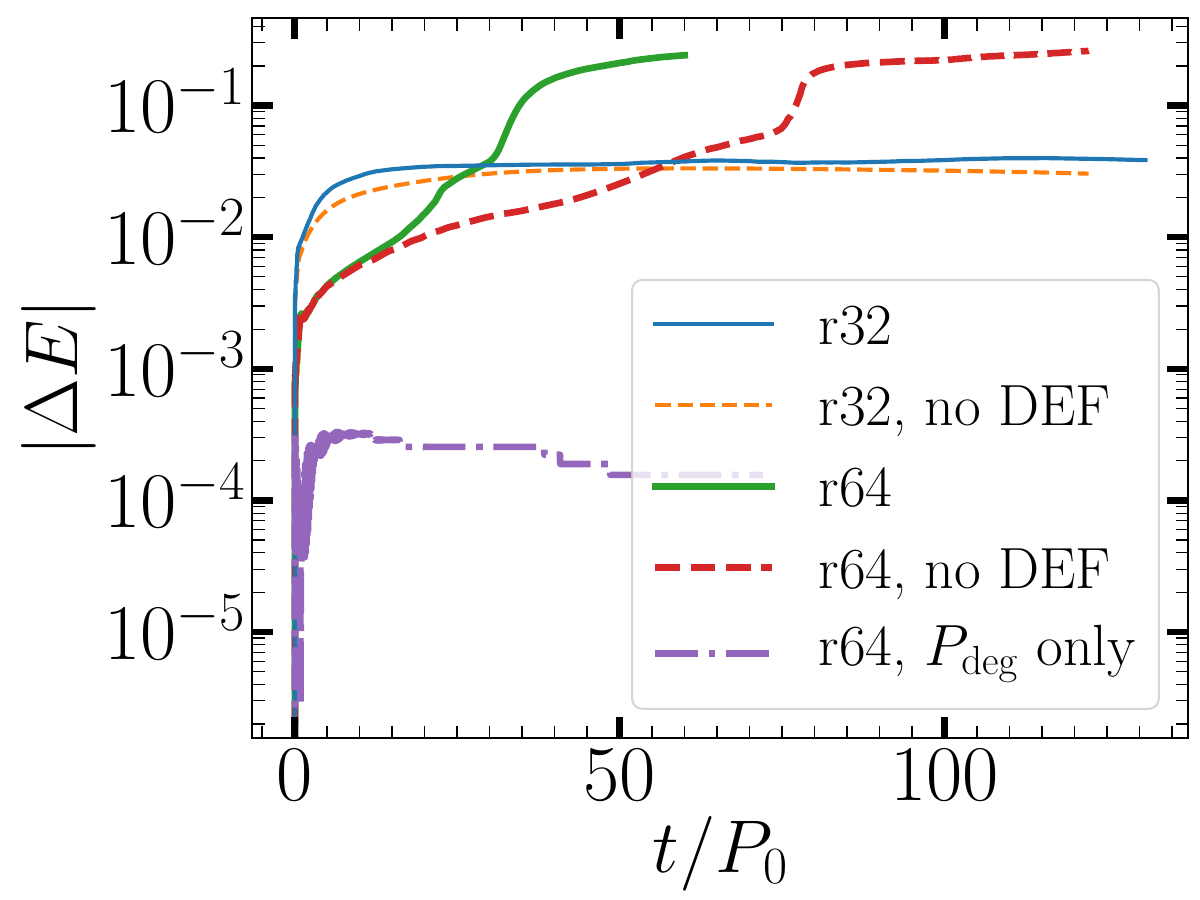}
    \caption{Stability tests of a single star with a ZTWD EoS. From left to right: The central density, the maximal radius of a density contour equal to $10^3~{\rm g/cm^3}$, and the relative energy error (see text) as a function of time. On the higher resolution, using the dual-energy formalism (DEF), the star starts to expand due to internal, convection-like instabilities, at about 40 initial periods $P_0$ of the $q=0.7$ binary (green solid thick line).  Disabling the DEF prolongs the stability of the star and postpones the sudden decrease in the central density to nearly 80~$P_0$ (or $\sim$2600 dynamical times of the single star; red dashed thick line). }
    \label{fig:single_ztwd_test}
\end{figure*}

In Figure \ref{fig:single_ztwd_test} we plot the central density (left), the maximal radial distance to the isopycnic surface $\rho=10^3~{\rm g/cm^3}$ (middle), and the deviation from conservation of energy (right): $\Delta E = \left( E(t) - E(0)\right) / E(0)$ as a function of time, for several runs using two spatial resolutions with and without DEF. Here $E(t) = \int\left( E_{\rm deg} + E_{\rm th} + E_{\rm kin} + \Phi\right) dV$ over the entire grid. In contrast with Figure~\ref{fig:q07_level12_rot_cons}, here we do not include outflows of energy in the calculation. Along with the effects of resolution and the usage of the DEF, we ran a case where only degenerate pressure is included (the thermal pressure is zero and the total pressure equal to the degenerate pressure).

The \flower\ method, (without the DEF; \emph{i.e.}, Eq. \ref{eq:flower_th}), maintains the star structure for longer, postponing the instabilities to a much later phase in the evolution. However, eventually, the star expands and the total energy deviation grows to 30 percent. The instabilities are completely avoided in the simulation that includes degenerate pressure only and the star remains intact throughout the evolution. This further confirms the role of the convection-like instabilities in the expansion and the decrease in the central density. It also pinpoints the problem of the way we add the contribution of the thermal pressure to the degenerate pressure. A similar stationary test with an ideal gas pressure only does not show the same instability (see the stationary star test in \citealt{Marcello2021}).
Moreover, an equivalent run performed with \flower\, including both degenerate and thermal pressure contributions, experiences the same convection-like instability, although much later, and results in the expansion of the star and the decrease of the central density. It is possible that using a Helmholtz EoS, which takes into account the varying degeneracy level of the electrons, rather than the ZTWD which only assumes a full degenerate electron pressure with thermal ions could resolve this problem. However, that is outside the scope of the current paper. In addition, it is important to note that previous grid-based simulations using the ZTWD (like in \citealt{Staff2012} or \citealt{Staff2018}), have never been run long enough to reveal this phenomenon. 

\section{The effect of disabling the dual energy formalism}
\label{app-def}

To check the effect of using the dual-energy formalism (DEF) on the short-driven simulations we performed a simulation where the DEF is disabled. The thermal energy in a cell is calculated then by subtracting the kinetic plus degenerate energies from the gas energy while forcing it to be non-negative, 
\begin{equation}
    E_{\rm th} = {\rm max} \left( E - E_{\rm deg} - E_{\rm kin},0 \right).
    \label{eq:flower_th}
\end{equation}

\begin{figure*}
\centering   
    \subfloat[]{\includegraphics[scale=0.28]{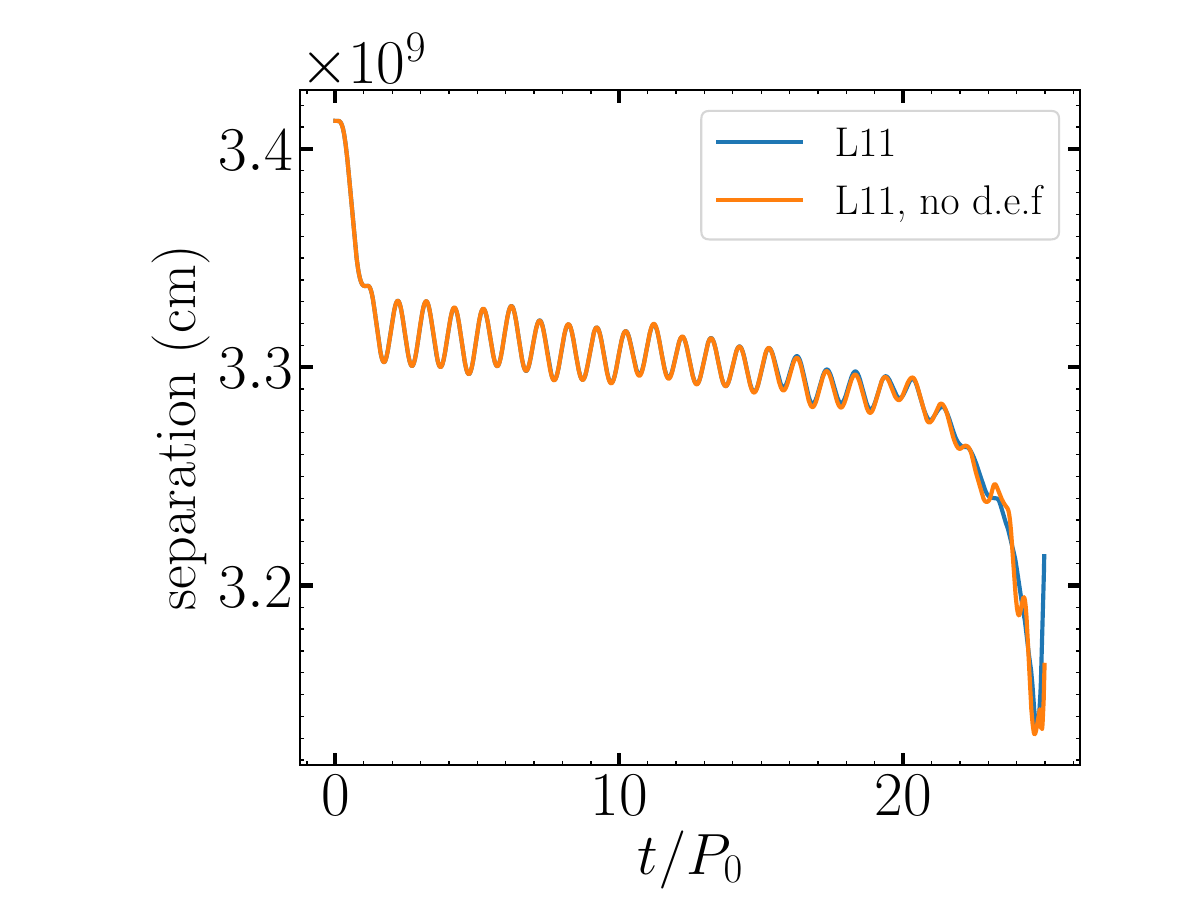}}    
    \subfloat[]{\includegraphics[scale=0.28]{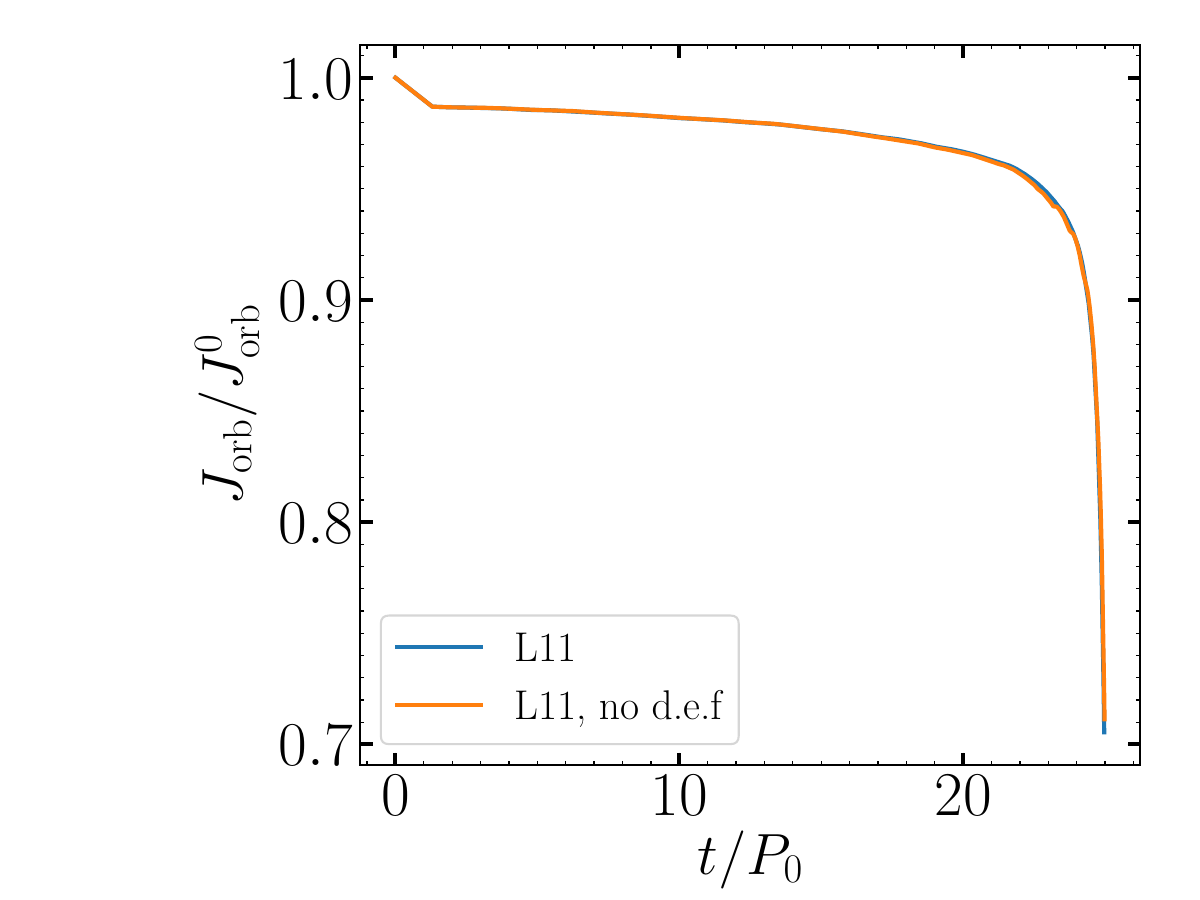}}%
    \subfloat[]{\includegraphics[scale=0.28]{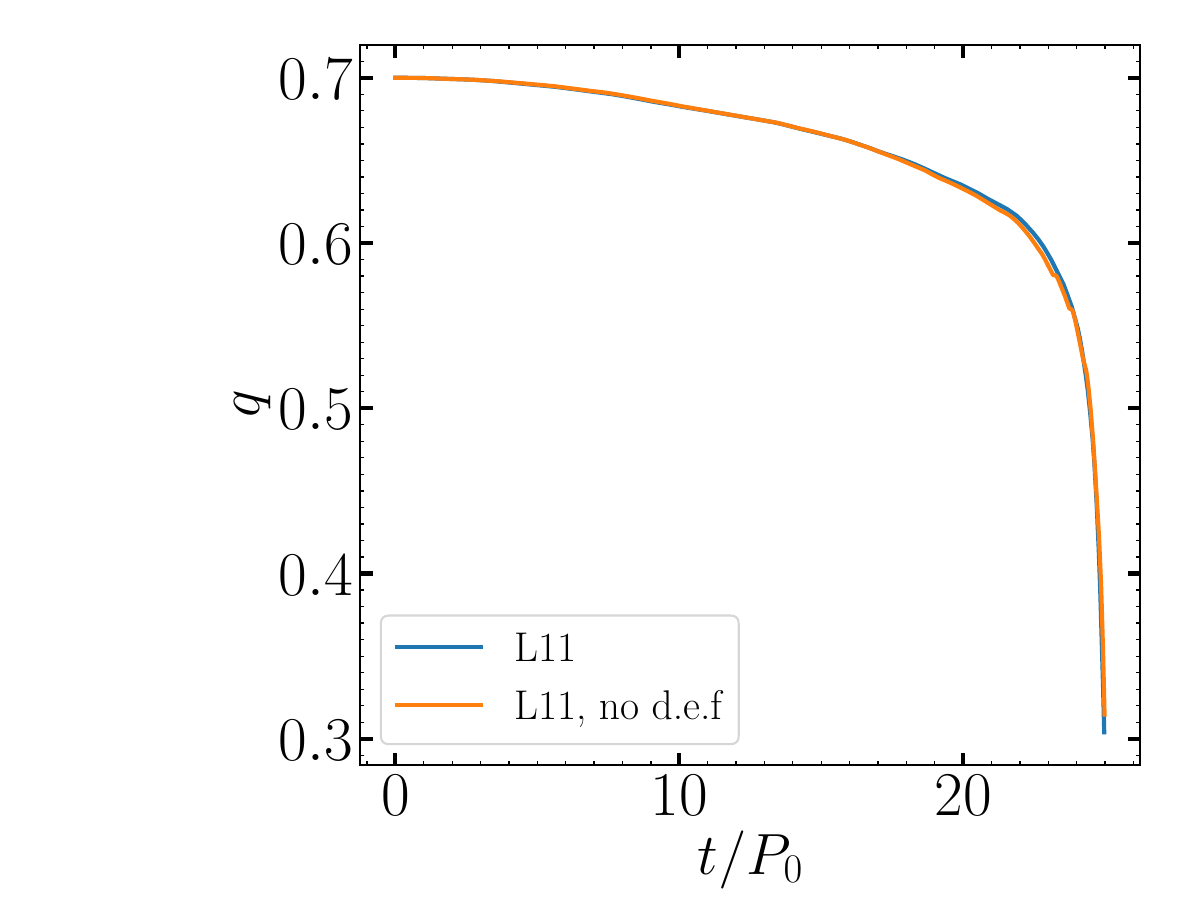}}
        \qquad        
    \subfloat[]{\includegraphics[scale=0.28]{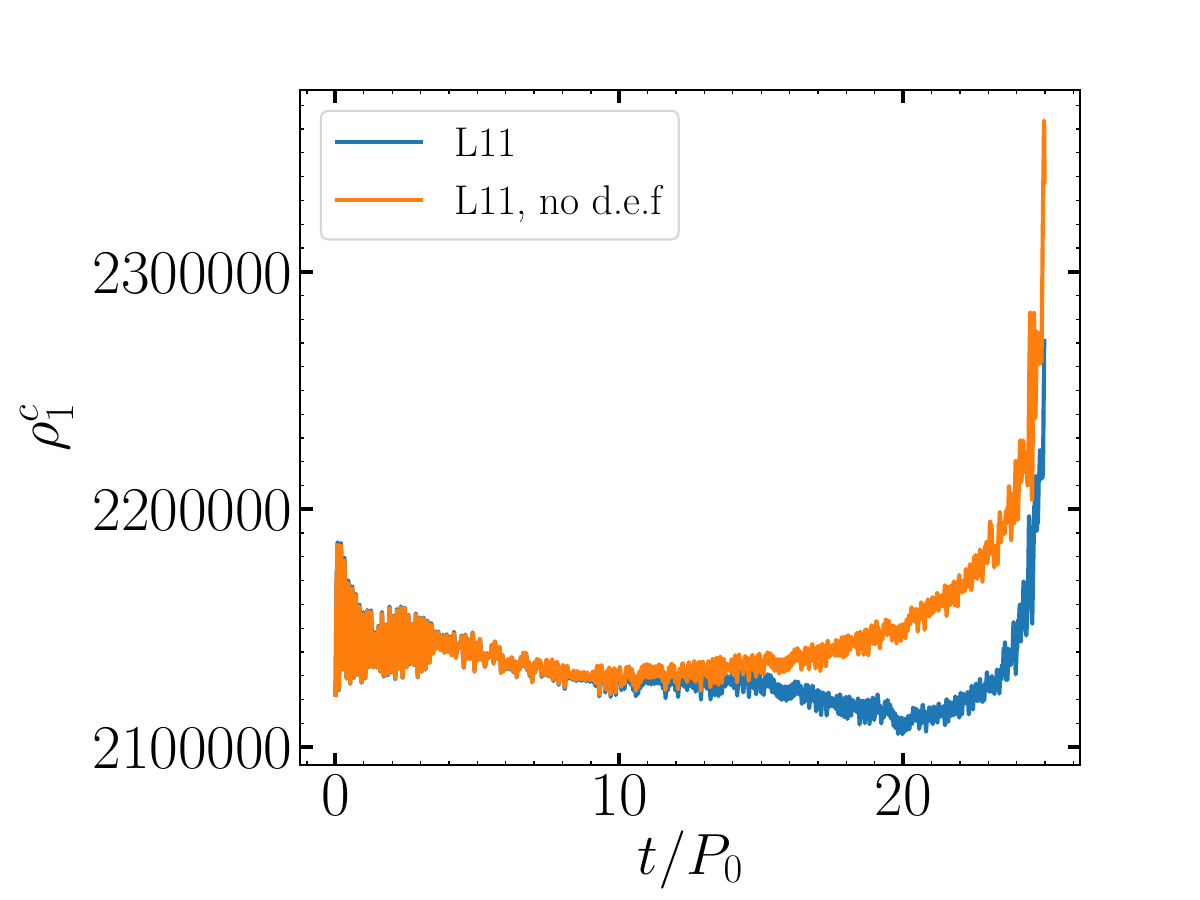}}    
    \subfloat[]{\includegraphics[scale=0.28]{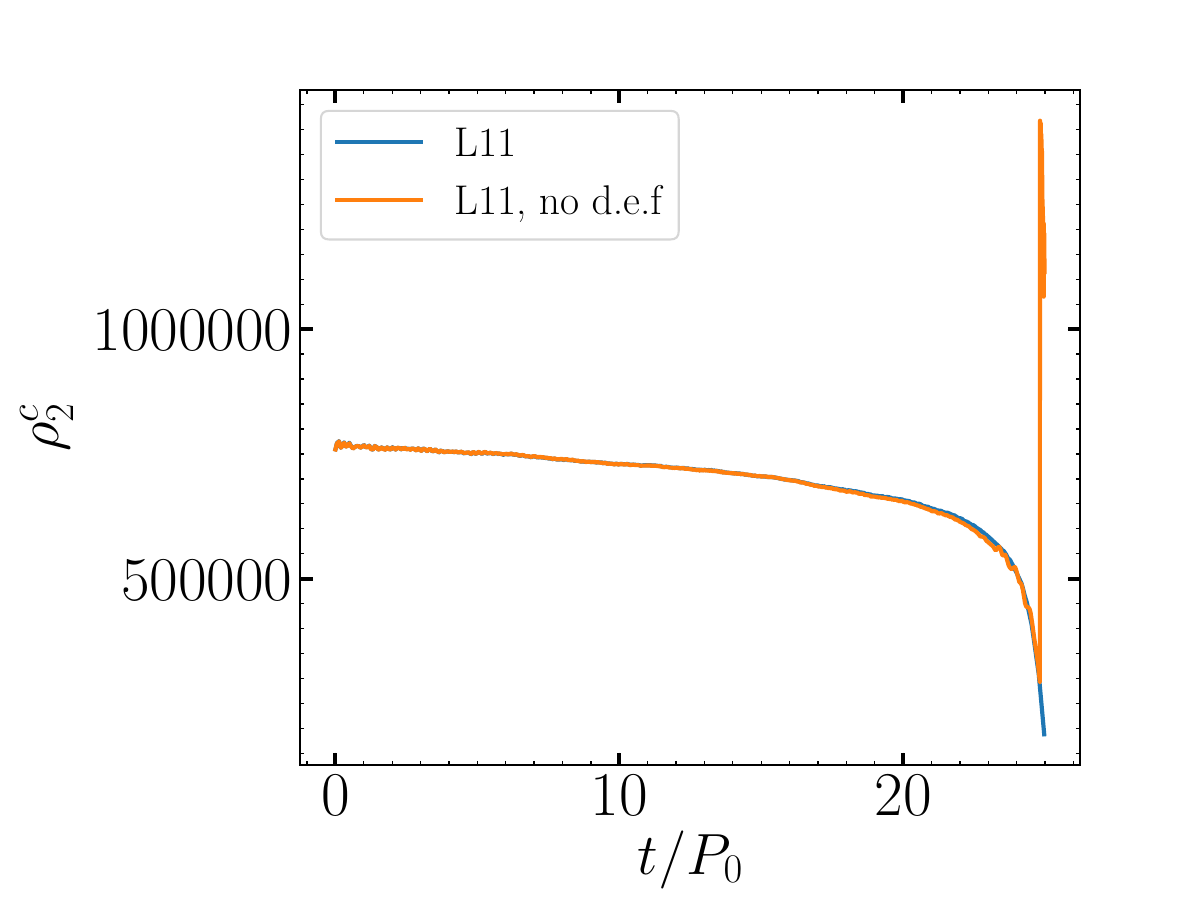}}%
    \subfloat[]{\includegraphics[scale=0.28]{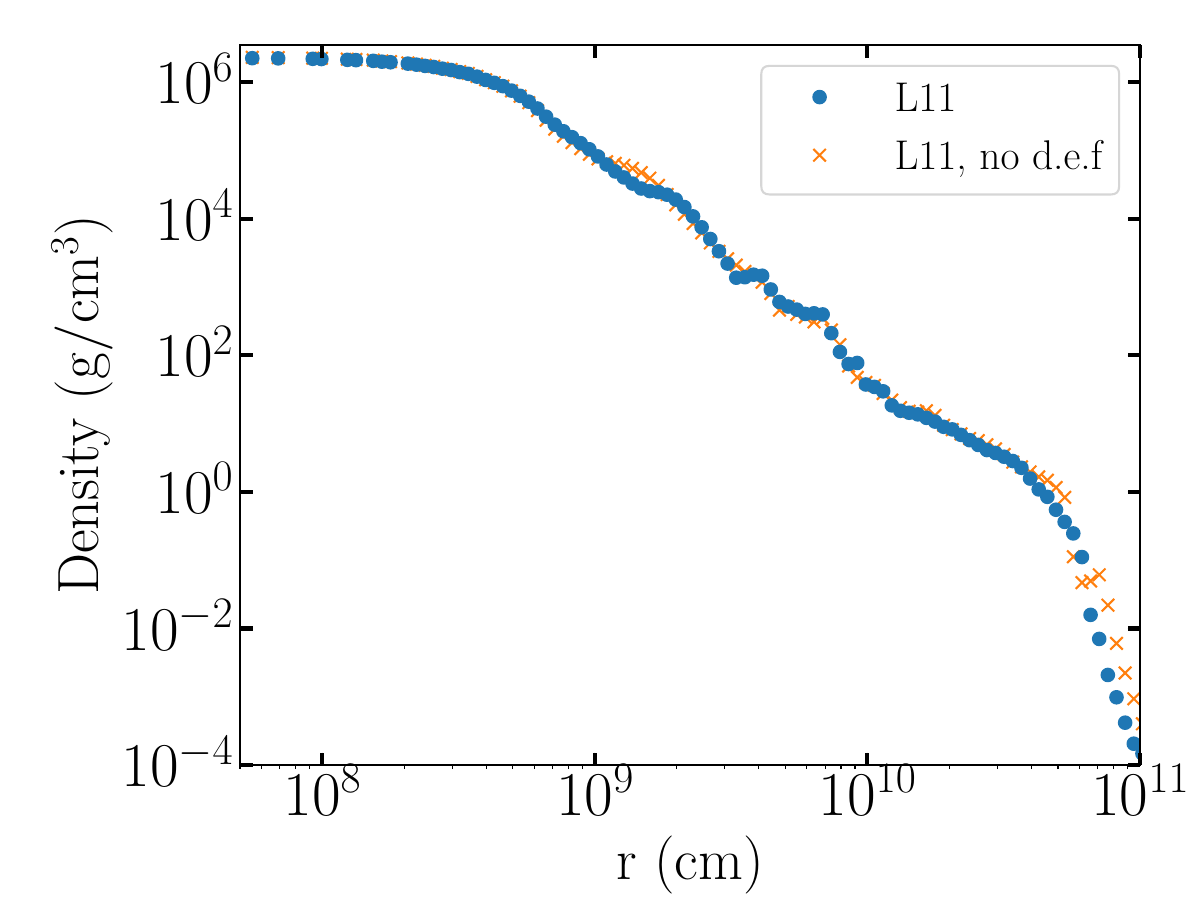}}  
        \qquad
    \subfloat[]{\includegraphics[scale=0.28]{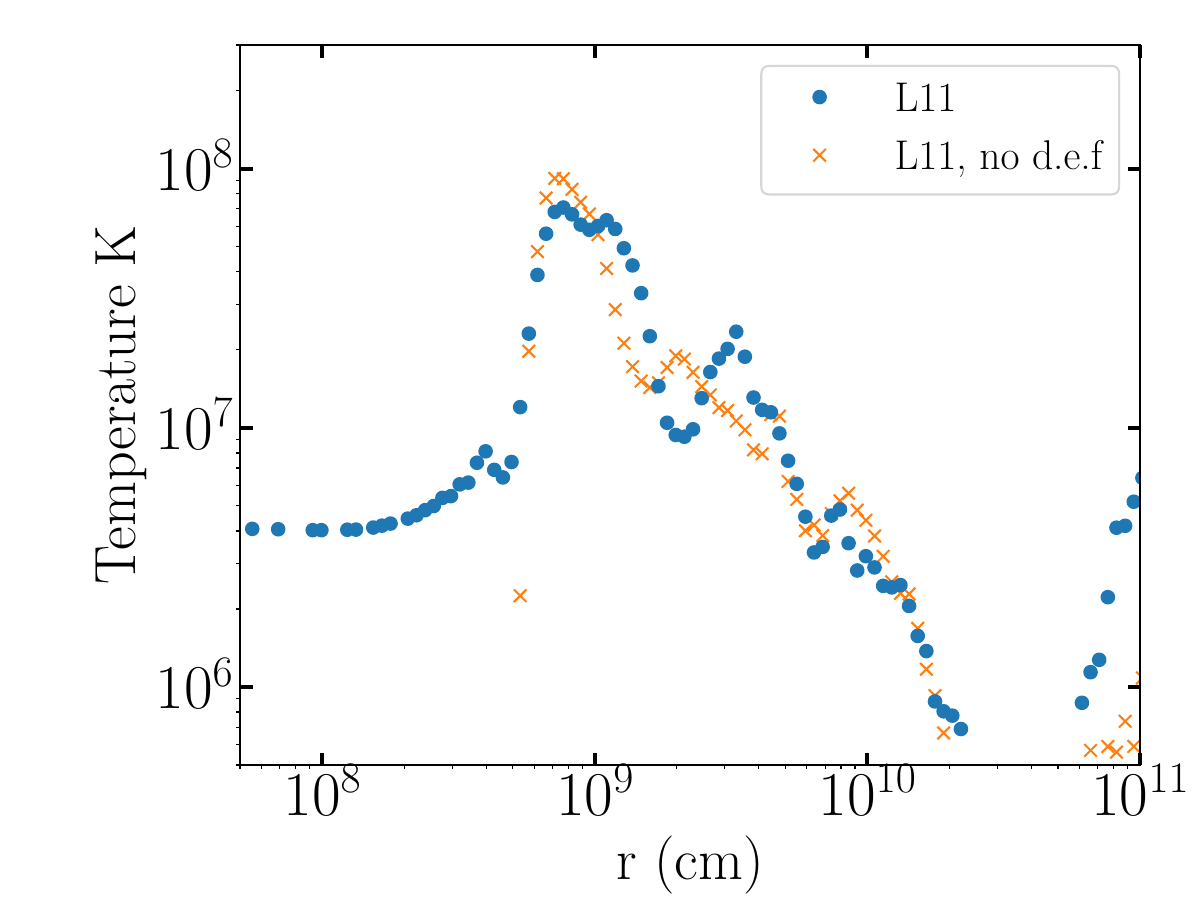}}    
    \subfloat[]{\includegraphics[scale=0.28]{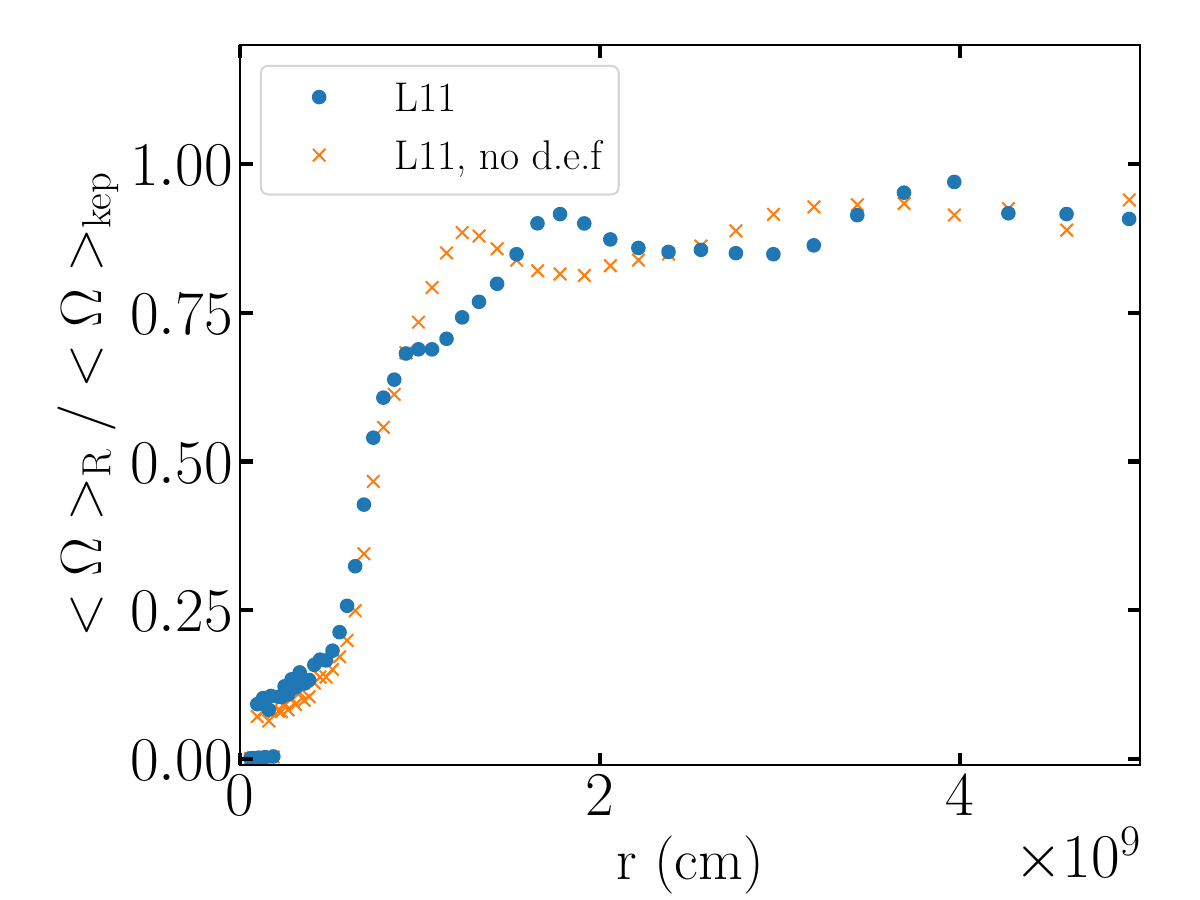}}%
    \subfloat[]{\includegraphics[scale=0.28]{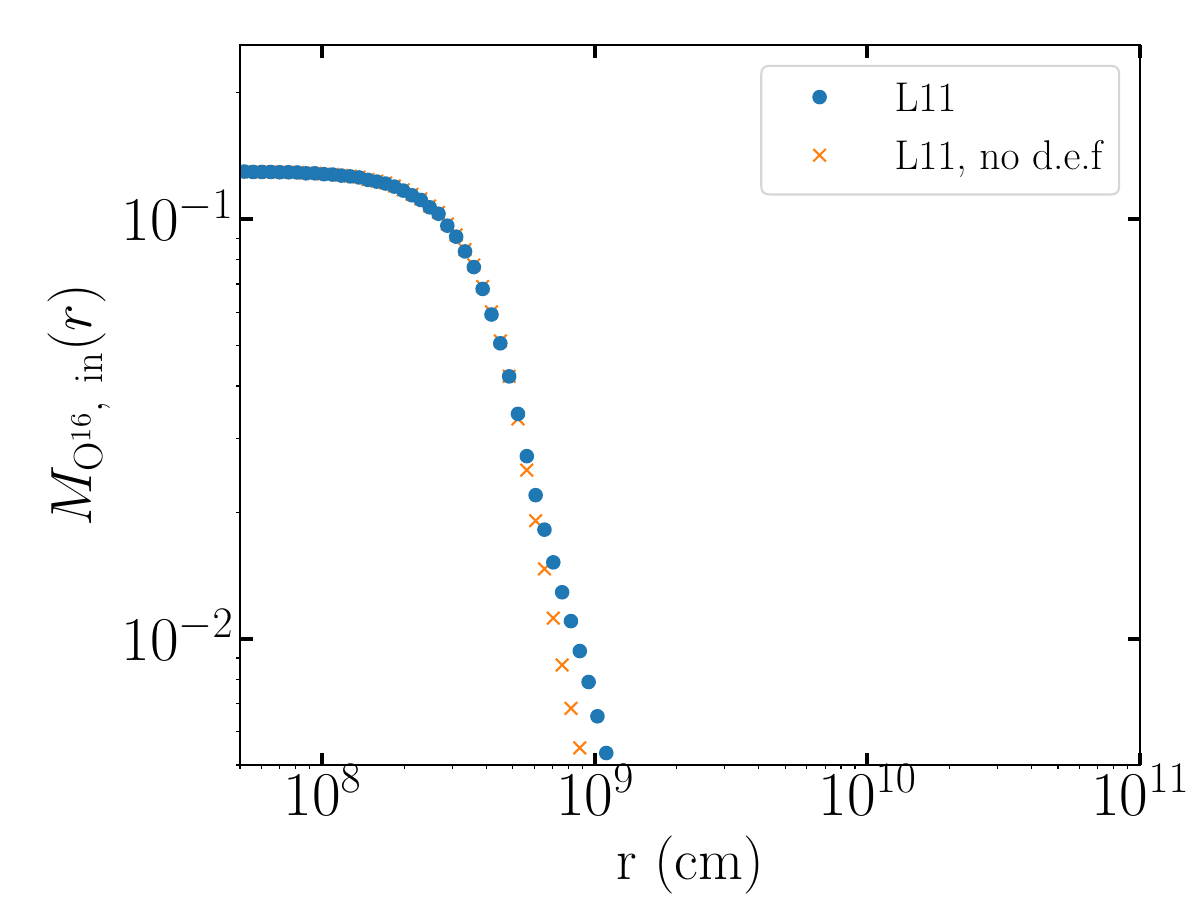}}
    \caption{Comparing two driven simulations with the same resolution but with different treatments for the derivation of the internal (thermal) energy. {\sc L11} is the Simulation shown throughout the paper in which the thermal energy is calculated according to Eq. \ref{eq:def_th}. We use this treatment in all the simulations shown in this paper. In the other simulation, we used instead Eq. \ref{eq:flower_th}, similar to what is being performed in \flower. Only very minor differences exist which indicate that the specific temperature prescription does not affect much the simulations.
     }
\label{fig:nodual_comp}
\end{figure*}

In this simpler treatment, the entropy tracer is disregarded and never used to calculate the thermal energy. This approach is used in all the \flower\ simulations shown in this paper.
We have run an equivalent of the {\sc L11} driven simulation just with the \flower\ approach (Eq. \ref{eq:flower_th}) and we demonstrate in Fig. \ref{fig:nodual_comp} that these simulations appear to be very similar. 

First, we note that the orbital properties of the system evolve with time almost identically in both of the simulations. The orbital separation (a), angular momentum (b), and the mass ratio (hence the mass-transfer rate) (c) evolve as described in \ref{ssec:evolution} and behave as expected, with their respective curves falling on top of each other most of the time. The donor central density (e) also behaves similarly, indicating that the donor star structure is not affected by the DEF treatment being used. The most notable difference is in the accretor's central density (d), which decreases more in the simulation where the DEF is used. This results in a slightly less dense core at the time of the merger. However, the averaged density structure of the merger five orbits after the merger (f), almost converges, with a small bump at $\sim~10^9~{\rm cm}$ as the only difference. The temperature and rotation profiles, (g) and (h), are also very close besides small variations. Finally, the accretor expands slightly further prior to the merger in the simulation with the DEF. This results in a little more diffusion of Oxygen-16 out (i) displaces the first peak in angular velocity slightly outward (h), although, overall, the effect is marginal.

\section{Scaling and code performance using NVIDIA GPUs}
\label{app-scaling}

\octo's support for GPU acceleration within the C\texttt{++} standard library for parallelism and concurrency (HPX)~\citep{kaiser2020hpx} was recently extended. 
Initially, GPU support was added (via CUDA) when \octo's gravity module was ported to the GPU and tested on CSCS's Piz Daint in \citep{10.1145/3295500.3356221}.
Recently, \octo's new hydro module using a higher order reconstruction was similarly ported and tested on ORNL's Summit in \citep{diehl2021octo}. For these simulations, we used HPX's native CUDA implementation. 

\begin{figure}
    \begin{tikzpicture}
    \begin{axis}[xlabel={\# nodes},ylabel={Processed sub-grids per second},title={},grid,anchor=north west,xmode=log,log basis x={2},xtick={1,2,4,8,16,32,64,128,256,512,1024,2048},ymode=log,log basis y={2},legend columns=3,legend style={at={(0.5,-0.2)},anchor=north}]
    \addplot[thick,mark=*,plot1] table [x expr=\thisrowno{0},y expr={7344*512/\thisrowno{1}/25}, col sep=comma] {perlmutter/level-10-gpu-close.csv};
    \addplot[thick,mark=*,plot2] table [x expr=\thisrowno{0},y expr={4796*512/\thisrowno{1}/25}, col sep=comma] {perlmutter/level-11-gpu-close.csv};
    \addplot[thick,mark=*,plot3] table [x expr=\thisrowno{0},y expr={10060*512/\thisrowno{1}/5}, col sep=comma] {perlmutter/level-12-gpu-close.csv};
    \addplot[thick,mark=square*,plot1] table [x expr=\thisrowno{0},y expr={7344*512/\thisrowno{1}/25}, col sep=comma] {perlmutter/level-10-cpu-close.csv};
    \addplot[thick,mark=square*,plot2] table [x expr=\thisrowno{0},y expr={4796*512/\thisrowno{1}/25}, col sep=comma] {perlmutter/level-11-cpu-close.csv};
    \addplot[thick,mark=square*,plot3] table [x expr=\thisrowno{0},y expr={10060*512/\thisrowno{1}/5}, col sep=comma] {perlmutter/level-12-cpu-close.csv};
    
    \legend{Level 10 (GPU), Level 11 (GPU), Level 12 (GPU),Level 10 (CPU), Level 11 (CPU), Level 12 (CPU)};
    \end{axis}
    \end{tikzpicture}
    
    \caption{Scaling on NERSC's Perlmutter for simulations for runs using CPU only and using CPU and GPU. All runs started on a single node, and we increased the number of nodes until we did not do sufficient work and the scaling flattened out. However, level 12 only fitted on eight nodes. The configuration on Perlmutter was as follows: We used four localities with 16 cores per locality for the CPU-only run. One NVIDIA\textsuperscript{\textregistered} A100 was added to each locality for the combined CPU and GPU runs.}
    \label{fig:scaling:summit:close}
\end{figure}
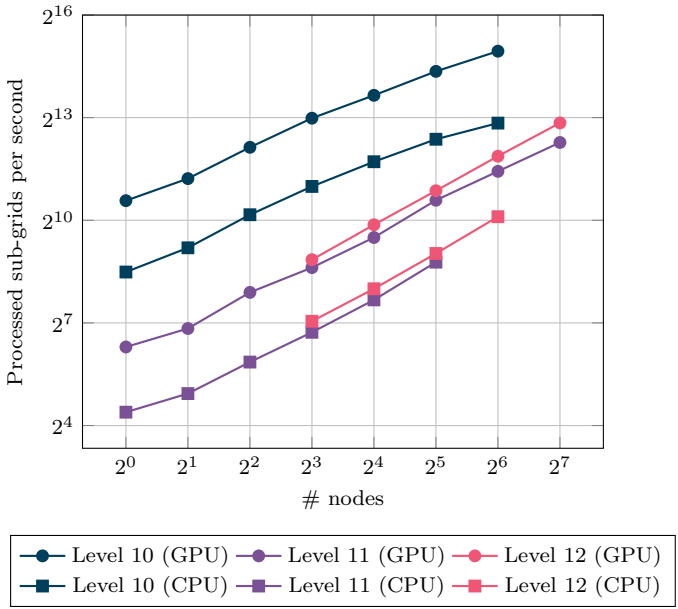

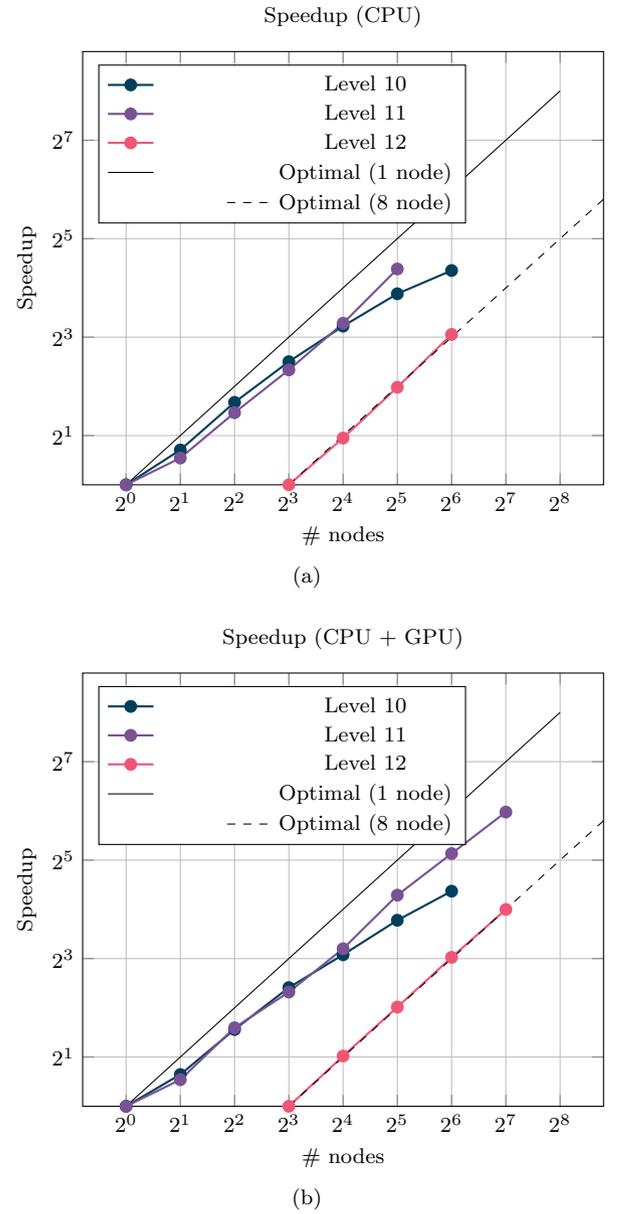
\begin{figure}
    \centering
    \subfloat[\label{fig:speedup:summit:close:cpu}]{
    \begin{tikzpicture}
    \begin{axis}[xlabel={\# nodes},ylabel={Speedup},title={Speedup (CPU)},grid,legend pos=north west,xmode=log,log basis x={2},xtick={1,2,4,8,16,32,64,128,256,512,1024,2048},ymode=log,log basis y={2},legend columns=1,ymin=1]
    \addplot[thick,mark=*,plot1] table [x expr=\thisrowno{0},y expr={419.563/\thisrowno{1}}, col sep=comma] {perlmutter/level-10-cpu-close.csv};
    \addplot[thick,mark=*,plot2] table [x expr=\thisrowno{0},y expr={4683.77/\thisrowno{1}}, col sep=comma] {perlmutter/level-11-cpu-close.csv};
    \addplot[thick,mark=*,plot3] table [x expr=\thisrowno{0},y expr={7790.52/\thisrowno{1}}, col sep=comma] {perlmutter/level-12-cpu-close.csv};
    \addplot[domain=1:256]{x};
    \addplot [domain=1:128,dashed,shift={(axis cs:8,1)},legend image post style={shift={(0,0)}}]{x};
    \legend{Level 10, Level 11, Level 12, Optimal (1 node), Optimal (8 node)};
    \end{axis}
    \end{tikzpicture}
    }
    \\
    \subfloat[\label{fig:speedup:summit:close:gpu}]{
    \begin{tikzpicture}
    \begin{axis}[xlabel={\# nodes},ylabel={Speedup},title={Speedup (CPU + GPU)},grid,legend pos=north west,xmode=log,log basis x={2},xtick={1,2,4,8,16,32,64,128,256,512,1024,2048},ymode=log,log basis y={2},legend columns=1,ymin=1]
    \addplot[thick,mark=*,plot1] table [x expr=\thisrowno{0},y expr={98.8544/\thisrowno{1}}, col sep=comma] {perlmutter/level-10-gpu-close.csv};
    \addplot[thick,mark=*,plot2] table [x expr=\thisrowno{0},y expr={1250.51/\thisrowno{1}}, col sep=comma] {perlmutter/level-11-gpu-close.csv};
    \addplot[thick,mark=*,plot3] table [x expr=\thisrowno{0},y expr={2238.85/\thisrowno{1}}, col sep=comma] {perlmutter/level-12-gpu-close.csv};
    \addplot[domain=1:256]{x};
     \addplot [domain=1:128,dashed,shift={(axis cs:8,1)},legend image post style={shift={(0,0)}}]{x};
    \legend{Level 10, Level 11, Level 12,  Optimal (1 node), Optimal (8 node)};
    \end{axis}
    \end{tikzpicture}
    }
    \caption{Speedup for the runs from close to the merger on NERSC's Perlmutter from one to 128 nodes. In \protect\subref{fig:speedup:summit:close:cpu} only the CPUs were used, and in \protect\subref{fig:speedup:summit:close:gpu} the NVIDIA\textsuperscript{\textregistered} A100 GPUs were added. Note that in the Phase 1, the maximal number of nodes was restricted to 128.}
    \label{fig:speedup:summit:close}
\end{figure}

Now, to investigate the speed-up brought by the NVIDIA A100 GPUs on NERSC's Perlmutter, we executed our simulations shortly before the merger. Thus, we have to have a very refined irregular grid. Each time, the code was run for 25 time steps for level 10 and level 11. For level 12, the runs on a single node without GPUs got very time-consuming, and we reduced the time steps from 25 to five time steps. Table~\ref{tab:summit:level} shows the number of cells and memory usage of each of the simulations. Figure~\ref{fig:scaling:summit:close} shows the scaling from close to the merger for 25 time steps. Figure~\ref{fig:speedup:summit:close} shows the corresponding speedup. 
We used four A100 GPUs per node and for each GPU, we used one MPI rank with 16 cores assigned. So in total, we used all 64 cores per node. Tables~\ref{tab:toolchain:summit} and~\ref{tab:hardware:perlmutter} show the software version and Perlmutter's hardware, respectively. The dependency cppuddle\footnote{\url{https://github.com/SC-SGS/CPPuddle}} is available on GitHub.

With the ongoing heterogeneity of acceleration cards in the latest supercomputers, \emph{e.g.}\ AMD GPUs in ORNL's Frontier, NVIDIA GPUs in NERSC's Perlmutter, and hopefully soon Intel GPUs in Aurora, we prepared \octo\ to support Kokkos~\citep{edwards2014kokkos} to target various GPU vendors. The Kokkos integration~\citep{daiss2021beyond} was tested on Riken's Supercomputer\ Fugaku~\citep{10196612} and Stony Brook's Ookami using Arm A64FX CPUS with SVE vectorization~\citep{daiss2022a}. Recently, SYCL support was added to Octo-Tiger to target Intel GPUs~\citep{daiss2023stellar}. Additionally, support for dynamic GPU work aggregation was added in~\cite{daiss2022b} to improve the GPU performance further.
These improvements and additions prepare \octo\ for a diverse set of acceleration cards. Significantly, the improved speedup using GPUs is promising since we have shown in this paper that extensive simulations are needed for future research. 

\begin{table}
    \centering
    \begin{tabular}{ccc}
    \toprule
     Level & Number of cells & Memory \\\midrule
     {\sc L10ND} & 3.8M & 11 GB \\
     {\sc L11ND} & 40.2M & 113 GB \\
     {\sc L12} & 257.3M & 724 GB \\\bottomrule
    \end{tabular}
    \caption{Number of cells and memory usage for simulations {\sc L10ND}, {\sc L11ND}, and {\sc L12} at the starting point of our scaling runs. We use a time earlier to the merger, when massive regridding takes place, as our starting point.}
    \label{tab:summit:level}
\end{table}

\begin{table}
    \centering
    \caption{Toolchain and \octo\'s dependencies on Perlmutter. }
    \begin{tabular}{ll|ll}\toprule
         gcc & 9.3.0 & hwloc & 1.11.12  \\
         cray-mpich & 8.1.11  & boost & 1.77.0  \\
         CUDA\textsuperscript{\texttrademark} & 11.4.0 & jemalloc & 5.1.0  \\
         hpx & 1.7.1 & silo & 4.10.2 \\
         hdf5 & 1.8.12 & cppuddle & \texttt{d32e50b} \\
      \bottomrule
    \end{tabular}
    \label{tab:toolchain:summit}
\end{table}

\begin{table}
    \centering
    \caption{NERSC's Perlmutter configuration}
    \begin{tabular}{ll}\toprule
     CPU    & AMD\textsuperscript{\textregistered} EPYC 7713 64-Core Processor\\
     GPU  &   4 $\times$ NVIDIA\textsuperscript{\textregistered} A100-PCIE-40GB  \\
     GPU driver & 450.162 \\
     Linux kernel  &  5.3.18 \\\bottomrule 
    \end{tabular}
    \label{tab:hardware:perlmutter}
\end{table}

\section*{Disclaimer}
The results on NERSC's Perlmutter were conducted in phase 1, and as such these results should not reflect or imply that they are the final results of the system. Numerous upgrades will be taking place for Phase 2 that will substantially change the final size and network capabilities of Perlmutter. 

\label{lastpage}

\end{document}